\documentclass[notitlepage,aps,prd,groupedaddress,nofootinbib]{revtex4-1}

\usepackage[sc]{mathpazo} 
\usepackage[T1]{fontenc} 
\usepackage{microtype} 

\usepackage[hmarginratio=1:1,top=32mm,columnsep=20pt]{geometry} 
\usepackage[hang, small,labelfont=bf,up,textfont=it,up]{caption} 
\usepackage{booktabs} 
\usepackage{float} 
\usepackage{hyperref} 
\usepackage{graphics}
\usepackage{graphicx}
\usepackage{float}
\usepackage{amssymb}
\usepackage{slashed}
\usepackage{fancybox}
\usepackage{amsfonts}
\usepackage{rotating}
\usepackage{amsmath}
\usepackage{amsmath}%
\usepackage{wasysym}%
\usepackage{braket}
\usepackage{lettrine} 
\usepackage{paralist} 
\usepackage[utf8]{inputenc}

\usepackage{fancyhdr} 
\allowdisplaybreaks

\begin{document}


\title{Comprehensive asymmetric dark matter model}


\author{Stephen J. Lonsdale}\email{lsj@student.unimelb.edu.au, corresponding author}
\author{Raymond R. Volkas}\email{raymondv@unimelb.edu.au}

\affiliation{ARC Centre of Excellence for Particle Physics at the Terascale, School of Physics,\\ The University of Melbourne, Victoria 3010, Australia}


                              \renewcommand \thesection{\arabic{section}}
\renewcommand \thesubsection{\arabic{section}.\arabic{subsection}}

\begin{abstract}
Asymmetric dark matter (ADM) is motivated by the similar cosmological mass densities measured for ordinary and dark matter.
We present a comprehensive theory for ADM that addresses the \emph{mass} density similarity, going beyond
the usual ADM explanations of similar \emph{number} densities. It features an explicit
matter-antimatter asymmetry generation mechanism, has one fully worked out thermal history and suggestions for other possibilities, and
meets all phenomenological, cosmological and astrophysical constraints. Importantly, it 
incorporates a deep reason for why the dark matter mass scale is related to the proton mass, a key
consideration in ADM models.
Our starting point is the idea of mirror matter, which offers an explanation for dark matter by duplicating
the standard model with a dark sector related by a $Z_2$ parity symmetry. 
However, the dark sector need not manifest as
a symmetric copy of the standard model in the present day. By utilising the mechanism of ``asymmetric 
symmetry breaking'' with two Higgs doublets in each sector, we develop a model of ADM
where the mirror symmetry is spontaneously broken, leading to an electroweak scale in the dark sector that is 
significantly larger than that of the visible sector. 
The weak sensitivity of the ordinary and dark QCD confinement scales to their respective electroweak scales leads to 
the necessary connection between the dark matter and proton masses. The dark matter
is composed of either dark neutrons or a mixture of dark neutrons and metastable dark hydrogen atoms.
Lepton asymmetries are generated by 
the $CP$-violating decays of heavy Majorana neutrinos in 
both sectors. These are then converted by sphaleron processes to produce the observed ratio of 
visible to dark matter in the universe. The dynamics responsible for the kinetic decoupling of the two sectors emerges as an important issue
that we only partially solve.

\end{abstract}
\maketitle 

\thispagestyle{fancy} 

\newpage
\tableofcontents
\newpage


                  \section{ \bf Introduction}\label{sec:Intro}
              
                  Cosmological observations have uncovered that the matter content
                  of the standard model is responsible for just $5\%$ of the energy or mass density of the present-day universe. 
                  The remaining energy density consists of the unknown dark matter (DM) and 
                  dark energy components, which according to the latest Planck results make 
                  up $27\%$ and $68\%$ respectively \cite{Ade:2015xua}.
                  Uncovering the nature of these missing components in our cosmos is one of the great challenges of modern physics. 
                  Since the evidence for DM comes from cosmological and astrophysical data, and the 
                  only confirmed mode of interaction is through gravity,  
                  it is common to imagine dark matter as being one or more new particles
                  that constitute a hidden or dark sector which, besides gravity, exhibits only minimal
                  interactions with the visible sector, or possibly none at all. 
                  Given our poor knowledge of the particle nature of 
                  dark matter and its interactions, it is tantalising to consider that the similar abundances of visible and dark matter,
                  which can be expressed in terms of the ratio of their critical 
                  densities, $\Omega_{\text{DM}} \simeq 5 \, \Omega_{\text{VM}}$, is a highly suggestive clue to the nature of the dark sector.
                  
                  Indeed this is the motivating principle of asymmetric dark matter (ADM) 
                  models, which seek to explain these similar densities
                  by connecting their origins (for reviews, see \cite{Davoudiasl:2012uw, Petraki:2013wwa, Zurek:2013wia}).
                  By the conservation of a quantum number that spans the two sectors, one can generate dark matter 
                  in the early universe in a way that is directly connected to the mechanism that 
                  generates the matter-antimatter asymmetry of the visible sector. 
                  The latter proceeds through out-of-equilibrium particle-number-violating and CP-violating processes which create an excess 
                  of baryons over antibaryons, after which the symmetric component 
                  annihilates, leaving a universe composed of matter rather than antimatter. 
                  In asymmetric dark matter models, the generation of dark matter is directly tied 
                  to this process through a dark-matter--dark-antimatter asymmetry and provides a way for the 
                  dark matter density to have similar magnitude to the present day baryon number density.
                  However the mass density, $\Omega_{\text{DM}}$, is a product of particle number density and the mass of individual particles, 
                  $\Omega_{\text{DM}} = n_{\text{DM}}\, m_{\text{DM}}$. The literature of asymmetric dark matter models has focused
                  almost entirely on explaining the 
                  similarity in number densities, while there are significantly fewer works that attempt to explain both similarities.
                  If we can explain the similarity in the masses of nucleons and dark matter in an asymmetric dark matter model, then the 1:5 
                  ratio truly finds a complete explanation.
                  
                  The discovery of the Higgs boson is a triumph for models of mass generation based on  
                  spontaneous symmetry breaking within gauge theories. 
                  It should be noted, however, that the vast majority of mass that the standard model contributes to the universe does not result from the sum of the 
                  masses of fundamental particles, but rather is almost entirely composed of the confinement 
                  energy of massive nucleons
                  after dynamical chiral symmetry breaking. 
                  The dominant form of visible mass in the universe is therefore the result of 
                  dimensional transmutation dynamics: the mass scale that is dynamically generated
                  when the asymptotically-free $SU(3)_C$ gauge coupling becomes non-perturbative in the low-energy regime.
                  We are thus strongly motivated to consider a dark sector that is similarly dominated by the mass of analogously bound states. It will turn out that the dark matter in our theory
                  can be either stable dark neutrons or a mixture of stable dark neutrons and metastable dark hydrogen atoms.
               
                  A direct connection to explain the similar mass densities can be obtained if the dark sector is identical to the visible sector with the two entwined under a
                  discrete $\Bbb{Z}_2$ mirror symmetry.
                  Models of mirror matter have the additional benefit of preserving parity when the mirror partners are chosen to be of opposite chiralities. 
                  The origin and development of such models can be followed in Refs.~\cite{Lee:1956qn, Kobzarev:1966, Pavsic:1974rq, Blinnikov:1982eh, Blinnikov:1983, Foot:1991bp, Foot:1991py, Foot:1995pa, Berezhiani:1995sp, Berezhiani:1995yi, Berezhiani:1995am, Berezhiani:1996sz, Foot:2000tp, Berezhiani:2000gw, Bento:2002sj, Ignatiev:2003js, Foot:2003jt, Berezhiani:2003wj,Berezhiani:2003xm, Foot:2004pq, Ciarcelluti:2004ik, Ciarcelluti:2004ip,Berezhiani:2005ek,Berezhiani:2005vv, Berezhiani:2006ac, Berezhiani:2008gi, Berezhiani:2008zza,Berezhiani:2008zz, Berezhiani:2009kx, Berezhiani:2011da, Foot:2014mia, Berezhiani:2015fba, Berezhiani:2015fba, Addazi:2015cua, Berezhiani:2016ong, Cerulli:2017jzz, Berezhiani:2017azg,Berezhiani:2017jkn}. References~\cite{Foot:2003jt} and \cite{Foot:2004pq} proposed that the visible/dark matter mass-density coincidence could be due
                  to the identical microphysics of the two sectors in models with exact mirror parity symmetry, with the two sectors being chemically related via lepton-portal effective operators. In this paper, we instead develop a new
                  kind of spontaneously broken mirror matter model.
                  In Ref.~\cite{Foot:2000tp} the possible VEVs that the Higgs and its mirror partner could 
                  gain and the implications for each sector were examined, and the possibility of spontaneously 
                  breaking the mirror symmetry through the Higgs potential was explored.
                  Our work draws on such previous models. Among them are Refs.~\cite{Lonsdale:2014wwa, Morrissey:2003sc} which examine the running 
                  couplings of gauge forces in mirror sectors.
                  Other works such as Refs.~\cite{Berezhiani:2000gw, Foot:2006ru} have explored
                  electroweak (EW) symmetry breaking in mirror sectors 
                  and the implications this can have on the little hierarchy problem and the 
                  existence of dark matter. We are also concerned with past works
                  on mirror models which explore the possible 
                  interactions between the sectors such as Ref.~\cite{Zhang:2013ama}. 
                  
                  In models of mirror matter the gauge group is $G \otimes G'$ where $G$ is, or contains, the SM gauge group, while $G'$ is its isomorphic dark partner.
                  Particles with SM representation $R$ are singlets under $G'$, transforming as $(R,1)$, while their mirror partners transform as $(1,R)$. 
                  By the mirror symmetry of the Lagrangian,
                  the gauge coupling constants are equal at energy scales where mirror symmetry is unbroken and in particular, for $G= SU(3)\otimes SU(2)\otimes U(1)$, we have $\alpha_3=\alpha'_3$.
                  If dark matter consists of dark baryons, then even if mirror symmetry is broken at low energy, the connected origin of
                  coupling constants at high energy can lead to a relation between the $SU(3)$ confinement scales of the 
                  two sectors.\footnote{While we only consider $G$ being the standard model gauge group in this paper, it is possible to extend the construction to mirror 
                  GUT theories; see, for example, Refs.~\cite{Lonsdale:2014yua, Lonsdale:2014wwa,Tavartkiladze:2014lla, Gu:2013nya, Gu:2012fg, Gu:2012in}.
                  Such models can be further motivated by the $E_8 \otimes E_8$ symmetry of the heterotic string model.}
                  In the simplest case of unbroken mirror symmetry, dark baryons have identical mass scales to visible baryons 
                  and the two sectors are generally taken to interact only minimally via kinetic and Higgs mixing~\cite{Foot:1991bp}. Of interest to this work is the 
                  idea that the physics of the dark sector can be differentiated from the visible sector yet connected 
                  by a mirror symmetry which is spontaneously broken at a higher energy scale.\footnote{The possibility that the mirror parity remains exact continues to be explored 
                  in the literature as the possible explanation of dark matter \cite{Foot:2003jt, Foot:2004pq, Foot:2012qc, Foot:2013uxa, Foot:2013vna, Foot:2013msa, Foot:2013nea, Foot:2013lxa, Foot:2014mia, Foot:2014uba, Foot:2014osa, Foot:2015sia, Foot:2015mqa, Foot:2015vva, Clarke:2015gqw, Foot:2016wvj, Clarke:2016eac, Chashchina:2016wle, Foot:2017dgx}. This remains a very interesting possibility, and it connects the mass scales of visible and dark matter in the most direct way conceivable.
                  The dark matter in this case is, by construction, just as complicated as ordinary matter, so determining its phenomenological status is very challenging. We do not hold an ideological position on whether
                  or not mirror parity should be broken, but rather explore the broken possibility here as a way to produce a form of asymmetric dark matter that is phenomenologically acceptable within a theory that
                  explains the origin of the dark matter mass scale despite the breaking of the mirror parity.}
                  This is also related to works that posit dark matter as a bound state of a 
                  dark sector containing at least a confining $SU(N)$ gauge group.
                  In recent years there have been numerous models of dark matter given by a dark
                  QCD with an emphasis on considering the masses of such dark bound states~\cite{Bai:2013xga, Boddy:2014yra}. 
                  These dark QCD states can in some cases further form dark atoms along with lighter charged fermions of the dark sector and 
                  these models can thus be compared to past work on atomic dark matter such as Refs.~\cite{CyrRacine:2012fz,vonHarling:2014kha, Pearce:2015zca, Petraki:2015hla,Petraki:2016cnz}.
                  In Ref.~\cite{Lonsdale:2017mzg} we explored the spectra of dark QCD in a hidden sector with an emphasis
                  on how the hadronic spectra changes with the confinement scale and the number of low mass dark quarks.
                  
                  Past work in Refs.~\cite{Lonsdale:2014yua, Lonsdale:2014wwa} demonstrated the 
                  concept of ``asymmetric symmetry breaking (ASB)'' where mirror models preserve mirror parity in the early universe but 
                  dynamically develop different properties at low energy. In general the two sectors can have 
                  different masses and gauge groups after spontaneous symmetry breaking has occurred.
                  We explore in this work a model of mirror sectors with two Higgs doublets in each sector 
                  where we can utilise the ASB mechanism to see just one of the two Higgs doublets 
                  responsible for mass generation in each sector. Critically, the doublets responsible for EWSB in each sector will \emph{not} be mirror partners. 
                  We then have that the Yukawa couplings that generate masses for fermions in the two sectors are independent of each other while,
                  above the scale of EWSB, the mirror symmetry imposes that any CP violation from the Yukawa Lagrangian will be the same. 
                  This results in equal lepton asymmetries being created in the visible and dark sectors with rapid sphaleron reprocessing then producing almost equal baryon asymmetries in the two sectors.

                  In this work we focus on the two stated aims of generating a similar abundance of visible and dark matter and explaining how the mass of dark matter has a similar mass scale to the proton.
                  For the first task, we use baryogenesis via leptogenesis resulting from the decay of heavy Majorana neutrino states
                  to realise the similarity in number density in the two sectors. For the second task we use ASB to explain the similarity of confinement scales.
                  In these two approaches our work follows from previous theories of leptogenesis 
                  in a mirror matter context. In particular, the use of a set of three right handed neutrinos and mirror partners was explored in Ref.~\cite{Cui:2011wk}
                  with the temperature difference between the sectors generated after electroweak symmetry breaking. 
                  In that work the temperature difference is brought about by the asymmetric reheating of the universe with the Higgs, mirror Higgs and a pure singlet scalar after they settle into the vacuum state.
                  In Ref.~\cite{Barbieri:2016zxn} a mirror model that relied on explicit breaking in different Yukawa couplings between the sectors was considered.
                  Similar models have considered simply a common set of singlet neutrinos shared between the 
                  sectors~\cite{Gu:2014nga, Gu:2013gic, An:2009vq}. 
                  In Ref.~\cite{An:2009vq}, thermal contact between the sectors is maintained indefinitely by the Higgs portal term and 
                  kinetic mixing in $U(1)$ gauge bosons and their mirror counterparts. In such equal-temperature scenarios, one must remove 
                  relativistic degrees of freedom from the mirror sector prior to the scale of big bang nucleosynthesis (BBN) to be consistent with the constraints on dark radiation as quantified by $\Delta N_{\text{eff}}$.
                  Recently Ref.~\cite{Farina:2015uea} suggested that in such models as the above, the two sectors 
                  could maintain thermal equilibrium down to a temperature range with a large enough difference in the 
                  degrees of freedom of the two sectors, after which the subsequent rate of cooling in each sector would generate a 
                  temperature difference sufficient to allow for some relativistic species in the dark sector to remain while being consistent with constraints from BBN. This can be deduced from entropy conservation. 
                  While the two sectors maintain equilibrium, if one sector undergoes a sudden drop in the number of degrees of freedom, much of its entropy density will be transferred to the opposite sector. 
                
                  The paper is structured as follows. In Sec.~\ref{sec:Themodel} we outline the field content of the model and the Yukawa structure. 
                  In Sec.~\ref{sec:ASB} we examine the scalar potential and spontaneous symmetry breaking in the model. 
                  Then in Sec.~\ref{sec:Neutrinos} we examine the neutrino mass matrix and comment on 
                  the ordering of massive neutrino states before and after mirror parity is broken.
                  Section~\ref{sec:Leptogenesis} analyses the generation of lepton and baryon asymmetries from the $CP$-violating decays 
                  of the heavy right-handed neutrinos. Following this, in Sec.~\ref{sec:Temp} we analyse the evolution of temperature in each of the two sectors, discuss their kinetic decoupling, and
                  examine the constraints on this model from astrophysical sources
                  and particle phenomenology. We also consider the process of nucleosynthesis in each sector to determine the composition of dark matter in the present-day universe.
                  We conclude in Sec.~\ref{sec:Conclusion}.

                  \section{\bf The model}\label{sec:Themodel}
                  
                  Past work on the mechanism of asymmetric symmetry breaking has examined how mirror symmetric grand unification groups 
                  such as $SU(5) \otimes SU(5)$~\cite{Lonsdale:2014wwa} and $SO(10) \otimes SO(10)$~\cite{Lonsdale:2014yua} can break to 
                  asymmetric sectors with different gauge symmetries.
                  In this work we will instead start with a high energy scale mirror symmetric $ [SU(3)\otimes SU(2)\otimes U(1)] \otimes [SU(3)'\otimes SU(2)'\otimes U(1)'] $ 
                  theory that duplicates the content  of the standard model. A discrete $\Bbb{Z}_2$ 
                  symmetry interchanges visible sector particles with dark sector counterparts. For fermions,
                  left-handed fields are interchanged with right-handed fields of the dark sector and vice versa,
                  \begin{equation}\label{eq:mirrormatter}
                  \phi \leftrightarrow \phi', \qquad G^\mu \leftrightarrow G'_\mu, \qquad f_L \leftrightarrow f_R',
                  \end{equation}
                  where $\phi$, $G^\mu$ and $f$ denote scalar, gauge and fermion fields, respectively.
                  This requires a mirror counterpart for every boson and fermion of the standard model. 
                  In addition to these we have three right handed singlet neutrinos, $N_R^i$, and the corresponding 
                  left handed states of the dark sector, $(N_L^i)'$. We also add a second Higgs doublet, $\Phi_2$, with 
                  its own $\Bbb{Z}_2$ partner, $\Phi_2'$. 
                  
                  This second Higgs doublet allows us to use the mechanism of 
                  asymmetric symmetry breaking to spontaneously break mirror symmetry and have $\Phi_2'$ be the 
                  instigator of electroweak symmetry breaking (EWSB ) in the dark sector,
                  while $\Phi_1$ takes on the usual role in our sector. While $\Phi_1$ and $\Phi_2$ carry the same quantum numbers,
                  their self couplings and the size of their Yukawa couplings to fermions will differ. 
                  In this respect our model is significantly different from Refs.~\cite{Cui:2011wk} and \cite{An:2009vq}
                  which used only the SM Higgs doublet and its 
                  mirror counterpart.  In Sec.~\ref{sec:Leptogenesis}, we will examine how this second
                  doublet also allows for successful thermal leptogenesis without potential naturalness concerns from heavy neutrino 
                  mass corrections to the squared Higgs boson mass.
                  The principle motivation for the two Higgs doublets is, however, our decision to 
                  implement the ASB mechanism into a complete model of ADM. By using ASB we can break the
                  mirror parity symmetry and give discernible differences to each of the 
                  sectors. The absolute minimum breaks the symmetry of the two sectors in different 
                  ways and we switch from a theory of two identical sectors
                  to a model composed of a visible sector, which carries the observable properties of
                  the standard model, and a dark sector with a phenomenology of its own
                  that yet retains an origin which guarantees that the mass content of the dark universe will ultimately be highly similar.
                  The total field content of the two sectors is listed in Table \ref{tab:fields}.
                  \begin{table}[h!]
                  \centering
                  \caption{Field content and their representations under the mirror symmetric gauge group, $ [SU(3)\otimes SU(2)\otimes U(1)] \otimes [SU(3)'\otimes SU(2)'\otimes U(1)'] $.}
                  \label{tab:fields}
                  \scalebox{1.4}{
                  \begin{tabular}{|l|l|}
                  \hline
                  $L_L^i \sim (1,2,-\frac{1}{2})(1,1,0)$   & $(L_R^i)' \sim (1,1,0)(1,2,-\frac{1}{2})$   \\ \hline
                  $e_R^i \sim (1,1,-1)(1,1,0)$             & $(e_L^i)' \sim (1,1,0)(1,1,-1)$       \\ \hline
                  $Q_L^i \sim (3,2,\frac{1}{6})(1,1,0)$    & $(Q_R^i)' \sim (1,1,0)(3,2,\frac{1}{6})$   \\ \hline
                  $u_R^i \sim (3,1,\frac{2}{3})(1,1,0)$    & $(u_L^i)' \sim (1,1,0)(3,1,\frac{2}{3})$     \\ \hline
                  $d_R^i \sim (3,1,-\frac{1}{3})(1,1,0)$   & $(d_L^i)' \sim (1,1,0)(3,1,-\frac{1}{3})$    \\ \hline
                  $N_R^i \sim (1,1,0)(1,1,0)$              & $(N_L^i)' \sim (1,1,0)(1,1,0)$      \\ \hline
                  $\Phi_1 \sim (1,2,\frac{1}{2})(1,1,0)$   & $(\Phi_1)' \sim (1,1,0)(1,2,\frac{1}{2})$ \\ \hline
                  $\Phi_2 \sim (1,2,\frac{1}{2})(1,1,0) $  & $(\Phi_2)' \sim (1,1,0)(1,2,\frac{1}{2}). $   \\ \hline
                  \end{tabular}}
                  \end{table}
                
                 \subsection{Yukawa couplings}\label{sec:TwoHiggs}
                 
                 The use of additional Higgs doublets in extensions to the SM has a history ranging
                 from minimal extensions of a single extra Higgs doublet to larger extensions in
                 which the observed Higgs state discovered at the LHC is the lightest 
                 state of a largely decoupled Higgs sector \cite{Branco:2011iw}.
                 Supersymmetric models require at least two Higgs doublets 
                 to generate mass for the fermions. 
                 In this work we will explore the implications of having two Higgs 
                 doublets in the visible sector and associated mirror partners in the dark sector.
                 Models of just two Higgs doublets are typically separated into three generic types based on the discrete symmetries they impose. 
                 Type I has just one of the doublets coupling to all fermions while
                 Type II models have one couple to the up-type quarks and the other the down-type.
                 Type III, which is the model that matches our own within the context of the visible sector,
                 gives both doublets the same quantum numbers and therefore allows each of them to
                 couple to all flavours of quarks and leptons.
                 In Type III models the fields can be expressed in a basis where only one doublet gains a non-zero 
                 VEV while the second doublet retains a vanishing VEV. This is known as the Higgs basis \cite{Branco:2011iw}.
                 This variety of model with a second Higgs doublet introduces highly 
                 constrained flavour changing neutral currents (FCNC) at tree level due to the fact that
                 couplings between the fermions and multiple Higgs doublets cannot in general be simultaneously diagonalised. 
             
                 Our work is distinct from other models of additional Higgs fields, such as inert Higgs doublet models \cite{Melfo:2011ie, Andreas:2009hj}, that identify dark matter with a Higgs state. 
                 In our case, the dark matter candidate instead arises from the
                 stable baryons of the dark sector. The roles of the additional Higgs fields are to facilitate the breaking of the mirror parity symmetry and the associated asymmetric gauge symmetry breaking. 
                 The second doublet is also a second source of CP-violating decays for heavy neutrino states in each sector at the scale of thermal leptogenesis. 
                 
                 The leptonic Yukawa couplings are
                 \begin{align}\label{eq:yukawa}
                 L_{\text{Lepton}} &= {\eta^l_{1}}_{ij}\, \overline{L^i_{L}} e^j_{R}\Phi_1      + \eta^l_{1ij}\, \overline{L^i_R}'  {e^j_L}' \Phi'_1             + {y_{1}}_{ij}\, \overline{L^i_L} N^j_R \tilde{\Phi}_1 + {y_{1}}_{ij}\,  \overline{{L^i_{R}}}' {N^j_L}' \tilde{\Phi}'_1  \nonumber\\ 
                           &+ {\eta^l_{2}}_{ij}\, \overline{L^i_{L}}e^j_{R}\Phi_2              + {\eta^l_{2}}_{ij}\, \overline{L^i_R}' {e^j_L}' \Phi'_2         + {y_{2}}_{ij}\, \overline{L^i_{L}} N^j_R \tilde{\Phi}_2 + {y_{2}}_{ij}\,  \overline{{L^i_{R}}}' {N^j_L}'\tilde{\Phi}'_2  \\ 
                           &+ {f_{1}}_{ij}\, \overline{L^i_{L}}  {{N^j_L}^c}' \tilde{\Phi}_1   + {f_{1}}_{ij}\,  \overline{L^i_R}' {N^j_R}^c \tilde{\Phi}_1'    + {f_{2}}_{ij}\, \overline{L^i_L}{{N^j_L}^c}'\tilde{\Phi}_2 + {f_{2}}_{ij}\,  \overline{L^i_R}' {N^j_{R}}^c \tilde{\Phi}_2' + \text{h.c.} \nonumber 
                 \end{align}
                 Mirror parity enforces that the coupling of the doublets is the same as their mirror counterparts. However, terms such as $y_1$ and $y_2$ will not
                 be the same and may differ by orders of magnitude. We make use of $\tilde{\Phi}= i \tau_2 \Phi^{*}$ in the above. In addition to these terms we have the ordinary Majorana mass terms along with a cross-sector mass term,
                 \begin{equation}\label{eq:majorana}
                 M_{ij}\left( \overline{{(N^i_R)^c}} N^j_R+\overline{{{(N^i_L)^c}}}' {N^j_L}'\right) + P_{ij}\overline{{N^i_R}}{N^j_L}'+ \text{h.c.}
                 \end{equation}
                 For the quark Yukawas we have,
                 
                    \begin{align}\label{eq:quark}
                L_{\text{Quark}} &= {\eta^u_{1}}_{ij}\, \overline{Q^i_L} u^j_R \Phi_1 + {\eta^u_{1}}_{ij}\, \overline{Q^i_R}' {u^j_L}'  \Phi'_1 + \eta^d_{1ij}\, \overline{Q^i_L} d^j_R \tilde{\Phi}_1 + {\eta^d_{1}}_{ij}\,  \overline{Q^i_R}'{d^j_L}'\tilde{\Phi}_1'  \nonumber\\ 
                           &+ {\eta^u_{2}}_{ij}\, \overline{Q^i_L} u^j_R \Phi_2 + {\eta^u_{2}}_{ij}\, \overline{Q^i_R}' {u^j_L}' \Phi'_2 + {\eta^d_{2}}_{ij}\, \overline{Q^i_L} d^j_R \tilde{\Phi}_2 + {\eta^d_{2}}_{ij}\,  \overline{Q^i_R}' {d^j_L}' \tilde{\Phi}_2'  + \text{h.c.}
                  \end{align}
               
                 We also consider the various other ways that the two sectors can interact. 
                 The photon-mirror photon kinetic mixing term,
                 \begin{equation}\label{eq:photonmix}
                 \epsilon_{\gamma \gamma'} F^{\mu \nu}{F_{\mu \nu}}',
                 \end{equation}
                 allows for visible sector states to have dark $U(1)$ millicharges. This parameter is constrained by orthopositronium decay 
                 experiments \cite{Badertscher:2006fm} to the degree of $ \epsilon_{\gamma \gamma'} \le 1.55\times 10^{-7}$. 
                 Such terms are naturally absent in GUT contexts at tree-level, although they will be radiatively generated if there are multiplets that simultaneously 
                 have non-trivial electroweak quantum numbers from each sector. 
                 There are also Higgs portal interactions. These will be explored further in the next section and later in the discussion of the thermal history.
                 Neutrino interactions that mix the two sectors are also present in Eqs.~\ref{eq:yukawa} and \ref{eq:majorana}; these will be further discussed in Sec.~\ref{sec:Neutrinos}. 
                 As we elaborate on in Sec.~\ref{sec:Temp}, the two sectors can only maintain rapid interactions until a decoupling temperature $T_{\text{Dec}}$. 
                 This limits the size of any portal terms in our model and these individual limits will depend on how the interaction rate scales with temperature.
                 This means that no interaction between the sectors can maintain thermal equilibrium indefinitely. This further constrains 
                 the kinetic mixing term $\epsilon_{\gamma \gamma'}$ due to the scaling of photon-mirror photon interactions with temperature. 
                  
                 While many models of dark sectors consider the possibility of the dark sector having a lower temperature than the visible sector from the 
                 outset (see, for example Ref.~\cite{Hardy:2017wkr}),
                 this work has the benefit that the separate Higgs doublets which facilitate EWSB in each sector can 
                 spontaneously break mirror symmetry which later causes a difference in the temperatures of the visible and dark sectors. 
                 This and the other differences between the two sectors is accomplished without the need for any soft symmetry breaking mass 
                 terms.\footnote{The cosmological domain-wall problem caused by the existence of our spontaneously broken $\Bbb{Z}_2$  symmetry may motivate soft breaking terms. 
                 A soft breaking that gives some separation in the squared mass terms of $\Phi_i$ and $\Phi'_i$ could be considered. 
                 Another possibility is a small difference between the Majorana masses of $N_R$ and $N'_L$ as in \cite{Cui:2011wk}.
                 However, there are other ways of obviating the domain-wall problem. One is to engineer the non-restoration at high temperatures of one or both EW symmetries (see Ref.~\cite{Bajc:1999cn} for a review), 
                 another is to embed the $\Bbb{Z}_2$  symmetry within a continuous symmetry, 
                 and a third is the possibility of breaking mirror symmetry prior to inflation in order to make the typical domain wall separation larger than the horizon distance \cite{Krauss:1992gf}.}

                  \section{\bf Asymmetric symmetry breaking}\label{sec:ASB}
                  
                  We build on past work in Ref.~\cite{Lonsdale:2014wwa} where the concept of asymmetric 
                  symmetry breaking was introduced. This involves two sectors that begin completely symmetric then spontaneously break in such a 
                  way that in the low energy regime, the gauge groups and masses of the two sectors 
                  are different. The original mechanism utilises the couplings of the Higgs multiplets of the two
                  sectors being in a particular region of parameter space such that the $\Bbb{Z}_2$ 
		  symmetry is spontaneously broken by giving nonzero VEVs to Higgs multiplets in each
		  of the two sectors such that their $\Bbb{Z}_2$ counterparts do \emph{not} gain nonzero VEVs. We expand on this idea in the following section.
                  
                  \subsection{Higgs potentials}\label{sec:decoupling}
                  
                  The scalar potential of our model is constructed from the four Higgs doublets $\Phi_1,\Phi'_1,\Phi_2,\Phi'_2$ with invariance under
                  $\Phi_1 \leftrightarrow \Phi'_1$ and $\Phi_2 \leftrightarrow \Phi'_2$ imposed. In the simplest version of asymmetric symmetry breaking, only one 
                  Higgs multiplet in a mirror pair gains a nonzero vacuum expectation value, with the sector containing that field switching from the first pair to the second pair, as we now review.
                  A toy-model potential suitable for transparently demonstrating asymmetric symmetry breaking is what we may call the `minimal asymmetric potential'. 
                  It is obtained by setting some parameters in the full potential to zero, and is given by
                  
                  \begin{eqnarray}\label{eq:higgspotentialA}
                  \begin{split}
                V_{\text{ASB}} &= \lambda_{1}\left(\Phi_1^{\dagger} \Phi_1 + {\Phi'}_1^{\dagger} {\Phi'}_1 - \frac{v^2}{2}\right)^2 + \lambda_{2}\left(\Phi_2^{\dagger} \Phi_2 + {\Phi_2'}^{\dagger} {\Phi_2'} - \frac{w^2}{2}\right)^2 +\kappa_{1} (\Phi_1^{\dagger} \Phi_1)( {\Phi_1'}^{\dagger} {\Phi_1'})  \\
                        &+   \kappa_{2} (\Phi_2^{\dagger} \Phi_2)( {\Phi_2'}^{\dagger} {\Phi_2'}) + \sigma_1 \left((\Phi_1^{\dagger} \Phi_1 )(\Phi_2^{\dagger} \Phi_2) + ({\Phi_1'}^{\dagger} {\Phi_1'} )({\Phi_2'}^{\dagger} {\Phi_2'})\right) \\
                        &+    \sigma_2 \left( \Phi_1^{\dagger} \Phi_1 + \Phi_2^{\dagger} \Phi_2 + {\Phi_1'}^{\dagger} {\Phi'}_1 + {\Phi_2'}^{\dagger} {\Phi_2'} - \frac{v^2}{2}- \frac{w^2}{2} \right)^2.  
                  \end{split}
                  \end{eqnarray}
                  In this simplest potential, the parameter space giving rise to asymmetric breaking is where each of the parameters, 
                  $[\lambda_1,\lambda_2,\kappa_1,\kappa_2, \sigma_1 ,\sigma_2]$, is real and positive. In this case
                  the potential is minimised when each of the non-negative terms in the potential is individually zero. This produces the global vacuum state,
                  \begin{align}\label{eq:vevs}
                  \braket{\Phi_1} =
\left[
\begin{array}{c}
0 \\
\frac{v}{\sqrt{2}}
\end{array}
\right]
,  \qquad  \braket{\Phi'_1} =  0, \nonumber\\            
\braket{\Phi_2} =  0,    \qquad \braket{\Phi'_2} =  \left[\begin{array}{c}0 \\
 \frac{w}{\sqrt{2}} 
\end{array}
\right] .
                  \end{align}
                  Going to unitary gauge and introducing the definitions,
                  \begin{align}
                  \Phi_1 = \left[ \begin{array}{c} 0 \\ \frac{1}{\sqrt{2}}(v + h) \end{array} \right],\qquad \Phi'_1 = \left[ \begin{array}{c} H_D^+ \\ \frac{1}{\sqrt{2}}( H_D + iA_D ) \end{array} \right], \nonumber\\
                  \Phi_2 = \left[ \begin{array}{c} H^+ \\ \frac{1}{\sqrt{2}}( H + iA ) \end{array} \right], \qquad \Phi'_2 = \left[ \begin{array}{c} 0 \\ \frac{1}{\sqrt{2}}(w + h_D) \end{array} \right],
                  \end{align}
                  we obtain the mass matrix
                  \begin{equation}\label{eq:higgsmass}
                  \frac{1}{2} \left( \begin{array}{cc} h\ ,\ h_D \ \end{array} \right) \left[ \begin{array}{cc} 2(\lambda_1 + \sigma_2) v^2 & \sigma_2 v w \\ \sigma_2 v w & 2(\lambda_2 + \sigma_2) w^2 \end{array} \right]
                  \left( \begin{array}{c} h \\ h_D \end{array} \right).
                  \end{equation}
                  Note that, in general, the eigenstate fields are admixtures of visible and dark states. The other mass eigenvalues are,
                  \begin{align}
                   m_H =m_A=m_{H^{\pm}}=&\sqrt{\frac{1}{2}\kappa_2 w^2 +\frac{1}{2} \sigma_1 v^2}, \\
                   m_{H_D}=m_{A_D}=m_{H^{\pm}_D}&=  \sqrt{\frac{1}{2}\kappa_1 v^2 +\frac{1}{2} \sigma_1 w^2} . 
                  \end{align}
                  
                  We can see in Eq.~\ref{eq:higgsmass} that we recover a standard model physical Higgs mass formula in the limit that $\sigma_2 \rightarrow 0$. 
                  This is the same limit that sets the off-diagonal mass term that couples the two sectors to zero.
                  If the EW scales of the two sectors are required to be very different, then it is necessary to take $\sigma_2$ to be sufficiently small so as to not mix the two energy scales.
                  In this limit, the asymmetric VEV is still the minimum and we can see that the massive states besides $h$ all have mass terms that scale with $w$. This is significant because the scale $w$, in the asymmetric configuration,
                  can raise the masses of all the scalars of both sectors except for the SM Higgs boson in the case that $w > v$. This is necessary to meet constraints from Higgs-induced FCNCs.
                  
                  We now consider this feature in the context of a full mirror symmetric potential with two Higgs doublets in each sector.
                  In contrast with past work in asymmetric symmetry breaking, which dealt exclusively with potentials that made use of the fact that the multiplets in each sector were in two different sized representations of the 
                  gauge group, in this work we are considering identical representations.
                
                  The most general potential in this case,
\begin{align}\label{eq:vmir}
V_{{\text{M2HDM}}} &= +m^2_{11}\, (\Phi_1^{\dagger} \Phi_1 + \Phi_1^{'\dagger} \Phi^{'}_1)  +m^2_{22}\, (\Phi_2^{\dagger} \Phi_2 + \Phi_2^{'\dagger} \Phi^{'}_2)\nonumber \\
  &+ \left(m^2_{12}\, (\Phi_1^{\dagger} \Phi_2 +\Phi_1^{'\dagger} \Phi^{'}_2  ) + h.c. \right) + \nonumber  \frac{1}{2}z_{1}\,\left( (\Phi_1^{\dagger} \Phi_1)^2 + (\Phi_1^{'\dagger} \Phi^{'}_1)^2 \right)  \\
  &+\frac{1}{2}z_{2}\,\left( (\Phi_2^{\dagger} \Phi_2)^2 + (\Phi_2^{'\dagger} \Phi^{'}_2)^2 \right) +  z_{3}( \Phi_1^{\dagger} \Phi_1\Phi_2^{\dagger} \Phi_2 + \Phi_1^{'\dagger} \Phi^{'}_1\Phi_2^{'\dagger} \Phi^{'}_2 )  \\
  &+ z_{4}\,( \Phi_1^{\dagger} \Phi_2\Phi_2^{\dagger} \Phi_1 + \Phi_1^{'\dagger} \Phi^{'}_2\Phi_2^{'\dagger} \Phi^{'}_1 ) + \nonumber \frac{1}{2}z_{5}\, \left( (\Phi_1^{\dagger} \Phi_2)^2 +      (\Phi_1^{'\dagger} \Phi^{'}_2)^2 + h.c.  \right)     \\
  &+\left[ (z_{6}\,     \Phi_1^{'\dagger} \Phi^{'}_1                                  +  z_{7}\,         \Phi_2^{'\dagger} \Phi^{'}_2     ) \Phi_1^{'\dagger} \Phi^{'}_2    + \nonumber   (z_{6}\,     \Phi_1^{\dagger} \Phi_1             +   z_{7}\,         \Phi_2^{\dagger} \Phi_2                           ) \Phi_1^{\dagger} \Phi_2        + h.c.\right] \nonumber \\
  &+z_{8}    \Phi_1^{\dagger} \Phi_1      \Phi_1^{'\dagger} \Phi^{'}_1      +z_{9}\,    \Phi_2^{\dagger} \Phi_2      \Phi_2^{'\dagger} \Phi^{'}_2   + \nonumber   (z_{10}\,   \Phi_1^{\dagger} \Phi_2          \Phi_1^{'\dagger} \Phi^{'}_2 + h.c.  )\nonumber \\
  &+    (z_{11}\,         \Phi_1^{\dagger} \Phi_2 \Phi_2^{' \dagger} \Phi^{'}_1     +h.c.                  )+ z_{12}(      \Phi_1^{\dagger} \Phi_1      \Phi_2^{'\dagger} \Phi^{'}_2      +   \Phi_1^{'\dagger} \Phi^{'}_1   \Phi_2^{\dagger} \Phi_2   ) \nonumber \\
  &+\left[ (z_{13}\,     \Phi_1^{'\dagger} \Phi^{'}_1          + \nonumber z_{14}\,         \Phi_2^{'\dagger} \Phi^{'}_2      ) \Phi_1^{\dagger} \Phi^{}_2    + \nonumber (z_{13}\,     \Phi_1^{\dagger} \Phi_1    + \nonumber z_{14}\,         \Phi_2^{\dagger} \Phi_2    ) \Phi_1^{' \dagger} \Phi^{'}_2         + h.c.\right],
\end{align}
     features a large number of new terms which we must consider carefully. 
     While such a potential increases the possible configurations giving the minima, 
     we can always perform individual basis rotations of the doublets in each sector to return the doublets to a 'Higgs basis' 
     where only one doublet has a non-zero VEV. What we then require in the context of asymmetric symmetry breaking
     is that the required transformation is different for the two sectors. 
     
     In this context we will consider the above general mirror 2HDM where the global minimum is given by
                     \begin{equation}
                    \braket{\Phi_i} =
\left[
\begin{array}{c}
0 \\
\frac{v_i}{\sqrt{2}}
\end{array}
\right]
,  \qquad  \braket{\Phi'_i} =  \left[\begin{array}{c}
0 \\
 \frac{w_i}{\sqrt{2}} 
\end{array}
\right] .
 \end{equation}
We then define
\begin{equation}
v=\sqrt{|v_1|^2 + |v_2|^2},\;\; w=\sqrt{|w_1|^2 + |w_2|^2}, \;\; \rho=\frac{w}{v}.
\end{equation}
We can then form what we term the `dual Higgs basis' through,
\begin{equation}
  H_1= \frac{v_1^*  \Phi_1 + v_2^* \Phi_2}{v}, \qquad H_2= \frac{-v_2 \Phi_1 + v_1 \Phi_2}{v},
 \end{equation}
in the visible sector, and
\begin{equation}
  H'_1= \frac{w_1^*  \Phi_1' + w_2^* \Phi_2'}{w}, \qquad H'_2= \frac{-w_2 \Phi_1' + w_1 \Phi_2'}{w},
 \end{equation}
 in the dark sector.  The fields $H_1$ and $H'_1$ have nonzero VEVs given by $v$ and $w$, respectively, while the orthogonal combinations have vanishing VEVs.
The two field rotations result in a potential that is not obviously mirror symmetric prior to symmetry breaking. The obvious mirror symmetry would, of course, 
reappear if we expand back to the original $\Phi$ basis. The new basis fields $H_1$ and $H'_1$ are not mirror partners and have different VEVs, masses and Yukawa couplings to fermions,
with $v_1 \neq w_1$ and $v_2 \neq w_2$ parameterising the spontaneous breakdown of the mirror parity symmetry.

To achieve asymmetric symmetry breaking, we require a vacuum configuration which was previously discussed in \cite{Lonsdale:2014yua}, given by the 
conditions
\begin{equation}\label{eq:minco}
 v_1 \gg v_2,\,\,\,\,\,\, w_2 \gg w_1,
\end{equation}
which breaks mirror symmetry. In general we will also be considering the case that 
$w_2 \gg v_1$ such that the VEVs in the dual Higgs basis obey $w \gg v$, or $\rho \gg 1$ in other words. The mirror symmetry is broken in such a way that not only is 
the EW scale of the dark sector at a higher value than its visible counterpart, but the fermion Yukawa couplings relevant to mass generation are independent. 
This is due to the fact that $H_1$ is mostly $\Phi_1$ while $H_1'$ is mostly $\Phi_2'$, thus approximating the configuration encountered in the `minimal asymmetric potential' toy model discussed above.
We define this to be what we mean by asymmetric symmetry breaking in the general case, and promote the ratios of the two 
quantities $\text{tan}(\beta)=v_2/v_1$ and $\text{tan}(\beta')=w_2/w_1$ to being pertinent parameters. 

The Yukawa terms relevant for generating quark masses, and those relevant for Higgs-induced FCNCs, are given in the Higgs basis by the respectively diagonal and non-diagonal matrices,
\begin{align}\label{eq:eta}
&\tilde{\eta}^Q_1= c_{\beta} V_L^Q \eta^Q_1 {V_R^Q}^{\dagger} + s_{\beta} V_L^Q \eta^Q_2 {V_R^Q}^{\dagger} \\ \nonumber
&\tilde{\eta}^Q_2= -s_{\beta} V_L^Q \eta^Q_1 {V_R^Q}^{\dagger} + c_{\beta} V_L^Q \eta^Q_2 {V_R^Q}^{\dagger},
\end{align}
where $Q = u,d$ and $V^Q_{L,R}$ are the left- and right-handed diagonalisation matrices.
For couplings in the dark sector we have different diagonal and non-diagonal matrices,
\begin{align}\label{eq:eta-prime}
&\tilde{\eta}^{Q'}_1= c_{\beta'} W_L^{Q} \eta^Q_1 {W_R^{Q}}^{\dagger} + s_{\beta'} W_L^Q \eta^Q_2 {W_R^Q}^{\dagger} \\ \nonumber
&\tilde{\eta}^{Q'}_2= -s_{\beta'} W_L^{Q} \eta^Q_1 {W_R^{Q}}^{\dagger} + c_{\beta'} W_L^Q \eta^Q_2 {W_R^Q}^{\dagger},
\end{align}
with $Q' = u',d'$ referring to dark quark flavours, and with $W$ denoting the diagonalisation matrices.
We can then see that in the limit of asymmetric VEVs in Eq.~\ref{eq:minco} that we have four different $\eta$ terms for each of the up and down type couplings. 
In particular the mass eigenvalues of the two sectors are almost entirely independent. The leptonic sector follows the same pattern.
The presence of the non-diagonal matrix for $H_2$ in the visible sector is the origin of FCNCs, as in that sector the model resembles the type III 2HDM.
This is a necessary part of our model as we use the mirror partner of the second doublet to play a similar role to the SM Higgs doublet in the dark sector.
The current constraints on such type III 2HDM impose significant restrictions on the masses of the additional scalars of the visible sector and the $\tilde{\eta}^Q_2$ Yukawa matrices.
The decoupling of the additional scalars in the visible sector from their sensitivity to the high scale $w$ seen in the minimal asymmetric model will occur in the full potential case as well and this will be critical in suppressing 
the size of the FCNCs. 
We can see already that it is the parameters $z_8$ and $z_9$ that play the role of $\kappa_1, \kappa_2$ from the minimal model. The $z_9$ term will mass-decouple the second doublet in the visible sector and thus
allow the FCNC constraints to be met. 
The parameter $z_{12}$, on the other hand, must be small in this basis to prevent the standard model Higgs from coupling to the higher mass scale, just as the parameter $\sigma_2$ was kept small in the toy model. 
It is important to note that these terms are cross-sector couplings. Because of this the limit that these couplings approach zero may be 
technically natural due to the enhancement of symmetry originating in the two sectors gaining independent Poincar\'{e} groups, as in Ref.\cite{Foot:2013hna}.
Note, however that $z_{8,9}$ must actually not be too small, so the naturalness issue is rather delicate in this model.

We can now examine how the mass scales originate from the potential of Eq.~\ref{eq:vmir}. We label the fields within the doublets in the original basis as
\begin{equation}
  \Phi_1= \left[
\begin{array}{c}
G_1^+ \\
\frac{1}{\sqrt{2}}(v_1 + \phi_1  +i G_1 )
\end{array}
\right], \qquad 
            \Phi_2= \left[
\begin{array}{c}
I_2^+ \\
\frac{1}{\sqrt{2}}(v_2 + \phi_2  +i a_2 )
\end{array}
\right]
 \end{equation}
 and
\begin{equation}
\Phi'_1= \left[
\begin{array}{c}
I_1^+ \\
\frac{1}{\sqrt{2}}(w_1 + \phi'_1  +i a_1 )
\end{array}
\right], \qquad 
            \Phi'_2= \left[
\begin{array}{c}
G_2^+ \\
\frac{1}{\sqrt{2}}(w_2 + \phi'_2  +i G_2 )
\end{array}
\right],
\end{equation}
            where the label $G$ denotes the states that dominate the admixtures defining the Goldstone bosons.
                  We then consider the $6 \times 6$  symmetric neutral mass matrix in Appendix A and six neutral mass eigenstates. In the limit of no mixing between the sectors, in a 2HDM
                  one typically labels these as $h^0, H^0, A^0$ and $H^{\pm}$ in the visible sector, as we did in the toy model. In the mirror sector
                  one would then have $h^0_D, H^0_D,A^0_D,$ and $H^{\pm}_D$. 
                  In our case, however, these are not mirror partners. Some of the mirror partners have in fact been been absorbed via the Higgs Mechanism into the massive gauge boson states.
                  Since we make use in the visible sector of the mass-decoupling limit of the second Higgs doublet, we have that all other
                  physical scalars acquire masses much larger than the SM Higgs state, $m_{H^{\pm}}, m_{A^{0}}, m_{H^{0}} \gg m_{h}$.
                  In this case the physics at low energy approaches that of just the SM Higgs state. Our approach to 
                  the mass-decoupling limit is different from both the minimal asymmetric case of the toy model, and the typical Type-III model, in a number of key ways. First, while a typical 
                  method in Type III would be to flip the sign of the $m^2_{22}$ mass squared parameter in order to give positive masses to all 
                  other scalars, in our model we are constrained to keep the sign the same as $m_{11}^2$ by the principles of asymmetric symmetry breaking and our mirror symmetry.
                  Second, the additional cross-sector terms in the full potential in Eq.\ref{eq:vmir} compared to the toy model will generate mixing between the scalars of the two sectors and in general mass eigenstates
                  will be composed of visible and dark interaction eigenstates. 
                  However, mass-decoupling still parallels the minimal model in that $w_2$ lifts the mass eigenvalues of all scalars except for one, which we identify with the physical Higgs. This mass-decoupling limit 
                  also generates the alignment limit in that this lowest mass eigenstate has couplings which align with SM values.
                  
                  We will label the six neutral mass eigenstates as $h^0_1, h^0_2, A^0_1, A^0_2, J^0_1, J^0_2$. We note that the first two have minimal cross-sector mixing.
                  It is important here that the field corresponding to the standard model Higgs boson not mix heavily with mirror states. However, we find it possible to have minimal visible-dark mixing for the low mass Higgs state with SM couplings
                  while the heavy additional scalars do, in fact, mix. This comes from terms such as
                  \begin{equation}
                   z_{10}\,   \Phi_1^{\dagger} \Phi_2  \Phi_1^{'\dagger} \Phi^{'}_2 + h.c.
                  \end{equation}
                  which in the limit of Eq.~\ref{eq:minco} only contains large mass terms of the form 
                  \begin{equation}
                   \frac{v_1 w_2}{2}  \left[\text{Im}(z_{10})(-a_1 \phi_2 +a_2 \phi'_1)  + \text{Re}(z_{10})(a_1 a_2  + \phi_2 \phi'_1) \right].
                  \end{equation}
                  Making this term large also does not interfere with the asymmetric minimum.
                  
                  In Table~\ref{tab:p1} we list a suitable set of parameters that give rise to an asymmetric symmetry breaking VEV configuration, and the EW scales and mass eigenvalues they imply. 
                  For this point in parameter space, it is also the case that the couplings of $\Phi_2$ induce masses for the visible sector eigenstates that are driven by the large dark sector VEV $w$, while keeping 
                  the state playing the role of the SM Higgs appropriately light, and maintaining the separation of the two EW scales, $v$ and $w$.
                  \begin{table}[h!]
                  \scriptsize
                  \centering
                  \caption{Higgs Potential parameter sample and associated masses of physical scalars in each sector. Masses are in units of $GeV$.} 
                  \label{tab:p1}
                  \begin{tabular}{ |l|l|l|l|l|l|l|l|l|l|l|l|l|l|l|l|l| }
      \bottomrule  $m^2_{11}=-87^2$ & $m^2_{22}=-2600^2$& $m^2_{12}=-90^2$  & $z_1=0.13$    & $z_2=0.13$              & $z_3=0.8$     & $z_4=0.01$        & $z_5=0.01$       & $z_6=0.01$ &$z_7=0.01$     \\ \hline     
                   $z_8=0.8$    & $z_9=0.8$ & $z_{10}=0.8$ & $z_{11}=0.01$& $z_{12}=10^{-8}$ & $z_{13}=0.01$ & $z_{14}=0.01$ &$m_h=125$   &  $m_{h_D}=3696$  &   $m_{A_1}=5965$   \\ \hline
                   $m_{A_2}=6912$   & $m_{J_1}=5965$   & $m_{J_2}=6512$     &   $m_{{H^+}}=5965$    &    $m_{{H^+}'}=6512$    & $v=246$ & $w=7276$ & $\text{tan}(\beta)=4 \times 10^{-5}$ & $\text{tan}(\beta')=18190$ & $\rho=30$ \\ 
                \toprule  
                \end{tabular}
                \end{table}
                  So, just as in the toy-model case, the additional scalars of the visible sector gain large masses driven by $w$.
                  
                  It is convenient to discuss the FCNC bounds in the dual Higgs basis, with the relevant couplings given in Eqs.~\ref{eq:eta} and \ref{eq:eta-prime}.
                  An estimate of the necessary mass scale for avoiding FCNCs is given in Ref.~\cite{Branco:2011iw} where $K-\overline{K}$ oscillation bounds are avoided if $m_{H_2} > 150\ \text{TeV}$. 
                  This assumes, however, that all Yukawas have similar magnitude to that of the top quark. In our case, 
                  the second Higgs doublet, $H_2$,  may have Yukawa couplings that are much smaller than the SM Higgs coupling to the top, and thus the lower bounds are less severe.
                  The sizes of these Yukawas are constrained by a number of sources in the model. Among them is the need to thermally decouple the two sectors in an acceptable way,
                  and the relevant scattering terms in the era of leptogenesis, as we discuss later.
                  In particular, as we will see in Fig.~\ref{fig:QCDS}, the necessary confinement scale of the dark sector may require that the most massive dark quark has a mass such that
                  its Yukawa coupling may be comparable to the SM up and down quarks, which are at least $10^4$ smaller than the top coupling. 
                  This constraint then directly limits the size of the couplings to $H_1'$ and therefore, by the parity symmetry, to $H_2$, such that FCNC bounds can be more easily satisfied.
                  All results to be presented later that depend on Yukawa couplings pertinent to FCNC processes have those couplings chosen so as to ensure that they meet the experimental bounds.
                  
                  We consider the mixing of neutral states by examining the rotation matrix $U$ in $U^T M^2 U =D^2$, where $D$ is diagonal. 
                  This transforms the $(\phi'_1, a_1,\phi'_2,\phi_2, a_2,\phi_1)$ basis into the $(J_2,A_2,J_1,A_1,h_2,h_1)$ mass-basis. 
                  For the parameter point in Table II we have,
                  
  \begin{equation}\label{eq:matII}
  U_{II}=
  \left(
\begin{array}{cccccc}
 -0.999996 & 0 & -0.00265105 & 0 & 0.000447771 & 0 \\
 0 & -0.999997 & 0 & -0.00260254 & 0 & 0 \\
 -0.000447834 & 0 & 0 & 0 & -1. & 0 \\
 -0.00265104 & 0 & 0.999996 & 0 & 0 & 0 \\
 0 & -0.00260254 & 0 & 0.999997 & 0 & 0 \\
 0 & 0 & 0 & 0 & 0 & 1. \\
\end{array}
\right),
 \end{equation}
                   which shows how the mixing $U_{\phi_1 h_1}$ can be 1, while the remaining states have larger mixing between the sectors.
                   This mixing can also allow for a Higgs portal term without affecting the couplings of the standard model Higgs boson and maintaining the decoupling limit for the additional scalar degrees of freedom. 
                   We will return to this possible connection between the two sectors in Sec.~6.
                   
                   In Table \ref{tab:p2} we list a set of parameters producing a larger ratio of the EW scales of the two sectors.

                  \begin{table}[h!]
                  \scriptsize
                  \centering
                  \caption{Higgs Potential parameter sample and associated masses of physical scalars in each sector. Masses are in units of $GeV$. This example point in parameter space for $\rho=3000$ will be used 
                  throughout this work to exemplify the large $\rho$ case.} 
                  \label{tab:p2}
                  \begin{tabular}{ |l|l|l|l|l|l|l|l|l|l|l|l|l|l|l|l|l| }
      \bottomrule  $m^2_{11}=-6.7^2$ & $m^2_{22}=-183150^2$& $m^2_{12}=-6.7^2$  & $z_1=0.13$    & $z_2=0.13$              & $z_3=0.8$     & $z_4=0.01$        & $z_5=0.01$       & $z_6=0.01$ &$z_7=0.01$     \\ \hline     
                   $z_8=0.01$    & $z_9=0.8$ & $z_{10}=0.8$ & $z_{11}=0.01$& $z_{12}=10^{-9}$ & $z_{13}=0.01$ & $z_{14}=0.01$ &$m_h=125$   &  $m_{h_D}=366300$  &   $m_{A_1}=590729$   \\ \hline
                   $m_{A_2}=645018$   & $m_{J_1}=590729$   & $m_{J_2}=645018$     &   $m_{{H^+}}=590729$    &    $m_{{H^+}'}=645018$    &   $v=246$ & $w=738 000$ & $\text{tan}(\beta)=10^{-4}$ & $\text{tan}(\beta')=10^{7}$ & $\rho=3000$ \\ 
                \toprule  
                \end{tabular}
                \end{table}
                 
                 The mixing matrix in the case of \ref{tab:p2} is then given by
                 \begin{equation}\label{eq:matIII}
  U_{III}=
                 \left(
\begin{array}{cccccc}
 -1. & 0 & 0 & 0 & 0.00002183 & 0 \\
 0 & 1. & 0 & 0 & 0 & 0 \\
 -0.00002183 & 0 & 0 & 0 & -1. & 0 \\
 0 & 0 & -1. & 0 & 0 & -0.000149029 \\
 0 & 0 & 0 & 1. & 0 & 0 \\
 0 & 0 & -0.000149029 & 0 & 0 & 1. \\
\end{array}
\right).
\end{equation}
                 
                 We can see that these potentials allow for ASB that can give significant differences to the two sectors while beginning with a mirror symmetry. 
                 In the next section we will focus on how the independent Yukawa couplings and the dark EW scale $w$
                 can vary without significantly changing the confinement scale of the $SU(3)'$ gauge group when compared to that of the visible $SU(3)$.
                 \newpage
                 \subsection{Dark confinement}
                  At high energy the mirror symmetry between the sectors imposes the condition that the gauge 
                  coupling of the $SU(3)$ and $SU(3)'$ groups are the same above the dark electroweak scale, after which
                  the mirror symmetry is broken and the couplings may become differentiated. In particular we have very different masses for the dark quarks compared to the ordinary quarks, which result from a combination of the larger electroweak VEV of 
                  $H_1'$ and Yukawa couplings which are almost entirely independent of the couplings to $H_1$.
                  This will in turn set the scale of quark mass threshold corrections in the running of $\alpha_3'$ such that it 
                  will become non-perturbative at an energy scale above that of the standard model. 
            
                  \begin{figure}[t!]
\begin{minipage}{\textwidth}
  \centering
  \includegraphics[width=.98\textwidth]{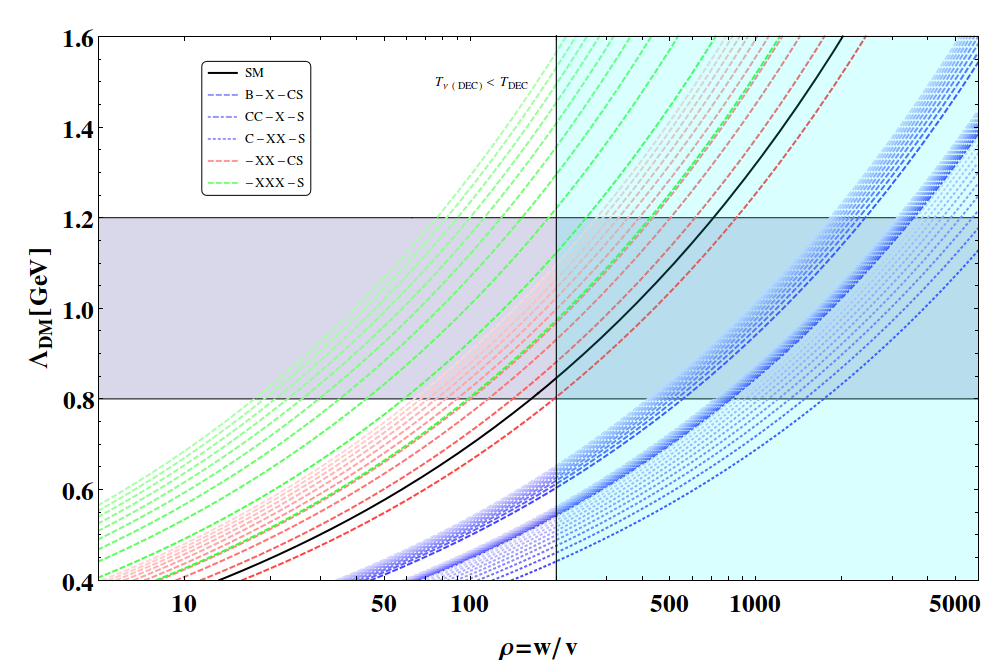}\quad 
  \caption{The confinement scale as a function of $\rho$, the ratio of EW scales, for a variety of Yukawa coupling hierarchies. The SM Yukawa range is shown for 
  reference in the black curve. Each of the sets of Yukawa couplings is divided into 
  ranges where one common coupling for one, two or three dark quark flavours varies between a 
  maximum at the SM Yukawa coupling constant listed first and a minimum at the SM Yukawa coupling constant listed last. 
  Yukawa coupling constants vary from left to right in each colour band in ten even divisions. For example, C-XX-S indicates that 
  the largest coupling is equal in magnitude to the SM charm Yukawa, then there are two identical Yukawa couplings that vary 
  between the SM charm and strange Yukawas, and a fourth Yukawa with a magnitude equal to that of the SM strange quark. 
  Each of these examples includes two light quarks which have masses far below the confinement scale. The red- and green-dashed cases have a 
  Yukawa coupling constant upper value set equal to that of the SM top quark.
  The blue shaded region indicates the divide between two different regions of parameter space that we explore, 
   one with light mirror neutrinos in the present day (left of vertical line), and one without. The grey shaded region indicates the
  desired range for the dark confinement scale to explain the similarity between dark matter and proton mass scales.}
  \label{fig:QCDS}
\end{minipage}\\[1em]
\end{figure}

                  At one-loop this relationship between confinement scale and coloured fermion mass thresholds was explored in Ref.~\cite{Lonsdale:2014wwa}. The relationship is
                  dependent on the scale at which the couplings begin to evolve to different values, 
                  which can be taken to be the mass of the heaviest dark quark in the one-loop case, and the number of dark quark flavours that switch off
                  in the running before the non-perturbative regime is reached. By solving the RGEs of each $SU(3)$ group we calculate the value of the dark confinement scale by finding $\alpha_3$ at 
                  the mass of the heaviest dark quark, then running $\alpha_3'$ to the point where it becomes asymptotic.

                  We limit ourselves to examining the cases where the number of these 'heavy' dark quarks is 1, 2, 3 or 4 in contrast to the SM where the $c$,$b$,$t$-quarks are heavy while $u$,$d$,$s$-quarks are light. 
                  The reason for this is to leave two dark quarks with small masses to form the dark baryons, a point we will return to in Sec.~\ref{sec:Temp}.
                  The principal goal of examining the quark Yukawa hierarchies is to examine which cases will allow for a dark confinement scale, $\Lambda_{\text{DM}}$, that is approximately five times that of the standard model value, $\Lambda_{\text{QCD}}$.
                  As Sec.~\ref{sec:Leptogenesis} will explore how to obtain near equal number densities of visible and dark baryons, this difference between confinement scales will be the cause of dark matter's 
                  larger role in the universe's mass density.
                  
                  In Fig.~\ref{fig:QCDS}, we plot the dark confinement scale as a function of the electroweak VEV ratio $\rho$ for different dark quark mass hierarchy cases chosen for illustrative purposes.
                  The equation for the central black line at one-loop can be found by taking Eq. 6 from Ref.~\cite{Lonsdale:2014wwa} and replacing each of ($m_1,m_2,m_3$) with ($m_c\, \rho, m_b\, \rho, m_t\, \rho$) along with replacing the scale $U$ with $m_t\, \rho$.
                  The horizontal grey shaded region shows the target area for Yukawa hierarchies that produce a $\sim 5 \, \text{GeV}$ dark baryon for a given VEV ratio.
                  This serves as a conservative estimate of the range of uncertainties in the baryon masses of our considered dark sectors combining the range of baryon mass scale values for different choices of interquark potentials in Ref.~\cite{Lonsdale:2017mzg}
                  and the possible contributions from more massive bare dark quark masses up to $100$ MeV.
                  As the size of the dark EW scale affects the entire dark quark mass  
                  spectrum, we see what the dark confinement scale is for a variety of hierarchies at each value of $\rho$.  The black solid curve shows the SM Yukawa hierarchy case for reference. 
                  As we will discuss in Sec.~\ref{sec:Temp}, depending on the assumptions of the model with regard to the thermal history, 
                  different limits on $\rho$ come from constraints on the abundance of  
                  charged dark baryons as well as from constraints on visible sector flavour physics and successful leptogenesis.

                \section{\bf Neutrino masses}\label{sec:Neutrinos}
                The neutrinos of our model consist of the left-handed states of the visible sector, heavy right-handed states with Majorana mass terms,
                along with all of the associated mirror counterparts: heavy left-handed states of the dark sector and light right-handed states. 
                Above the mirror breaking scale we have only the mixing between the heavy Majorana states given by Eq.~\ref{eq:majorana}. 
                The cross-sector mixings, given by the parameters $P_{ij}$, must be suppressed in order
                to prevent too much mixing between light neutrinos in each sector once the heavy degrees of freedom are integrated out. 
                This requirement is consistent with technical naturalness from independent Poincar\'{e} symmetries, as discussed in Sec.~\ref{sec:ASB}. If the matrix $P$ is the product of 
                a small dimensionless cross-sector coupling and a right-handed neutrino mass scale, $\sim M$, that develops near the GUT scale, then we expect $M \gg P$.
                However it must be noted that any non-zero value of $P$ will induce maximal mixing for the heavy mass eigenstates as mirror parity commutes with the Hamiltonian. 
                The heavy states that undergo $CP$-violating decays to produce the lepton asymmetry of both sectors are thus equal admixtures of 
                visible and mirror matter states. This is similar to the case in Ref.~\cite{Gu:2014nga} of an $SO(10)\otimes SO(10)$ model. 
                
                Below the scale of symmetry breaking, we consider both the heavy and light neutrino states of the theory.
                For the Lagrangian given by Eqs.~\ref{eq:majorana} and \ref{eq:yukawa}, we have mass terms in the basis $(\overline{\nu^{\phantom{c}}_L}, \overline{\nu_R^{c}}', \overline{N_R^c},\overline{N_L}')$ given by
                \begin{equation}\label{eq:matrixeqn}
                \mathbb{M}=
                \begin{pmatrix}
                0 & 0 &  y_1 v& f_1 v               \\
                0 & 0 & f^*_2 w & y^*_2  w          \\
                y_1^{\dagger} v & f_2^{T} w & M & P \\
                f_1^{\dagger} v& y_2^{T}w & P^{T} & M 
                \end{pmatrix}.
                \end{equation}
                 Above the scale at which parity is broken, where the mass eigenstates must also 
                 be parity eigenstates, the matrix only contains the lower right $2 \times 2$ matrix and, with inter-family mixing switched off,
                 we can write the heavy Majorana states for each family as maximal linear combinations of visible and dark sector flavour states \cite{Berezinsky:2003fb, Foot:1999ph, Bell:2000bu},
                \begin{equation}
                 N^{\pm}= \frac{1}{\sqrt{2}} \left({(N'_L)^C \pm N_R} \right).
                \end{equation}
                 These have masses $M_N^{\pm}= M \pm P$. By taking the cross-sector mass $P$ to be small, they become near degenerate at a scale $M_N \simeq M$. In this way, for 
                 each generation, we gain two near-degenerate mass eigenstates with masses $M_N=(M_1, M_2, M_3)$.
                 We see directly that taking this cross-sector mass to be small also sets the mixing of light visible and mirror neutrinos to be minimal once the heavy neutrinos are integrated out.
                 
                 In the general three-generation case, above the scale of mirror symmetry breaking, the heavy Majorana mass eigenstates, $N_i$ will be divided into parity-even and parity-odd states,
                \begin{align}
                 &N^{+}_{i}=  \frac{1}{\sqrt{2}} \left[ \alpha^+_i(N_{R_1} + N'_{L_1}) +\beta^+_i(N_{R_2} + N'_{L_2}) +\gamma^+_i(N_{R_3} + N'_{L_3}) \right]     \\ \nonumber      
                 &N^{-}_{i}=  \frac{1}{\sqrt{2}} \left[ \alpha^-_i(N_{R_1} - N'_{L_1}) +\beta^-_i(N_{R_2} - N'_{L_2}) +\gamma^-_i(N_{R_3} - N'_{L_3}) \right],
                 \end{align}
                 with $\left| \alpha^{\pm}_i \right|^2 + \left| \beta^{\pm}_i\right|^2 +\left| \gamma^{\pm}_i\right|^2=1$ and $i=1,2,3$. We will consider the case of hierarchical thermal leptogenesis
                 where the near equal masses $M_1 \pm \frac{P}{2}$ of the lightest heavy states will be relevant to the temperature where
                 out of equilibrium $CP$-violating decays generate lepton asymmetries in each sector.
                
                \subsection{Small cross-sector coupling case}
                
                We first consider the case where the cross-sector Yukawa couplings in Eq.~\ref{eq:yukawa} satisfy $y_i \gg f_i$. 
                Below the scale of symmetry breaking, where mirror parity has been broken, the mass eigenvalues effectively result in two independent seesaw mechanisms \cite{Chacko:2016hvu} with masses given by
                \begin{equation}
                 m_\nu \simeq \frac{(y_1 v)^2}{M_N}\left(1+ \mathcal{O}\left(\frac{P}{M_N}\right)\right) ,   \qquad   m_{\nu'} \simeq \frac{(y_2 w)^2}{M_N}\left(1+ \mathcal{O}\left(\frac{P}{M_N}\right)\right).
                \end{equation}
                So, the light neutrino states of the dark sector gain larger masses than their visible sector counterparts because of the larger Dirac mass terms, 
                while being suppressed by the same Majorana mass scale. The Dirac masses will differ on two accounts: 
                the larger electroweak scale $w$ and the different couplings of $y_2$ compared to $y_1$, with the dark Dirac mass being the larger unless $y_2$ is very much smaller than $y_1$.
                The limit on the number of effective relativistic degrees of freedom during big bang nucleosynthesis and from the cosmic microwave background will limit how many dark neutrino flavours can be relativistic.
               
                 As we have two seesaw mechanisms, one in each sector, that rely on different Yukawa couplings we now turn to parameterising the couplings and masses of the three light flavours.
                 The visible sector has mass matrix $m_\nu= v^2 y_1 D^{-1}_M y_1^T$ while in the dark sector we have $m_{\nu'} \simeq w^2 y_2 D_M^{-1} y_2^T$, where $D_M$ is the 
                 diagonalised Majorana mass matrix of the heavy sector. Because we are considering the regime $P \ll M$, the dark sector matrix $D_{M'} \simeq D_M$.
                 Each of these is independently diagonalised into $D_m$ and $D_{m'}$ by different matrices which
                 we label $U$ and $U'$, respectively. Following the Casas-Ibarra \cite{Casas:2001sr} parametrisation, we then have in the exact asymmetric vacuum configuration that
                 \begin{equation}\label{eq:casa}
                  y_1= \frac{1}{v} U D^{\frac{1}{2}}_m R D^{\frac{1}{2}}_M\quad \text{and}\quad y_2= \frac{1}{w} U' D^{\frac{1}{2}}_{m'} R' D^{\frac{1}{2}}_{M}
                 \end{equation}
                 where $R$ and $R'$ are independent complex, orthogonal matrices. In particular $U'$ will contain phases from the rotation of charged mirror leptons which gain mass from $H_2$ such that
                 \begin{equation}
                  U' = (W^l_L)^\dagger U^{\nu'},
                 \end{equation}
                 where $U^{\nu'}$ diagonalises the low energy $m_{\nu'}$. $W^l_L$ also depends strongly on the texture of $\eta^l_2$. In solving the Boltzmann equations for the related Yukawa couplings above, sample $\eta^l_2$ matrices are chosen such that the mass eigenstates 
                 are in accordance with the thermal history assumptions for dark lepton masses in that region of parameter space while $U^{\nu'}$ is an unconstrained unitary matrix as the PMNS matrix of the dark sector is not known. 
                 
                 \emph{A priori}, the dynamics of thermal leptogenesis could be dominated by the $y_1$ coupling, or $y_2$, or they could play comparable roles.
                 However, it turns out to be advantageous to focus on the parameter space where
                 $y_2$ is the dominant coupling for the creation of a population of heavy neutrino states and their subsequent decay when $T < M_1 $. 
                 It will then also be the major contribution to the lepton asymmetry created in both the visible and the dark sectors. 
                 In this case, successful thermal leptogenesis imposes more constraints on the dark sector's low temperature light neutrino parameter space, in contrast with the situation of ordinary thermal leptogenesis. 
                 In the next section, we examine how an $L$ asymmetry can form in both sectors during the mirror symmetric temperature regime.
                 
                 \subsection{Significant cross-sector coupling case}
                 
                 We now consider the case where the cross-sector terms $f_i$ are comparable to the other Yukawa couplings. These will create Dirac mass terms 
                 that induce some mixing between the light neutrinos of the two sectors.
                 We will see in the next two sections that the bounds on the number of effective relativistic  degrees of freedom in the dark sector 
                 require one of two possibilities. First, the dark EW scale $w$ can be small enough 
                 that any relativistic neutrino species of the dark sector remain in thermal equilibrium until the eventual temperature drop of the dark sector, and 
                 therefore fall to a lower temperature together with the dark photons. (Recall that dark radiation bounds require that the temperature of the dark sector be sufficiently lower than that of the visible sector by the time of BBN.)
                 
                 The second possibility is that the dark EW scale is large enough so that the dark neutrinos become non-relativistic and subsequently decay from the plasma prior
                 to the temperature shift. In this latter case we will 
                 require that the dark neutrinos are heavy enough to decay to SM species with a short enough lifetime.
                 We examine the masses and mixing in this case. Points in parameter space such as 
                 \begin{equation}\label{eq:para1}
                   f_1 = 0.05,\; f_2=0.0005, \;\rho=3000,\; y_2=0.005, \;y_1=10^{-6}, \;M_{1}= 10^7\ \text{GeV} 
                 \end{equation}
                 allow for heavy states that are still approximately equal admixtures of $N$ and $N'$, but there now exists a small amount of mixing between the light states of the
                 two sectors. This parameter point will be used for illustration throughout the rest of this paper for the large $\rho$ situation.
                 With the much larger EW scale of the dark sector we can have all three flavours of dark neutrino in the $100 \, \text{MeV}$ range 
                 and mixing between the two light species of order $\delta_{\nu \nu'}=10^{-3}$ where $\tilde{\nu}= \nu + \delta_{\nu \nu'} \nu'$. This is similar to the case of Ref.~\cite{An:2009vq}, except that in our case
                 it is mixing between the light neutrinos of one sector with the heavy states of the opposing sector that causes the dark sector light neutrinos to be more massive than the visible sector light neutrinos. 
                 The light neutrinos of the dark sector then 
                 decay to visible sector species via $\nu' \rightarrow \nu e^+ e^-$ just prior to BBN and the only remaining light degrees of freedom in the dark sector ares from the dark photons.
                 With these larger $f_i$ couplings we need to examine the subsequent additional terms in the Boltzmann equations in the era of thermal leptogenesis.
                 Since we will consider at that temperature a population of mass eigenstates $N^{\pm}$, this will amount to an extra pair of terms contributing to the 
                 total decay rate, $\Gamma(N \rightarrow \phi l)$, in each sector and modify the washout rates and $CP$ asymmetry parameter. We explore this and the first case in detail in the next section.

                 \section{\bf Symmetric leptogenesis}\label{sec:Leptogenesis}
                 
                 We now consider how the visible and dark sectors generate comparable amounts of baryon asymmetry and dark-baryon asymmetry. 
                 As with all models of baryogenesis we require the fulfilment of the Sakharov conditions, that is, a $B$-violation 
                 mechanism, $C$- and $CP$-violation, and a departure from thermal equilibrium. 
                 Mirror symmetry then provides an associated mechanism in the dark sector, though the exact process will differ once mirror symmetry is broken. 
                 Ultimately it is necessary for the two sectors to be different to explain why there is more dark matter than visible matter by mass in the universe. 
                 Either DM is more massive than the proton or the dark sector contains a higher concentration, or it is a combination of each,
                 to such a slight degree that the mass densities remain within the same order of magnitude.
                 The dark sector must also have a mechanism for the symmetric components of the plasma to annihilate, for example through dark photons.
                 
                 In our model, the $B$ and $B'$ asymmetries are created through 
                 sphaleron effects that partially convert lepton asymmetries generated through high scale thermal leptogenesis from the decays of $N_1^\pm$, the lightest of the heavy neutral lepton pairs. 
                 As most of the lepton asymmetry will develop at a scale above that of mirror parity breaking,
                 the generation of the mirror lepton asymmetry will proceed almost identically. Following this the sphaleron effects in 
                 each sector will convert these lepton asymmetries into similar amounts of 
                 $B$ and $B'$ asymmetries. The amount of $B$ and $B'$ will not be identical as sphaleron effects in the dark sector will switch off at a higher temperature than in
                 the visible sector, due to $w \gg v$ causing the dark EW phase transition to occur before the visible counterpart. 
                                  
                 Since the second Higgs doublet in the visible sector can 
                 provide a way to generate a sufficient amount of lepton asymmetry while not contributing to the
                 Higgs squared-mass corrections usually seen in the thermal leptogenesis case, the associated naturalness bound~\cite{Vissani:1997ys} can be avoided. This can be compared to recent 
                 work in Ref.~\cite{Clarke:2015hta} which examined how the tension between Higgs mass corrections and the requirements of standard leptogenesis
                 can be solved with a second Higgs doublet.
                 The Higgs squared-mass correction from the right-handed neutrino mass scale is
                 \begin{equation}\label{eq:Vissanibound}
                 \delta \mu^2 \approx \frac{1}{4 \pi} y_1^2 M_N^2  ,
                 \end{equation}
                 which is independent of $y_2$.
                 The mass corrections to the heavier bosons will be larger if $y_2$ is larger, however since 
                 the mass scale of the dark Higgs, the dark EW scale, and the other decoupled scalars are orders of magnitude larger also, 
                 the quantum corrections have the capacity to still be natural. If one adopts the criterion that such corrections should be 
                 no larger than of order $1 \, \text{TeV}^2$ \cite{Vissani:1997ys} for the SM Higgs, 
                 then Eq.~\ref{eq:Vissanibound} translates to an upper bound of $\sim 3 \times 10^7 \, \text{GeV}$ for the Majorana mass in the single flavour case for the small $\tan(\beta)$ parameter space
                 we will always consider. Note that our representative large $\rho$ parameter point of Eq.~\ref{eq:para1} meets this naturalness bound.
                
                While we compute the asymmetries directly through numerical solution of the Boltzmann equations (see later), it is useful to also compare those solutions to
                a standard analytic approximation, valid in the `strong washout' regime for the standard leptogenesis case.
                For our situation of maximally mixed heavy neutrino states, $N_i^{\pm}$, in the hierarchical $M_1 \ll M_{2,3}$ regime, the generated 
                 $\mathcal{N}_{B-L}$ and $\mathcal{N}'_{B-L}$ asymmetries are conventionally expressed as 
                 \begin{equation}
                 \mathcal{N}'_{B-L} = \mathcal{N}_{B-L} = -\frac{3}{4} \kappa_f \epsilon_1,
                 \end{equation}
                 where $\kappa_f$ and $\epsilon_1$ are called the `efficiency factor 'and `$CP$-asymmetry parameter', respectively. The former quantifies the extent of asymmetry washout, while the latter obviously
                 scales with the strength of $CP$ violation.
                 The equal values for the two asymmetries is due to the maximal mixing
                 between the heavy visible and dark neutrinos. The $\mathcal{N}$ quantities are particle numbers in a reference co-moving volume containing one photon.
                 
                 For the case of a single doublet, the $CP$-asymmetry factor for the lightest right handed state $N_1$ is
                 \begin{equation}
                 \epsilon_1= \frac{\Gamma (N_1 \rightarrow l \phi^*)- \Gamma (N_1 \rightarrow l^c \phi)}{\Gamma (N_1\rightarrow l \phi*)+ \Gamma (N_1 \rightarrow l^c \phi)}.
                 \end{equation}
                 Adding the second doublet allows for two relevant $CP$-asymmetry factors, from $y_1$ and $y_2$. 
                 Including the mirror sector adds decay channels through $f_1$ and $f_2$ that can generate an asymmetry in the opposite sector.
                
                 In the standard seesaw model it is useful to define a `decay parameter' as per
                 \begin{equation}\label{eq:decayparameter}
                 K = \frac{\Gamma_D(z=\infty)}{H(z=1)},
                 \end{equation}
                 where $z \equiv M_1/T$, and $\Gamma_D(z=\infty)$ is the $N_1$ decay width. This ratio provides a measure of whether $N_1$ is in 
                 thermal equilibrium when it begins to go non-relativistic ($K > 1$) or not ($K < 1$), which in turn defines the strong- and weak-washout regimes, respectively.
                 This parameter is typically used to set bounds on the light neutrino masses in most models. In our case, however, some of the $CP$ asymmetry 
                 is being generated by the coupling of heavy neutrino states to $\Phi_2$ and $\Phi'_2$ and 
                 so, in the visible sector, the Yukawa coupling relevant to mass is not important in terms of the generation of the lepton asymmetry. The efficiency factor $\kappa_f$ is related to $K$, as we discuss below.
                 
                 We consider in this work the generation of a $B-L$ asymmetry in the one-flavour approximation. 
                 While this approximation is typically only completely accurate at temperatures of lepton asymmetry 
                 generation above the scale at which the tau charged lepton Yukawa interactions are in equilibrium, so that 
                 the value of $M_1$ violates the Vissani bounds on naturalness, in this case the role of flavour effects will be more complicated. 
                 In particular the thresholds for when charged lepton Yukawa interactions are in equilibrium will depend on the couplings to $\Phi_2$, as well as to $\Phi_1$.
                 Additionally the charged lepton Yukawa coupling matrices to $\Phi_1$ and $\Phi_2$, which contribute to the thermal masses, cannot in general be simultaneously diagonalised.  
                 A full analysis of possibly significant flavour effects in this case would need to carefully consider the relative 
                 rates among six interactions of lepton doublets $l$ involving $(N \Phi_1, N \Phi_2, N' \Phi_1, N' \Phi_2, \Phi_1 e_R, \Phi_2 e_R)$ as well as the 
                 Hubble rate at each temperature range in order to determine the coherent state evolution.
                 For the present analysis, we make the simplifying assumption that the combination of $y_1, y_2, f_1, f_2$ channels for washout is similar in magnitude for each 
                 individual flavour and so the unflavoured approximation may yield a more accurate result for the total $B-L$ asymmetry with $T < 10^{12} \,\text{GeV}$ in this 
                 case than ordinary Type 1 see-saw thermal leptogenesis models.
                 
                 We follow most of the literature by favouring the strong-washout regime, since predictability is increased by making the final visible and dark asymmetries insensitive to
                 the initial distributions of the heavy neutral lepton states $N_i^{\pm}$. In particular, all the plots below of numerical solutions to the Boltzmann equations are for strong-washout cases.
                 
                 \subsection{Small cross-sector coupling case}
                 
                 We first consider the case with $y_i \gg f_i$, within the hierarchical case for the Majorana masses, $M_2,M_3 \gg M_1$. 
                 The heavy states decay to visible and dark leptons with equal rates and the asymmetry in these rates is also the same.
                 We must also account for the fact that each of $\Phi_1$ and $\Phi_2$ can appear in the self energy and vertex diagrams.
                 We then obtain the expressions for the CP parameters $\epsilon_1^{\phi_1}$ and  $\epsilon_1^{\phi_2}$ 
                 from the decays, $N_1 \rightarrow l \phi_1$ and $N_1 \rightarrow l \phi_2$. These are given by             
                 \begin{align}\label{eq:epsilons}
                  \epsilon_1^{\phi_1} \simeq \sum _{i=1,2} \sum_{k \ne 1} \frac{1}{16 \pi}\frac{M_1}{M_k} \left[\frac{\text{Im}\left[  (y_1^\dagger y_i)_{k1} {(y_i^\dagger y_1)}_{k1}+2(y_1^\dagger y_1)_{k1} {(y_i^\dagger y_i)}_{k1}           \right]}{ {(y_1^\dagger y_1)}_{11}   }  \right],\\ \nonumber
                  \epsilon_1^{\phi_2} \simeq \sum _{i=1,2} \sum_{k \ne 1} \frac{1}{16 \pi}\frac{M_1}{M_k} \left[\frac{\text{Im}\left[  (y_2^\dagger y_i)_{k1} {(y_i^\dagger y_2)}_{k1}+2(y_2^\dagger y_2)_{k1} {(y_i^\dagger y_i)}_{k1}           \right]}{ {(y_2^\dagger y_2)}_{11}   }  \right].
                 \end{align}
                 Note that there is an additional term proportional to $M_1^2/M_k^2$ which is neglected in the hierarchical case we are considering. Such terms are ordinarily exactly zero after summing over lepton flavours, 
                 however $y_2$ allows one of these to survive \cite{Covi:1996wh}. 
                  From the Yukawa couplings, we have $CP$-violating decays 
                 both in $N \rightarrow \phi_1 l$ and $N \rightarrow \phi_2 l$
                 and in the mirror sector processes $N \rightarrow \phi'_1 l'$ and $N \rightarrow \phi'_2 l'$. 
              
                 The tree-level total decay rates of the heavy states are given by
                 \begin{align}\label{eq:fytoo}
                 \Gamma_{N_i}&= \frac{1}{8 \pi} \left[ (y_1^\dagger y_1)_{ii}+(y_2^\dagger y_2)_{ii}+(f_1^\dagger f_1)_{ii}+(f_2^\dagger f_2)_{ii}\right]M_{i}  \simeq 
                 \frac{1}{8 \pi} \left[ (y_1^\dagger y_1)_{ii}+(y_2^\dagger y_2)_{ii} \right]M_{i},
                 \end{align}
                 which may be used to compute the decay parameter $K$ for $N_1$ from Eq.~\ref{eq:decayparameter}.
                 We now consider each of the strong- and weak-washout cases. In the weak-washout regime ($K<1$), inverse decays create an abundance of heavy neutrino states
                 which surpasses the equilibrium number density and their subsequent out-of-equilibrium decays produce an asymmetry which depends on the initial conditions. 
                 The size of the initial population also determines the opposite-sign asymmetry produced during inverse decays
                 as the equilibrium density is reached. The final asymmetry is then a combination of the asymmetry produced in each phase. 
                 In the strong-washout regime ($K>1$), the large coupling will bring any initial population of heavy $N$ states quickly to the equilibrium density. 
                 The strong coupling leads to high rates for decays and inverse decays that wash out 
                 any initial asymmetry, after which a final asymmetry is produced when inverse decays become suppressed as $T < M_1$ and the non-relativistic heavy neutrinos undergo $CP$-violating decays. 
              
                 It is useful to consider the limiting case of taking the contribution from $y_1$ terms to be  small enough that the dominant contribution to the $CP$-violating decays comes from just $y_2$.
                 In this case, the asymmetry is
                 \begin{equation}\label{eq:y2dominantN}
                 \mathcal{N}_{B-L}= - \frac{3}{4} \kappa_f \epsilon_1^{\phi_2} ,
                 \end{equation}
                  and we can use the usual approximate analytical expression for the  strong-washout efficiency factor \cite{Buchmuller:2004tu},
                 \begin{align}\label{eq:kappaf}
                  \kappa_f &\simeq \frac{2}{K z_B} \left[ 1- \text{exp}\left(-\frac{1}{2} z_B K \right)   \right],\\
                  z_B(K)   &\simeq 1 + \frac{1}{2} \text{ln}\left[ 1+ \frac{\pi K^2}{1024}\left(\text{ln}\left(\frac{3125 \pi K^2}{1024}  \right)  \right)^5\right].
                  \label{eq:zB}
                 \end{align}
                 We have checked that Eqs.~\ref{eq:epsilons}, \ref{eq:y2dominantN}-\ref{eq:zB} give an asymmetry value consistent with our numerical solutions to the Boltzmann equations (to be given below) 
                 for the case where $y_2$ dominates. This serves as a useful check of our numerical routine. Note, however, that in the numerical solutions presented below, we actually do not neglect $y_1$ effects.
            
            \subsection{General cross-sector coupling case}
            
            \subsubsection{Boltzmann equations}
                 
                 The general case, where the $f_i$ couplings constants are not necessarily negligible and $y_1$ contributions are included, 
                 must be analysed through numerical solution of the Boltzmann equations. In these equations, we must consider
                 decays, inverse decays, scattering processes and washout processes. The dynamical variables are the particle numbers $\mathcal{N}_1^+$ and $\mathcal{N}_1^-$ and 
                 the visible and dark asymmetries $\mathcal{N}_{B-L}$ and $\mathcal{N}'_{B-L}$. The considered decays/inverse-decays are the four processes
                 \begin{equation}
                 N_1^{\pm} \leftrightarrow \phi_1^* l, \quad \phi_1 l^c,\quad \phi_2^* l,\quad \phi_2 l^c ,
                 \end{equation}
                 and their mirror analogues
                 \begin{equation}
                 N_1^{\pm} \leftrightarrow {\phi'_1}^* l', \quad \phi'_1 {l'}^c,\quad {\phi'_2}^* l',\quad \phi'_2 {l'}^c .
                 \end{equation}
                
                 For scattering processes, it is known that the dominant ones in the standard leptogenesis case are leptons with top quarks, so
                 in our situation we must consider if the Yukawa couplings to quarks of the second Higgs doublet are in a similar range. We thus incorporate the following
                 processes into the Boltzmann equations:
                 \begin{equation} 
                 N_1^{\pm} l^c \leftrightarrow t \bar{q}, \quad N_1^{\pm} q \leftrightarrow t l ,
                 \end{equation}
                 driven by virtual $\phi_1$ exchange in the $s$- and $t$-channels, and the $\phi_2$-driven counterparts,
                 \begin{equation}
                 N_1^{\pm} l^c \leftrightarrow u_i \bar{q},\quad N_1^{\pm} q \leftrightarrow u_i l ,
                 \end{equation}
                 where $q=d, s, b$ and $u_i$ is any charge $+2/3$ quark flavour that has a significant Yukawa coupling to $\phi_2$.  The mirror analogues of these processes also must be included, of course.
                 For the strong-washout regime, processes such as $\phi_1 l \leftrightarrow \phi_1^* l^c$ and the like can be small due to the large mass of the exchanged virtual heavy neutral lepton compared
                 to the temperature at which the asymmetries are generated.
                 
                 In the strong washout regime, the relevant Boltzmann equations are~\cite{HahnWoernle:2009qn}
                 \begin{align}
                 \frac{d\mathcal{N}_{B-L}}{dz} = &-(\epsilon^{\phi_{1}}_{1^+} D^{\phi_1}_+ + \epsilon^{\phi_{2}}_{1^+} D^{\phi_2}_+) (\mathcal{N}_{N^+_1}- \mathcal{N}_{N^\text{Eq}_{1}})  \nonumber \\
                 &-(\epsilon^{\phi_1}_{1^-} D^{\phi_1}_{-}+ \epsilon^{\phi_2}_{1^-} D^{\phi_2}_{-}) (\mathcal{N}_{N^-_1}- \mathcal{N}_{N^\text{Eq}_{1}})- W_T\mathcal{N}_{B-L},   \nonumber  \\
                 \frac{d\mathcal{N}'_{B-L}}{dz} = &-(\epsilon^{\phi_{1}}_{1^+} D^{\phi_1}_+ + \epsilon^{\phi_{2}}_{1^+} D^{\phi_2}_+) (\mathcal{N}_{N^+_1}- \mathcal{N}_{N^\text{Eq}_{1}})  \nonumber \\
                 &-(\epsilon^{\phi_1}_{1^-} D^{\phi_1}_{-}+ \epsilon^{\phi_2}_{1^-} D^{\phi_2}_{-}) (\mathcal{N}_{N^-_1}- \mathcal{N}_{N^\text{Eq}_{1}})- W_T\mathcal{N}'_{B-L},   \nonumber  \\
                 \frac{d\mathcal{N}_{N^+_1}}{dz}   = &- (D_+ +S_+)  (\mathcal{N}_{N^+_1}- \mathcal{N}_{{N^\text{Eq}_{1}}}), \\
                 \frac{d\mathcal{N}_{N^-_1}}{dz}   = &- (D_- +S_-)  (\mathcal{N}_{N^-_1}- \mathcal{N}_{{N^\text{Eq}_{1}}}).\nonumber
                 \end{align}
                 We now define the various coefficients on the right-hand sides of these equations.

                 The $CP$ parameters for $N_1^{\pm}$ can be written
                 in terms of $Y_i=y_i+f_i, F_i=y_i-f_i$ as per,
                 \begin{align}\label{eq:cpwithf}
                  \epsilon_{1^+}^{\phi_1} = \sum _{i=1,2} \sum_{k} \frac{1}{16 \pi} &\left[\text{R}(M^+_{i},M^-_{k})\frac{\text{Im}\left[2\left((Y_1^\dagger F_1)_{k1} {(Y_i^\dagger F_i)}_{k1}\right) +  2\left( (Y_1^\dagger Y_1)_{k1} {(Y_i^\dagger Y_i)}_{k1} \right)   \right]}{ {(Y_1^\dagger Y_1)}_{11}   }    \right. \nonumber  \\
                      &\qquad +  l\left(\frac{M_1}{M_k}\right)\frac{ \left. \text{Im}\left[ (Y_1^\dagger Y_1)_{k1} {(F_i^\dagger F_i)}_{k1} +(Y_1^\dagger Y_i)_{k1} {(Y_i^\dagger Y_1)}_{k1} \right. \right.}{ {(Y_1^\dagger Y_1)}_{11}   } \nonumber\\
                       &\qquad \qquad \qquad \left.\frac{\left.+(F_1^\dagger F_i)_{k1} {(Y_i^\dagger Y_1)}_{k1}+(Y_1^\dagger Y_1)_{k1} {(F_i^\dagger Y_i)}_{k1}     \right]}{ {(Y_1^\dagger Y_1)}_{11}   }  \right],\\
                           \epsilon_{1^+}^{\phi_2} = \sum _{i=1,2} \sum_{k} \frac{1}{16 \pi} &\left[\text{R}(M^+_{i},M^-_{k})\frac{\text{Im}\left[2\left((Y_2^\dagger F_2)_{k1} {(Y_i^\dagger F_i)}_{k1}\right)+  2\left( (Y_2^\dagger Y_2)_{k1} {(Y_i^\dagger Y_i)}_{k1} \right)   \right]}{ {(Y_2^\dagger Y_2)}_{11}   }    \right. \nonumber  \\
                      &\qquad+  l\left(\frac{M_1}{M_k}\right)\frac{ \left. \text{Im}\left[   (Y_2^\dagger Y_2)_{k1} {(F_i^\dagger F_i)}_{k1} +(Y_2^\dagger Y_i)_{k1} {(Y_i^\dagger Y_2)}_{k1} \right. \right.}{ {(Y_2^\dagger Y_2)}_{11}   } \nonumber\\
                     &\qquad \qquad \qquad\left.  \frac{\left.+(F_2^\dagger F_i)_{k1} {(Y_i^\dagger Y_2)}_{k1}+(Y_2^\dagger Y_2)_{k1} {(F_i^\dagger Y_i)}_{k1}     \right]}{ {(Y_2^\dagger Y_2)}_{11}   }  \right], \nonumber
                  \end{align}
                 where 
                 \begin{equation}
                  l(x)=x\left[1-(1+x^2)\text{ln}\left(\frac{1+x^2}{x^2}\right)+\frac{1}{1-x^2}\right],
                 \end{equation}
                 and 
                 \begin{equation}
                  R(M^+_i,M^-_k) =\frac{{M^+_i}({M^+_i}^2-{M^-_k}^2)}{({M^+_i}^2-{M^-_k}^2)^2 +{M^+_i}^2 \Gamma_{N^-_k}^2 }.
                 \end{equation}
                 The  $\epsilon_{1^-}$ $CP$-parameters are obtained simply by interchanging $F$ and $Y$. 
                 Note that the sum now extends to $k=1$ and we have a non-zero $CP$ parameter for some of these terms.\footnote{These expressions can be compared to the case of the 
                 double seesaw mechanism of leptogenesis \cite{Gu:2010xc}.} 
                 The tree level, loop and vertex decay diagrams are shown in Fig.~\ref{fig:interferncediagrams} while the relevant scattering diagrams are listed in Appendix B in Figs.~\ref{fig:ss1} and \ref{fig:ss2}.
                 The decay terms in the above equations are given by
                 \begin{align}
                 &D^{\phi_1}_+=z \frac{(Y_1^\dagger Y_1)_{11}}{H(z=1)} \frac{K_1(z)}{K_2(z)}\,,\,\,\,\, D^{\phi_2}_+=z \frac{(Y_2^\dagger Y_2)_{11}}{H(z=1)} \frac{K_1(z)}{K_2(z)},\\ \nonumber
                 &D^{\phi_1}_-=z \frac{(F_1^\dagger F_1)_{11}}{H(z=1)} \frac{K_1(z)}{K_2(z)}\,,\,\,\,\, D^{\phi_2}_-=z \frac{(F_2^\dagger F_2)_{11}}{H(z=1)} \frac{K_1(z)}{K_2(z)},
                 \end{align}
                 and
                 \begin{equation}
                 D_{\pm} = 2 ( D^{\phi_1}_{\pm} + D^{\phi_2}_{\pm} )
                 \end{equation}
                 with $K_{1},\, K_{2}$ being modified Bessel functions of the second kind of order one and two, respectively. 
                 
                 For scatterings, we have that
                 \begin{equation}
                 S_\pm=2(S^s_\pm + 2 S^t_\pm)
                 \end{equation}
                 where
\begin{equation}\label{eq:Sbolt}
S^{s,t}_\pm=\frac{\Gamma^{s,t}_\pm}{H z},
\end{equation}
with $s, t$ denoting $s$- and $t$-channel process, respectively~\cite{HahnWoernle:2009qn}. The scattering rates are 
\begin{equation}\label{eq:Srate}
\Gamma^{s,t}_\pm = \frac{M_1}{24 \zeta(3)g_N \pi^2}\frac{I^{s,t}_\pm}{K_2(z)z^3},
\end{equation}
where
\begin{align}\label{eq:IST}
 I^{s,t}_\pm&=\int_{z^2}^{\infty} d \Psi \, \hat{\sigma}^{s,t}_\pm(\Psi) \,\sqrt{\Psi}\, K_1(\sqrt{\Psi}), \nonumber \\
\hat{\sigma}^{s,t}_+&=\frac{3}{16 \pi} \left[ (\eta_1^{t})^2(Y_1^\dagger Y_1)_{11}+(\eta_2^{{u_i}})^2(Y_2^\dagger Y_2)_{11}\right] \chi^{s,t}(x),\\
\hat{\sigma}^{s,t}_-&=\frac{3}{16 \pi} \left[ (\eta_1^{t})^2(F_1^\dagger F_1)_{11}+(\eta_2^{{u_i}})^2(F_2^\dagger F_2)_{11}\right] \chi^{s,t}(x), \nonumber
\end{align}
 and the functions $\chi^{s,t}$ are given in Eqs.~(4.7) and (4.8) of Ref.~\cite{HahnWoernle:2009qn}. The parameter $\eta_1^t$ is the top Yukawa coupling constant, while $\eta_2^{{u_i}}$ is the Yukawa
 coupling constant for any charge $+2/3$ quark to $\phi_2$ that is comparable in magnitude to $\eta_1^t$. Note that the label $t$ has, awkwardly, two meanings in these expressions.
 The washout from inverse decays is then given by 
\begin{equation}\label{eq:WID}
W_{ID}=\frac{1}{2}(D^{\phi_1}_+ +D^{\phi_2}_+ +D^{\phi_1}_-  +D^{\phi_2}_- )\frac{\mathcal{N}_{{N^\text{Eq}_{1}}}}{\mathcal{N}_{{l^\text{Eq}}}},
\end{equation}
while
\begin{equation}\label{eq:WT}
 W_T=W_{ID}\left[ 1 + \frac{\left(2 \left(\frac{\mathcal{N}_{N^+_1}}{\mathcal{N}_{{N^\text{Eq}_{1}}}} S^s_+ + \frac{\mathcal{N}_{N^-_1}}{\mathcal{N}_{{N^\text{Eq}_{1}}}}S^s_-\right) + 4( S^t_+ +S^t_-)  \right)}{(D^{\phi_1}_+ +D^{\phi_2}_+ +D^{\phi_1}_-  +D^{\phi_2}_- )}\right]
\end{equation}
is the total washout rate. We now comment on some important features of the solutions of these equations and the various coefficients.

\begin{figure}[!ht]
                \begin{minipage}{\textwidth}
                \centering
                \includegraphics[width=.26\textwidth]{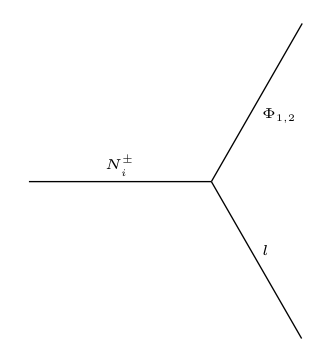}\hspace{3.5cm}
                  \includegraphics[width=.26\textwidth]{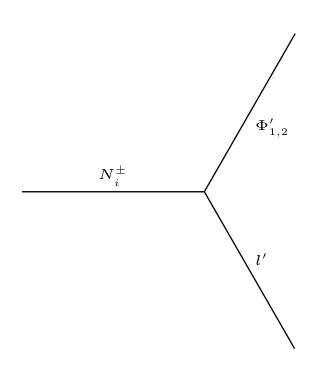}\\ \vspace{0.01cm}
                   \includegraphics[width=.44\textwidth]{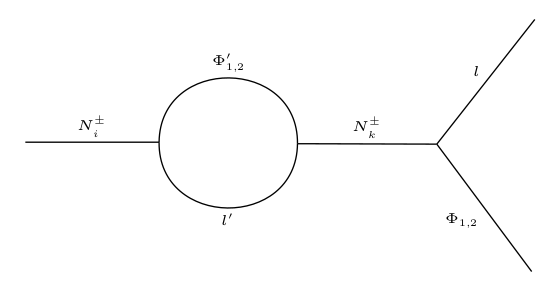}\hspace{0.5cm}
                  \includegraphics[width=.44\textwidth]{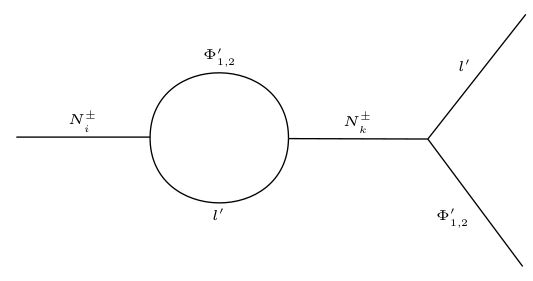}\\ \vspace{0.6cm}
                   \includegraphics[width=.44\textwidth]{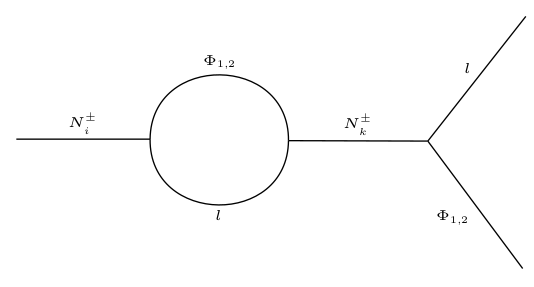}\hspace{0.5cm}
                  \includegraphics[width=.44\textwidth]{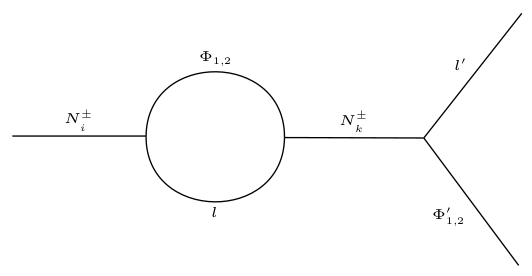}\\ \vspace{0.6cm}
                   \includegraphics[width=.44\textwidth]{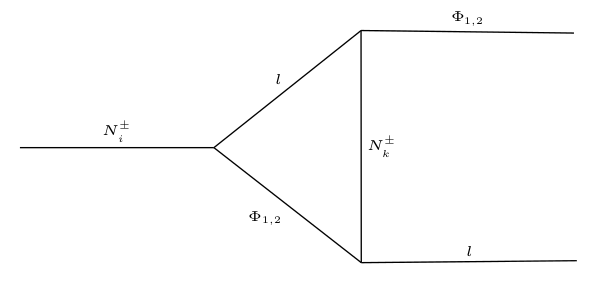}\hspace{0.5cm}
                  \includegraphics[width=.44\textwidth]{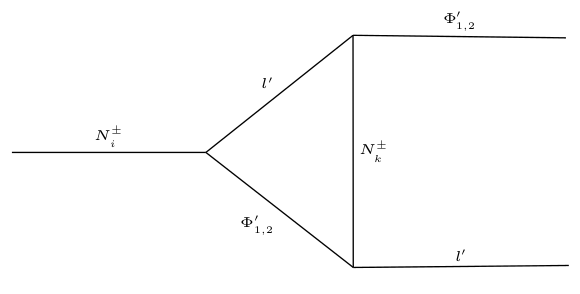}\\ \vspace{0.6cm}
                        \caption{The interference of all diagrams in the left column for a fixed final state $\Phi_1$ and initial state $i=1^+$ contributes to $\epsilon^{\phi_1}_{1^+}$ while fixing $\Phi_2$ and $i=1^+$ allows us to calculate $\epsilon^{\phi_2}_{1^+}$.
                        Choosing $i=1^-$ yields $\epsilon^{\phi_{1,2}}_{1^-}$.
                        Likewise the right hand column gives the interference among mirror counterpart diagrams in the dark sector.}
                \label{fig:interferncediagrams}
                \end{minipage}\\[1em]
                \end{figure} \clearpage

\subsubsection{Small $\rho$ versus large $\rho$}

                 An important issue is the scale of EWSB in the dark sector. We will consider two separate regions of parameter space that are compatible with 
                 all of the other features in the model. We can classify these as the small $\rho$ regime with $\rho<200$ and the large $\rho$ regime, as defined by the vertical line in Fig.~\ref{fig:QCDS}.
                 These different cases require different treatments for a number of key reasons.
                 
\begin{figure}[!ht]
                \begin{minipage}{\textwidth}
                \centering
                \includegraphics[width=.49\textwidth]{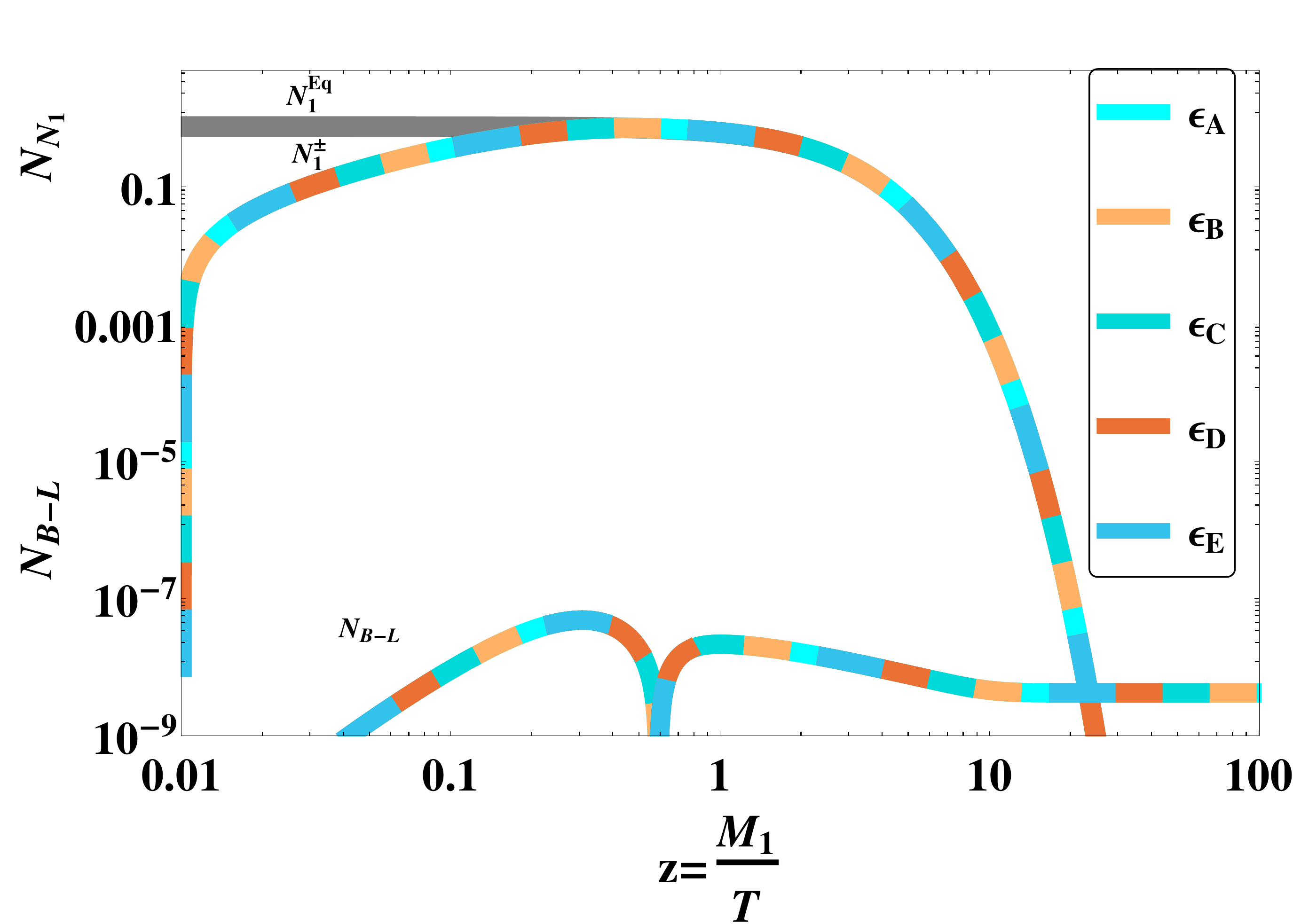}\
                \includegraphics[width=.49\textwidth]{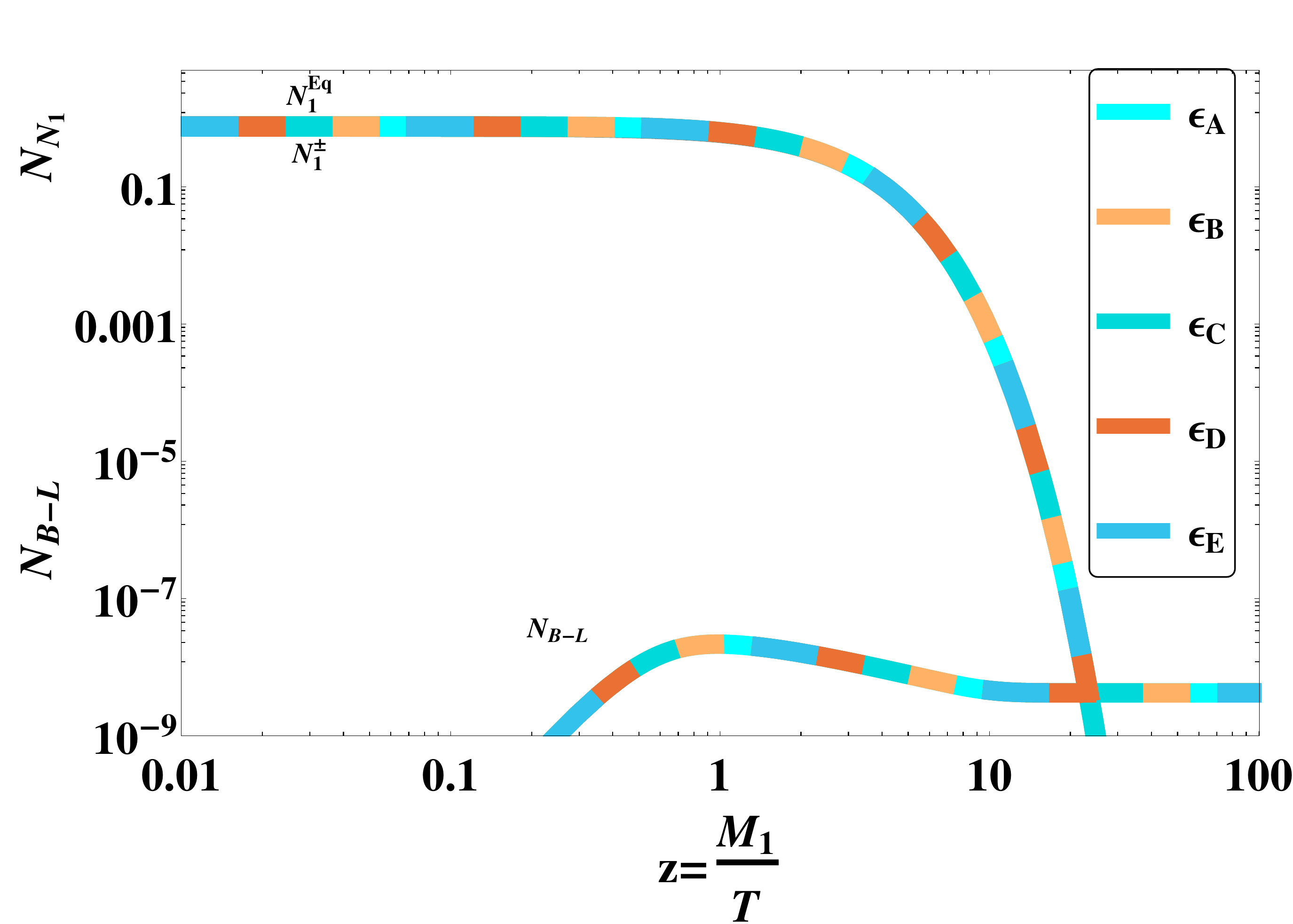}
                \caption{
                 Evolution for the $N_1$ number density (the sum of the densities of $N_1^+$ and $N_1^-$ divided by two) and the $B-L$ asymmetry for the small $\rho$ case for 
                 different choices of the mass of the heaviest dark light neutrino.  The left panel has
                 zero initial $N_1$ abundance, while the right panel has an initial thermal abundance. In all cases, the observationally required $B-L$ asymmetry is produced at the
                 low temperature end, and is independent of the initial abundance because we work in the strong washout regime.
                 Case A corresponds to 1.1 eV, case B to 3.2 eV, while cases C-E have 5.6 eV for the dark neutrino mass upper limit. 
                 Each curve illustrates an allowed choice for the Casas-Ibarra $R$ and $R'$ matrices with $\rho=30$ and $M_1=10^9$ GeV.
                 The small $\rho$ case allows for larger quark scattering rates to $\phi_2$ and is comparable to past works in thermal leptogenesis. 
                 The limits on the $CP$ parameters $\epsilon_{A-E}$ come from limiting the amount of dark matter composed of hot dark neutrinos. }
                \label{fig:sub1}
                \end{minipage}\\[1em]
                \end{figure}

                 In the small $\rho$ regime, dark EW interactions are still in equilibrium at the time of the kinetic decoupling of the two sectors. 
                 This means the light neutrinos of the dark sector will undergo a temperature change along with the dark photons, 
                 as they have not yet decoupled. Because of this, it is possible to have light neutrinos in the dark sector and still satisfy the constraints on $N_{\rm eff}$, as we will discuss in Sec.~\ref{sec:Temp}. 
                 The light neutrinos of the dark sector have their masses constrained by the requirement that they do not significantly contribute as a hot dark matter candidate. This imposes limits on the size of $y_2$ in this case. 
                 These constraints will also depend on the temperature of the light mirror neutrino states which will be lower than the temperature of SM neutrinos.
                 In Sec.~\ref{sec:Temp}, we will show that these limits on $m_{v'}$ in the low $\rho$ case depend on the thermal history assumptions. 
                 We consider illustrative cases of the heaviest dark light neutrino limited by upper bounds of $1.1\ \text{eV} $, $3.2\ \text{eV}$ and $5.6\ \text{eV}$. Sample $R$ and $R'$ matrices for the two seesaw mechanisms are then 
                 explored in cases $A$ and $B$ in Fig.~\ref{fig:sub1} for the first two bounds while the largest temperature difference that allows $5.6\ \text{eV}$ has three sample pairs of matrices $R$ and $R'$, denoted cases $C$ to $E$.
                 All these examples give rise to the correct low-temperature baryon asymmetry.

\begin{figure}[!ht]
                \begin{minipage}{\textwidth}
                \centering
                \includegraphics[width=.49\textwidth]{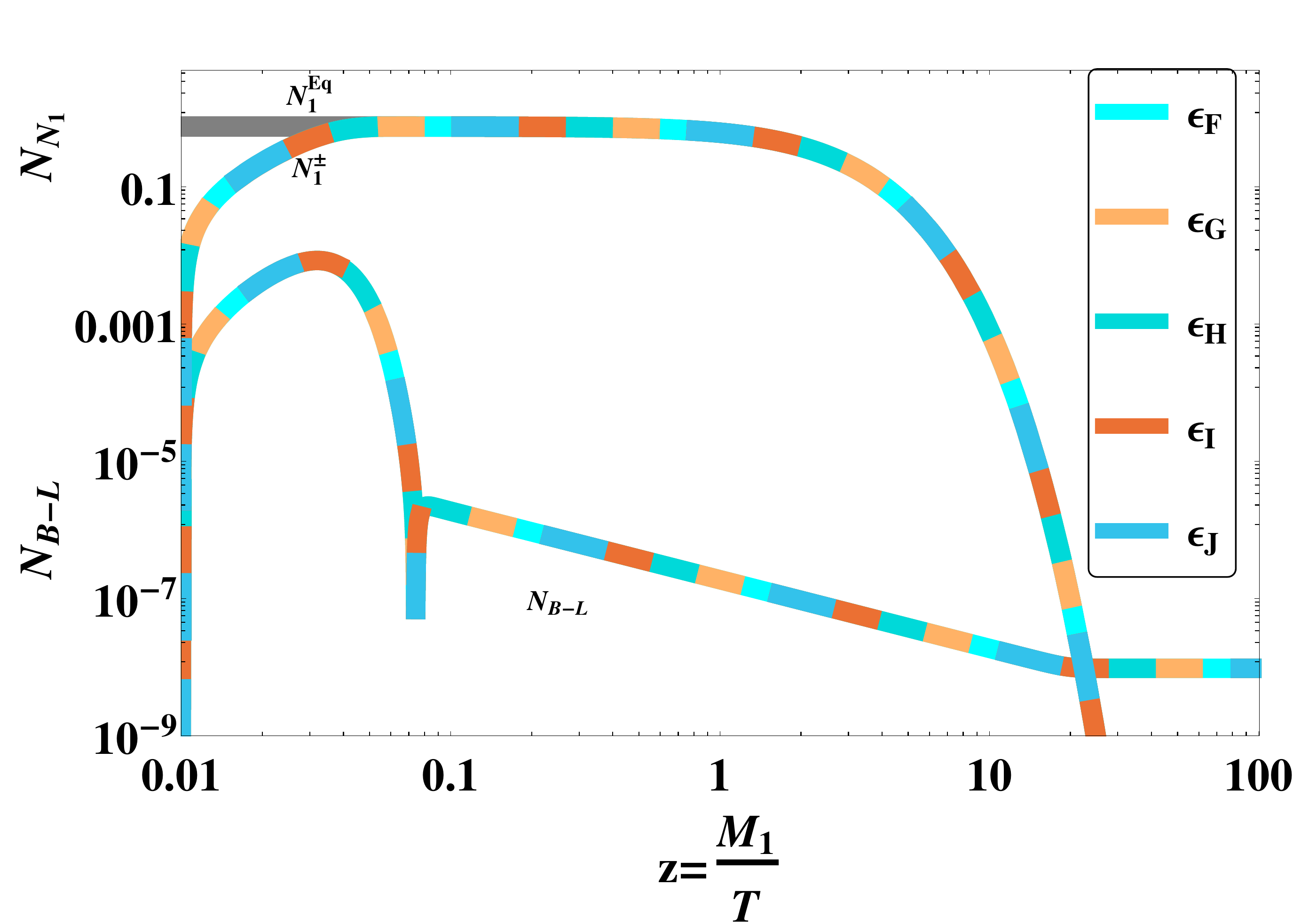}\
                \includegraphics[width=.49\textwidth]{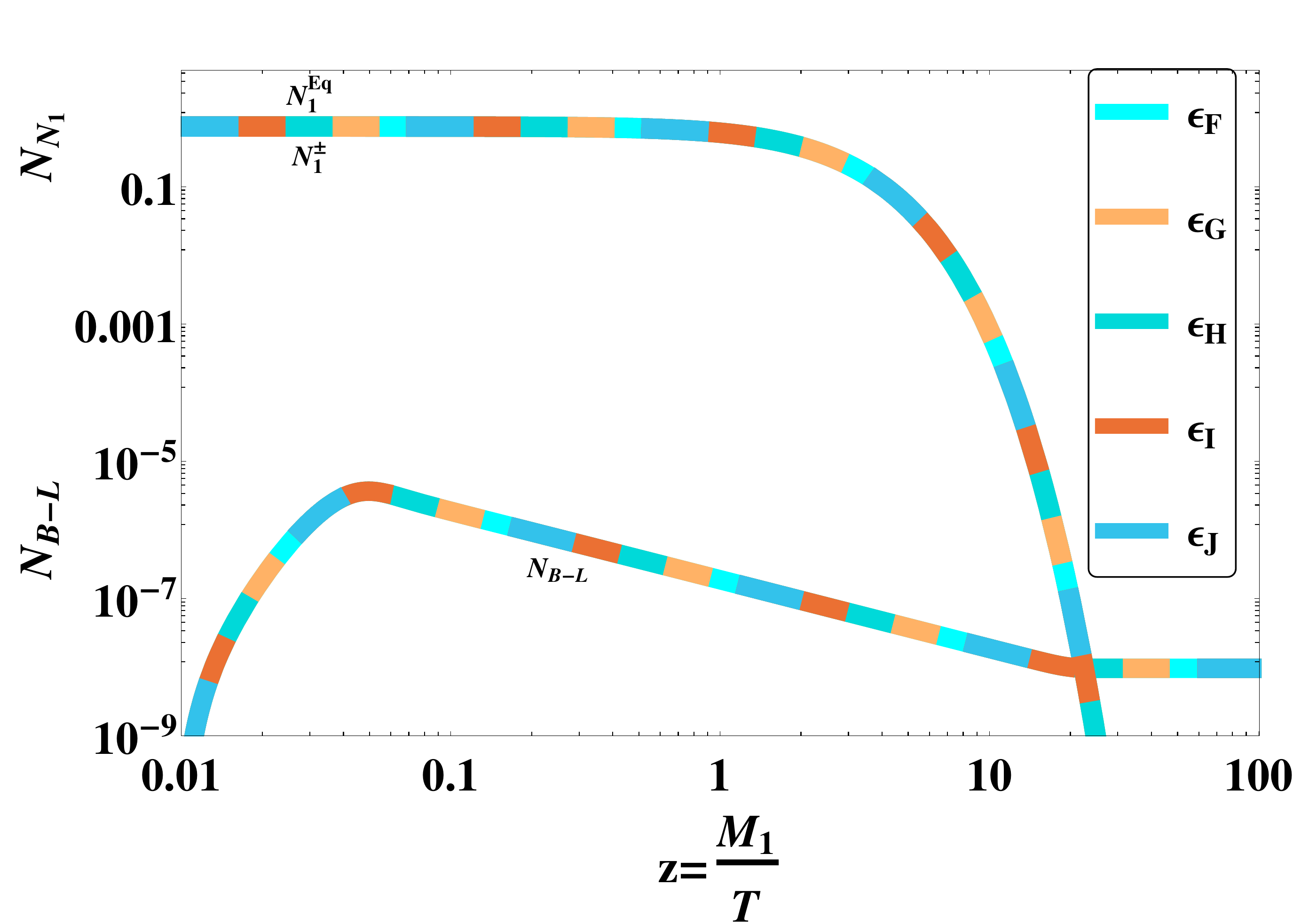}\\
                \caption{$B-L$ asymmetry and $N$ abundance in the visible sector for the large $\rho$ case for the parameter point of Eq.~\ref{eq:para1}. The $N_1^+$ and $N_1^-$ number densities are almost indistinguishable.
                Each graph displays five distinct choices for the matrices $y_1,y_2,f_1$ and $f_2$,  labeled $F-J$. These solutions gain an 
                enhanced $B-L$ from the resonance effect 
                between the $N^{+}$ and $N^{-}$ mass eigenstates. The $CP$ parameter additionally varies from the coupling to both Higgs doublets.
                Each case has different initial abundances for $N_1$: vanishing (left panel) and thermal (right panel).}
                \label{fig:sub2}
                \end{minipage}\\[1em]
                \end{figure}

                 This is to be contrasted with the large $\rho$ case. This scenario has fewer constraints from FCNCs, however above $\rho=200$ the neutrinos of the
                 dark sector decouple prior to the visible and dark sectors becoming thermally decoupled, so all of the light neutrinos of the dark sector must be non-relativistic.
                 Clearly multiple things must change. In the large $\rho$ case, the requirements on the dark confinement scale
                 will necessitate smaller $\phi_2$ quark couplings than the SM in general, as illustrated in Fig.~\ref{fig:QCDS}. In the case that all three of the light neutrinos are $\sim 100\ \text{MeV}$  and $\rho>200$, 
                 the constraints on parameters such as in Eq.~\ref{eq:para1} from washout require that the $CP$ parameter be much larger.
                 At the same time, with $f_i$ and $y_i$ comparable and $P$ non-zero such an enhancement is immediate, as we discuss below.  Figure~\ref{fig:sub2} is the analogous plot for the large $\rho$ case
                to Fig.~\ref{fig:sub1}, for the parameter point of Eq.~\ref{eq:para1}.  
                 The different cases $F-J$ describe different sample points in parameter space of the matrix in Eq.~\ref{eq:matrixeqn} that satisfy the condition that non-relativistic light neutrinos 
                 of the dark sector have masses of order  $100\, \text{MeV}$ and decay into SM species prior to BBN.

\subsubsection{Resonant enhancement of $CP$ violation}

\begin{figure}[h!]
                 \begin{minipage}{\textwidth}
                 \centering
                 \includegraphics[width=.78\textwidth]{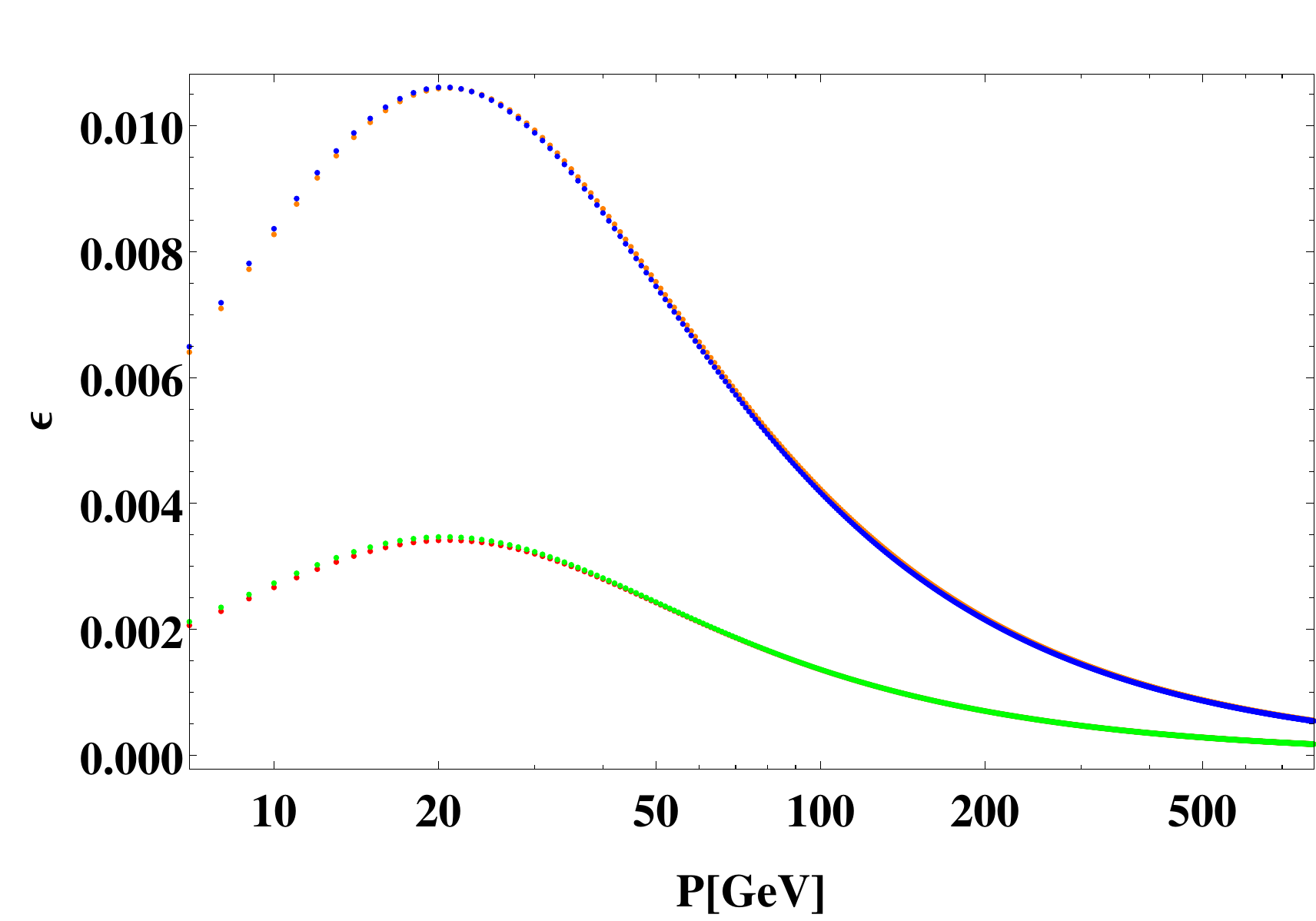}\ %
                 \caption{The resonant increase in the CP parameters $\epsilon^{\phi_{1}}_{1^+},\epsilon^{\phi_{2}}_{1^+},\epsilon^{\phi_{1}}_{1^-},\epsilon^{\phi_{2}}_{1^-}$ (blue, orange, red, green, respectively) 
                 as a function of cross-sector neutrino mass term $P$ in GeV. The other relevant parameters are as given in Eq.~\ref{eq:para1}.}
                 \label{fig:res}
                 \end{minipage}\\[1em]
                 \end{figure}

Similar to double seesaw mechanisms~\cite{Gu:2010xc} we have in our case the fact that $P$, the cross-sector masses, being small automatically puts us in a 
parameter space of near resonance between $N_1^{+}$ and $N_1^{-}$. 
                 This comes with the non-zero $CP$ violation originating from the interference of diagrams that involve both $N^{+}$ and $N^{-}$.
                 With $P/M_N \approx 10^{-9}$ we are within the range of the resonance effect through the interference in the self-energy diagrams with $N_1^+$ and $N_1^-$ for the 
                 illustrative Yukawa coupling constants in Eq.~\ref{eq:para1}. 
                 In Fig.~\ref{fig:res} we display that resonant increase in the $CP$-parameters for the large $\rho$ case of Eq.~\ref{eq:para1}. Note that this feature differs from \cite{An:2009vq} in that 
                 the resonance comes from adjusting the single small parameter $P$ which was previously constrained to be small rather than adjusting $M_1$ and $M_2$ to bring their difference close to the decay width.
                 The resonance boost can allow for a smaller value of $M_N$ such that in the high $\rho$ case we can bring $M_N$ down to $\sim 10^7\ \text{GeV}$.
                 
 \subsubsection{Comparison of decay, scattering and washout rates}
                 
\begin{figure}[]
                 \begin{minipage}{\textwidth}
                 \centering
                 \includegraphics[width=.49\textwidth]{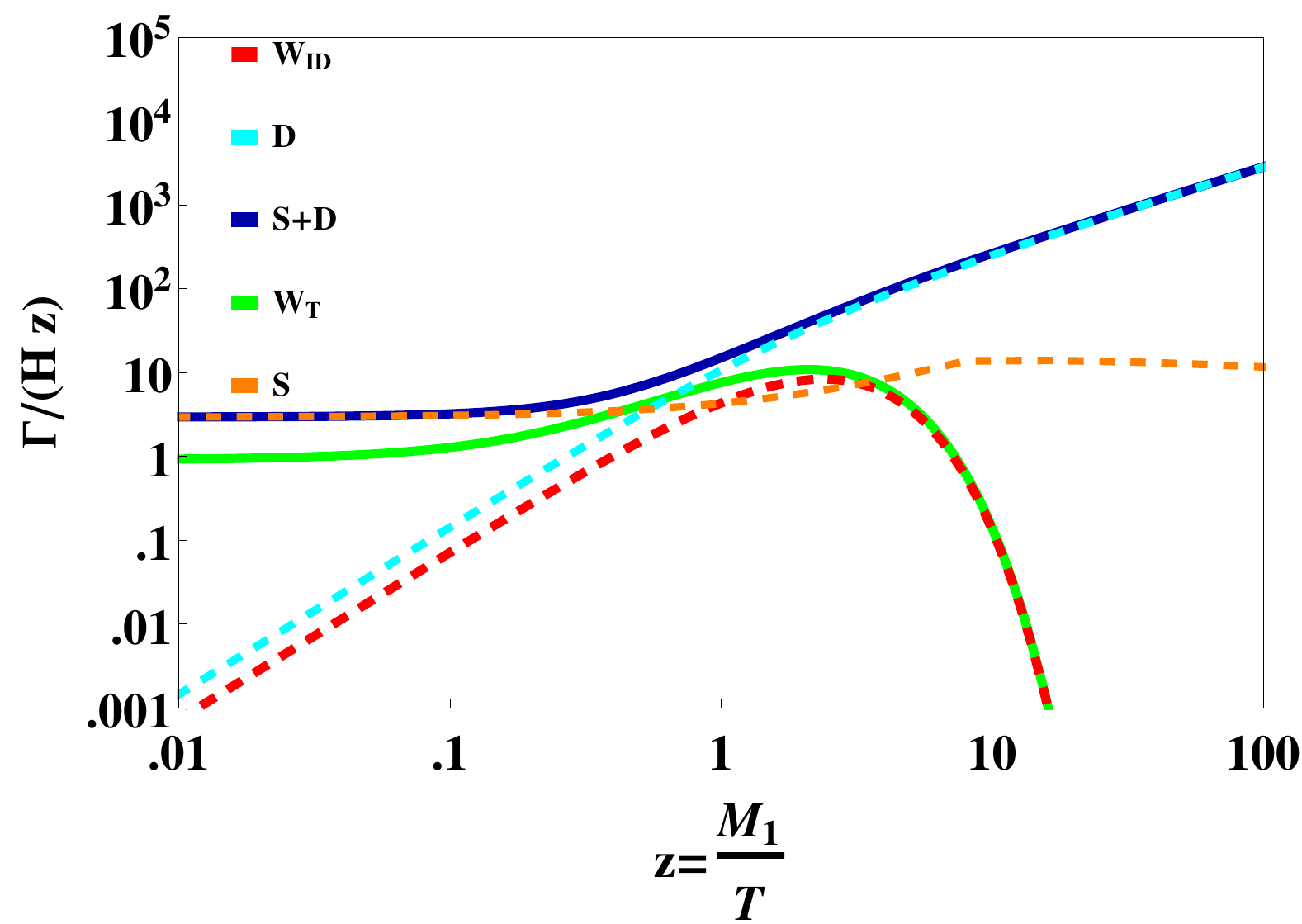}\ %
                 \includegraphics[width=.49\textwidth]{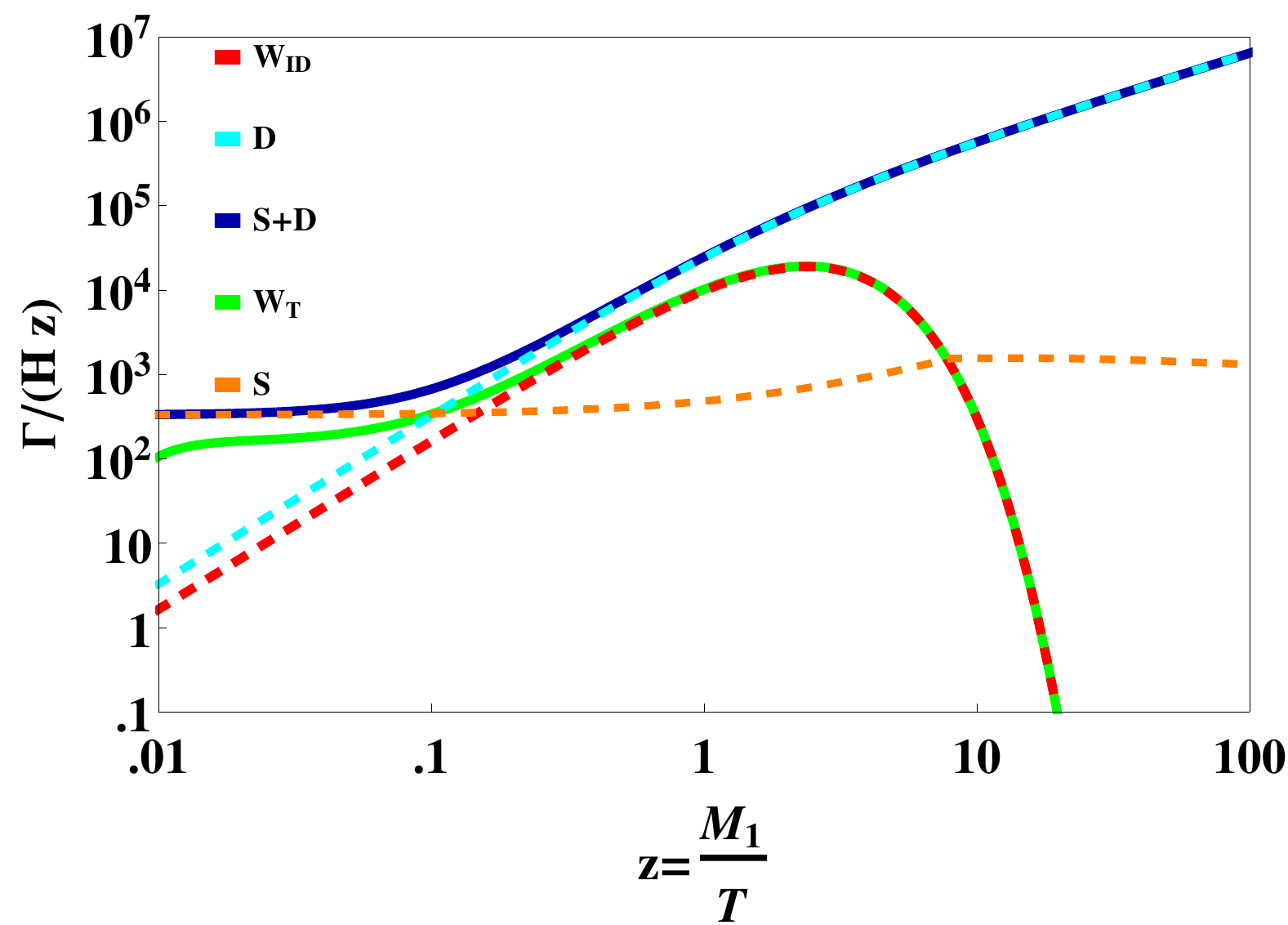}\\   %
                 \caption{Scaled Rates ($D_+,S_+,W_{ID},W_T, S_++D_+$) in the Boltzmann equations for the cases of small (left panel) and large (right panel) $\rho$. The parameter points are as for
                 Fig.~\ref{fig:sub1} and Fig.~\ref{fig:sub2}, respectively.
                  As the quark Yukawa couplings to the second doublet $\Phi_2$ can be smaller than those to $\Phi_1$ in the large $\rho$ case, the $(qt\leftrightarrow \phi _2l)$ scattering rates in the Boltzmann equations can be minimal.
                  In the strong washout regime, $(K>1)$ we see in the large $\rho$ case significant washout rates, the natural resonance that can arise in the CP parameter can allow 
                  the necessary asymmetry to survive at the point the washout rates become ineffective.}
                 \label{fig:rates}
                 \end{minipage}\\[1em]
                 \end{figure}

In Fig.~\ref{fig:rates} we can examine the decay, scattering rate and total washout rates in the different cases of quark mass hierarchies in the dark sector. The left panel corresponds to the small $\rho$ regime, and the right
panel to the large $\rho$ regime.  As per Fig.~\ref{fig:QCDS}, the large $\rho$ case requires the Yukawa coupling constants to the second doublet $\Phi_2$ to be smaller than those to $\Phi_1$ in order to achieve
the correct dark confinement scale and thus the right dark matter mass scale. This in turn makes the scattering rate correspondingly smaller, relative to the decay rate and, for a temperature range, to the washout rate.

                  \section{\bf Cosmological history}\label{sec:Temp}
                  
                  We now examine the consequences of this model for the thermal histories of the visible and dark sectors. In particular, we examine how the two sectors can be consistent with successful 
                  BBN and the constraints on the number of relativistic degrees of freedom. In the dark sector we consider how, after the breaking of mirror parity, a clear dark matter candidate can form which satisfies all of the current
                  astrophysical bounds. 
                  
                 \subsection{Sphaleron reprocessing in the visible and dark sectors}
                 
                 The lepton asymmetry of each sector is converted to a baryon asymmetry by sphaleron processes beginning at the temperature at which the $B-L$ asymmetry is produced by thermal leptogenesis. 
                 Immediately after this, the relation between the $B$, $L$ and $B-L$ asymmetries of each sector are related by 
                 \begin{align}\label{eq:BL}
                 \mathcal{N}_B = \frac{24 + 4 N_H}{66 + 13 N_H} \mathcal{N}_{B-L}\ ,\\
                 \mathcal{N}_L=-\frac{42 + 9 N_H}{66 + 13 N_H} \mathcal{N}_{B-L}\ ,\nonumber
                 \end{align}
                 where $N_H$ is the number of Higgs doublets that remain in equilibrium \cite{Harvey:1990qw}.
                 Following the EWPT of the dark sector, sphaleron processes in that sector are no longer rapid and the relation between $B'$, $L'$ and $(B-L)'$ is fixed at the values as in Eq.~\ref{eq:BL} with $N_H=2$.
                 At this point, one of the Higgs doublets of the visible sector gains a large positive squared mass value sufficient to decouple it while
                 sphaleron effects of the visible sector are still rapid. The $B-L$ asymmetry in the visible sector is then reprocessed to satisfy Eq.~\ref{eq:BL} with $N_H=1$ and this will be the value that
                 remains when visible sphaleron processes are no longer rapid following the visible sector's EWPT.
                 The present-day baryon asymmetries are therefore given by 
                 \begin{equation}
                  \mathcal{N}_B= \frac{28}{79} C \mathcal{N}_{B-L}\, , \;\;\;  \mathcal{N}'_B= \frac{8}{23} C \mathcal{N}'_{B-L}\,
                 \end{equation}
                 where $C=g^*(T_0)/g^*(T)$ accounts for the variation in photon density between the onset of leptogenesis and now.
                 We therefore have an abundance ratio from the symmetric leptogenesis phase that is slightly less than one which 
                 still suggests a mass ratio of dark baryons that are $\sim 5.57$ that of the proton. In Fig.~\ref{fig:QCDS} this is considered as the 
                 range for the dark confinement scale, with the exact mass of any heavy baryon having a range depending on the dark coloured quark masses.
                 Further discussion of such hidden QCD models and the exact relationship between confinement scale and baryon masses was explored in Ref.~\cite{Lonsdale:2017mzg}. 
                 We now consider some of the broader consequences of the dark sector.

                \subsection{Thermal history}
                
                We focus on the case where the two sectors decouple within the temperature region between the confinement scales of the two sectors. This means the decoupling
                occurs after the dark quark-hadron phase transition but before the visible quark-hadron phase transition (QHPT).
                By having this arrangement, we use dark confinement to reduce the number of degrees of freedom of the dark sector participating in thermal processes while the two sectors are still in thermal contact and thus transfer 
                the majority of the entropy and energy density of the universe to the visible sector. 
                This then allows for the dark sector to cool at a faster rate than the visible sector and thus acquire a lower temperature at the time of BBN when constraints on the number of
                effective neutrino species are stringent. This idea was explored in past works on dark QCD models, in particular in Refs.~\cite{Farina:2015uea, Lonsdale:2017mzg}. 
                In dark-sector theories, the exact relationship between the temperatures of each sector has important limits imposed by constraints on dark radiation energy density from BBN and CMB.
                This is usually quantified through an effective excess neutrino number defined by 
                \begin{equation}\label{eq:neff}
                 N_{\text{eff}}= 3\left( \frac{11}{4}\right)^{4/3}\left( \frac{T_{\nu}}{T_{\gamma}}\right)^{4} +\frac{8}{7}\left( \frac{11}{4}\right)^{4/3}\frac{g^*_D}{2} \left( \frac{T_{D}}{T_{\gamma}}\right)^{4},
                \end{equation}
                with the entire second term constituting $\Delta N_{\rm eff}$. The terms $g^*_D$ and $T_D$ are the degrees of freedom of the dark sector and their temperature.
                Recent measurements have obtained the bound
                $\Delta N_{\rm eff} =  0.11 \pm 0.23$  at the 68\% confidence limit level~\cite{Ade:2015xua}. 
                Using the conservation of entropy density, it is easy to show that the ratio of the temperatures of the two sectors, $T_V$ and $T_D$, at the scale of BBN is a 
                function of the degrees of freedom in each sector compared to the value at the time of decoupling \cite{Petraki:2011mv},
                \begin{equation}
                \frac{T_V^3}{T_D^3}= \frac{g^*_D}{g^*_V} \left(\frac{g^*_V}{g^*_D}\right)_{\text{Dec}}.
                \end{equation}
                In mirror symmetric models where the two sectors remain in thermal contact, 
                such as in Ref.~\cite{An:2009vq}, the constraints on the dark degrees of freedom are strong enough that it is necessary that all mirror relativistic particles be removed prior to BBN. 
                If the temperature of the dark sector is less than the visible sector then the BBN constraint may be satisfied 
                with at least a massless photon and possibly additional species still present in the dark sector, as discussed in Ref.~\cite{Lonsdale:2017mzg}.
               
                In order to decouple the two sectors in the energy range between $\Lambda_{QCD}$ and $\Lambda_{DM}$, we require a particle interaction that proceeds 
                fast enough at high energy and becomes ineffective shortly after the dark sector becomes confining. 
                We then require that either one or more of the species involved must become Boltzmann suppressed, or the rate of the reaction as a function of temperature must drop below the Hubble rate. 
                There are multiple distinct cases in which this can work in our model, depending on the value of $\rho$ and the masses of the particle species in the dark sector.
              
                \subsubsection{Large $\rho$ case}
\begin{figure}[!t]
                \centering
                \includegraphics[angle=0,width=0.68\textwidth]{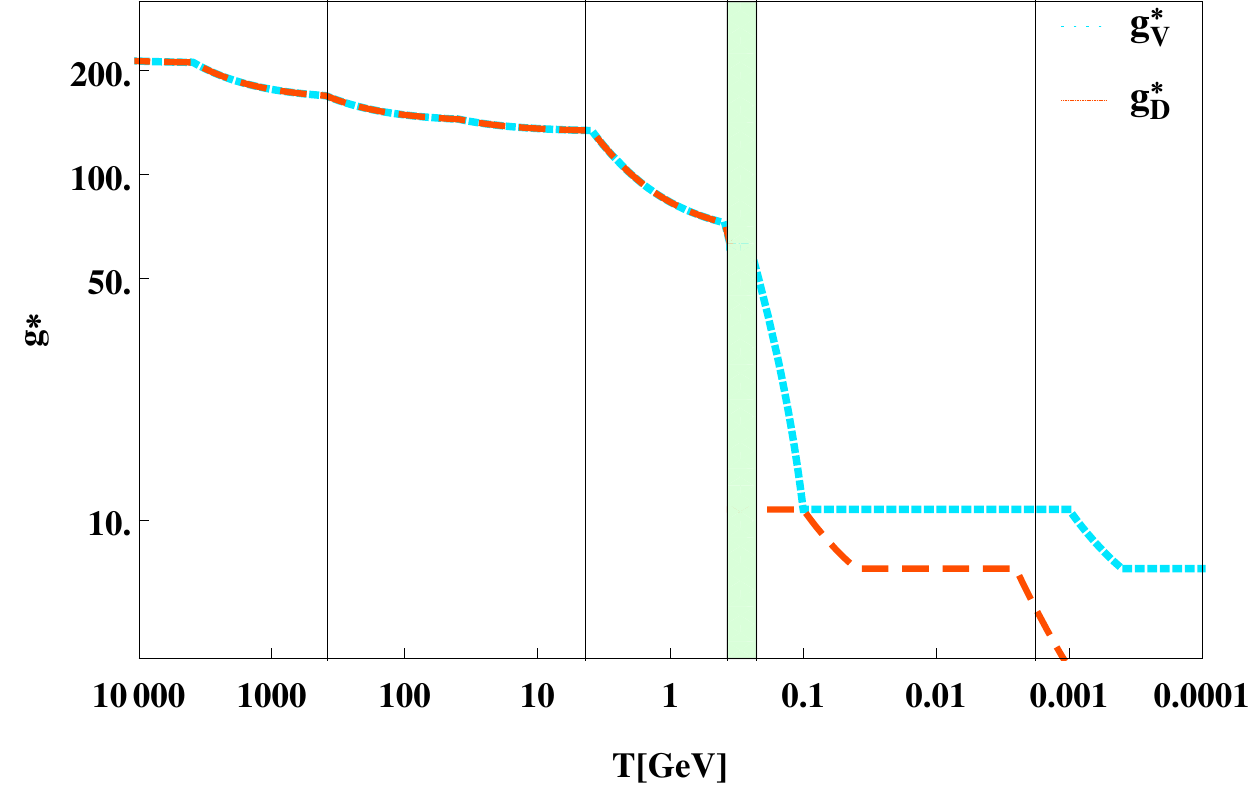}
                \caption{The degrees of freedom in the two sectors as a function of $T$. Initially they constitute one thermally-equilibrated sector. 
                After the thermal decoupling, the two sectors evolve independently and the degrees of freedom have separate histories. 
                For $ \Lambda_{\text{QCD}} <  T_{\text{Dec}}< \Lambda_\text{DM}$ (green shaded region) the ratio of temperatures after decoupling can fall to $T_D < 0.5 T_V$
                as the entropy density of the dark sector is shifted to the visible sector. The effective $g^*_{D}$ contributing to the expansion rate is then suppressed by this temperature difference.
                The arrangement of dark quark masses is not significant to this picture. Any dark quarks that become suppressed prior to $\Lambda_{\text{DM}}$ will contribute to the difference in $g^*$, however any that
                are still in equilibrium will hadronize at $\Lambda_{\text{DM}}$ and thus their degrees of freedom will be removed in any case. 
                However, the temperature shift can be affected by mass scales of dark leptons and dark pions.
                }
                \label{fig:degs}
                \end{figure}

                In this case all three of the dark sector's light neutrino species have large masses and thus become
                 non-relativistic prior to the thermal decoupling of the two sectors. 
                 This is possible in our model given the independence of $y_2$ and with $w$ sufficiently larger than $v$ such that $m_{\nu_1'},m_{\nu_2'}, m_{\nu'_3} \sim 100\ \text{MeV}$. 
                 We are thus led to only the relativistic dark photon contributing to the dark radiation density 
                 in the region of parameter space where $\rho > 200$, as mentioned previously. In this case, with the temperature difference arising from the 
                 difference in the number of degrees of freedom seen in Fig.~\ref{fig:degs}, we obtain a value for the $\Delta N_{\text{eff}}$ component of Eq.~\ref{eq:neff} given by
                 \begin{equation}
                  \Delta N_{\text{eff}}= \frac{2}{0.45}\left(\frac{T_D}{T_V}\right)^4 \simeq 0.17,
                 \end{equation}
                 where the factor of 2 counts the degrees of freedom of the single dark photon and the temperature ratio is $T_D/T_V\simeq 0.44$. This results from degrees of freedom at decoupling, $g_V^*=61.75$ and $g_D^*=5.5$, that is, counting only the dark electron and dark photon.  This value of $\Delta N_{\rm eff}$  is well within the observationally allowed limits. The thermal history of the universe in this case is summarised in Table~IV.
                 
                 \subsubsection{Small $\rho$ case}
                
                The contrasting small case has all three dark neutrinos being relativistic and has a sufficient temperature difference to allow for all four species to contribute to dark radiation.
                In order for neutrinos to undergo the temperature change from the above mechanism, it is critical 
                that the lightest neutrino population not decouple from the dark plasma prior to the thermal decoupling of the two sectors. 
                As the dark weak interaction rate scales as $1/\rho^2$ with the increasing mass of the electroweak gauge bosons, an upper bound of 
                $\rho < 200$ can be placed assuming that the maximum value of $T_\text{Dec}$ is just below the dark confinement scale.
                While in Refs.~\cite{Farina:2015uea, Lonsdale:2017mzg} the large change in $g^*$ following the dark QHPT is only large enough to allow 
                for one light neutrino species in the dark sector, the independent Yukawa couplings of the dark sector in the model considered here allows for a more complex thermal history.
                In particular, any difference between the two sectors in whether a species annihilates into photons before or after $T_\text{Dec}$ can further the temperature difference.
                \begin{itemize}
                \item Consider, for example, the ${\mu^+}'+{\mu^-}'\leftrightarrow \gamma' + \gamma'$ annihilations after the temperature of the plasma drops below the dark muon mass, but at a temperature above $T_\text{Dec}$.
                The energy density is shared between the two sectors. After $T_\text{Dec}$ when $T_V$ falls below the mass of SM muons the visible plasma is heated but the dark plasma is not. 
                This asymmetry in muon annihilation together with the entropy shift between the confining phase transitions, and with 
                three light mirror neutrinos leads directly to a temperature ratio of $T_D/T_V\simeq 0.42$ such that $\Delta N_\text{eff}=0.50$, which pushes the upper boundary of acceptability at about the $2\sigma$ level.
               
                \item We can take this further by considering the dark pions. If these have a mass that is above $\Lambda_{QCD}$ and above $T_\text{Dec}$ then they too will release their entropy density partially into the visible sector.
                The pion mass depends significantly on the bare masses of the lightest dark quarks and in particular will increase with more massive bare dark quarks. 
                With the asymmetric photon injection from both dark pions and dark muons one obtains $T_D/T_V\simeq 0.39$ and $\Delta N_\text{eff}=0.387$.
                
                \item If in addition to dark muons we additionally have that all of the dark electron energy is shared between the sectors, this yields $T_D/T_V\simeq 0.35$ and $\Delta N_\text{eff}=0.229$, which is well within the current constraints.
                We can also have a thermal portal that decouples the two sectors when the temperature falls below the mass of dark electrons, due to dark electrons being reactants in the portal term.
                In this case only a fraction of dark electron energy will be transferred to SM species while the thermal portal is still active, so that $\Delta N_\text{eff}$ will therefore be between $0.229$ and $0.387$.
                
                \end{itemize}
                Each of these thermal histories for relativistic light dark neutrinos sets a different final temperature for the dark neutrino species and therefore contributes a different limit to how 
                massive the light neutrinos of the dark sector can be before contributing too much hot dark matter to the model. Taking the limit that $\Omega_\text{HDM}<0.011$ we obtain for the above three cases a limit of 
                $\sum m_{\nu'} < 1.1\, \text{eV} , 3.2\, \text{eV}, 5.6\, \text{eV}$, respectively, the values used in Fig.~ \ref{fig:sub1}. 
                This then sets a limit on how large a role $y_2$ can take in the low $\rho$ regime of thermal leptogenesis.

                          \begin{table}[h!] 
   \caption{The major epochs of the universe including dark matter in the large $\rho$ thermal history. This timeline matches the degrees-of-freedom plot in Fig.~\ref{fig:degs}. \vspace{0.7cm}}
   \footnotesize
        \centering
         \tabcolsep=0.11cm
         \label{tab:tabletimeline}
         \fbox{
  \begin{tabular}{|l|l|l|}
\bottomrule
&$\,\,\,\,\,\,\,\,\,\,\,\,\,\,\,\,\,\,$&\\ 
$10^{19}\,$ GeV& $ t \sim 10^{-43}s\, $& $\bullet$ Planck scale era. Mirror symmetric sectors. $E_8\times E_8$? \\
                  &$\,\,\,\,\,\,\,\,\,\,\,\,\,\,\,\,\,\,$& \\
$10^{15}\,$ GeV& $ t \sim 10^{-38}s\, $& $\bullet$ Inflation ends, grand unified symmetry breaking scale.\\
             &$\,\,\,\,\,\,\,\,\,\,\,\,\,\,\,\,\,\,$& \\
$10^{12}\,$ GeV& $ t \sim  10^{-30}s\,$& $\bullet$ Majorana neutrinos begin to populate both sectors.\\
             &$\,\,\,\,\,\,\,\,\,\,\,\,\,\,\,\,\,\,$& \\
$10^9\,$ GeV&  $ t \sim  10^{-24}s\,$& $\bullet$ Thermal leptogenesis produces $B-L$ asymmetries in the \\
             &$\,\,\,\,\,\,\,\,\,\,\,\,\,\,\,\,\,\,$& visible and dark sectors.\\
              &$\,\,\,\,\,\,\,\,\,\,\,\,\,\,\,\,\,\,$& $\bullet$ Visible and dark sphalerons create the baryon asymmetries.\\
               &$\,\,\,\,\,\,\,\,\,\,\,\,\,\,\,\,\,\,$& \\
$10^{5}\,$ GeV&  $ t \sim 10^{-14}s\, $& $\bullet$ Universe has cooled to allow the Higgs fields\\ 
            &$\,\,\,\,\,\,\,\,\,\,\,\,\,\,\,\,\,\,$&of the dark sector to attain a nonzero \\
            &$\,\,\,\,\,\,\,\,\,\,\,\,\,\,\,\,\,\,$& vacuum expectation value, triggering the dark electroweak\\ 
            &$\,\,\,\,\,\,\,\,\,\,\,\,\,\,\,\,\,\,$&phase transition. Dark fermions gain mass.\\
           &$\,\,\,\,\,\,\,\,\,\,\,\,\,\,\,\,\,\,$& $\bullet$ Mirror symmetry is broken. Thermal interactions \\
           &$\,\,\,\,\,\,\,\,\,\,\,\,\,\,\,\,\,\,$&between the two sectors are maintained. \\
         &$\,\,\,\,\,\,\,\,\,\,\,\,\,\,\,\,\,\,$& $\bullet$ One of the visible doublets gains a large positive \\
         &$\,\,\,\,\,\,\,\,\,\,\,\,\,\,\,\,\,\,$&squared mass term from the cross-sector couplings.\\
            &$\,\,\,\,\,\,\,\,\,\,\,\,\,\,\,\,\,\,$& \\
$10^2\,$ GeV&  $ t \sim 10^{-10}s\, $& $\bullet$ Asymmetric symmetry breaking VEV breaks \\
            &$\,\,\,\,\,\,\,\,\,\,\,\,\,\,\,\,\,\,$&the EW symmetry of the visible sector. \\
            &$\,\,\,\,\,\,\,\,\,\,\,\,\,\,\,\,\,\,$& $\bullet$ Visible fermions gain mass.\\
            &$\,\,\,\,\,\,\,\,\,\,\,\,\,\,\,\,\,\,$& \\
$1\,$ GeV   & $ t \sim  10^{-3}s\, $& $\bullet$ The gauge coupling of the dark $SU(3)$ becomes \\
              &$\,\,\,\,\,\,\,\,\,\,\,\,\,\,\,\,\,\,$&non-perturbative, breaking \\
              &$\,\,\,\,\,\,\,\,\,\,\,\,\,\,\,\,\,\,$& chiral symmetry and confining all free dark quarks into hadrons.\\
               &$\,\,\,\,\,\,\,\,\,\,\,\,\,\,\,\,\,\,$& $\bullet$ The resulting difference in degrees of freedom effects a \\
                   &$\,\,\,\,\,\,\,\,\,\,\,\,\,\,\,\,\,\,$&transfer of entropy to the visible sector. \\
                    &$\,\,\,\,\,\,\,\,\,\,\,\,\,\,\,\,\,\,$& \\
$\sim 500 \,$ MeV   & $ t \sim  10^{-3}s\, $& $\bullet$ Thermal decoupling. The interactions between the two sectors\\
              &$\,\,\,\,\,\,\,\,\,\,\,\,\,\,\,\,\,\,$&fall below the expansion rate. Each sector evolves independently. \\
              &$\,\,\,\,\,\,\,\,\,\,\,\,\,\,\,\,\,\,$& \\
$200\,$ MeV         &$t \sim  10^{-3}s\,$&$ \bullet$ Visible quarks form nucleons following the QHPT. \\
                &$\,\,\,\,\,\,\,\,\,\,\,\,\,\,\,\,\,\,$& \\
$1 \,$MeV &  $ t \sim 180 s\,$& $\bullet$ Dark hadrons persist as individual nucleons. \\
              &$\,\,\,\,\,\,\,\,\,\,\,\,\,\,\,\,\,\,$& $\bullet$ The universe's expansion becomes dominated\\
             &$\,\,\,\,\,\,\,\,\,\,\,\,\,\,\,\,\,\,$& by matter over radiation.\\
              &$\,\,\,\,\,\,\,\,\,\,\,\,\,\,\,\,\,\,$& \\
$1 \,$eV & $ t \sim  10^{12}s\, $& $\bullet$ Dark neutrons and dark Hydrogen atoms dominate the \\
              &$\,\,\,\,\,\,\,\,\,\,\,\,\,\,\,\,\,\,$&hadrons of the dark sector forming dark matter and \\
              &$\,\,\,\,\,\,\,\,\,\,\,\,\,\,\,\,\,\,$&seeding visible structure formation. \\
              &$\,\,\,\,\,\,\,\,\,\,\,\,\,\,\,\,\,\,$& $\bullet$ Visible atoms coalesce into stars. \\
       \toprule
\end{tabular}
}
 \end{table}
                 
                \subsubsection{Thermal decoupling mechanism}
                
                We now come to the crucial issue of the thermal decoupling mechanism. The crucial aspect is our need to have the decoupling occur between the dark and visible QHPTs that occur at
                temperatures of about $1\ \text{GeV}$ and $200\ \text{MeV}$, respectively.
                We present three possibilities, in decreasing order of conceptual attractiveness and increasing order of quantitative certainty.
                
                \vspace{3mm}
                
                \noindent
                \emph{Neutron--dark-neutron portal:}\ In order to make our scenario completely natural, we would need kinetic equilibrium to be maintained at relatively low temperatures by
                a portal that naturally switches off at a temperature scale just below
                the dark QHPT. 
                Possibly the most appealing example is a dimension-9 quark interaction that constitutes a neutron--dark-neutron portal operator. An example is
                \begin{equation}
                \frac{1}{M^5}\;  \overline{u} \overline{d} \overline{d} \; u' d' s' + h.c.
                 \label{eq:neutronportal}
                \end{equation}
                Here, $u$ and $d$ are the usual up and down quarks, while $u'$ and $d'$ are the two light dark-quarks that constitute the dark-neutron dark matter: $u' d' d'$. The
                state $s'$ is another (dark) charge $-1/3$ dark quark that is sufficiently light to still be present in the plasma at $T \sim 1\ \text{GeV}$, but such that the dark $\Lambda^0$, $u'd's'$, is unstable
                to dark-weak decay to the dark neutron. The parameter $M$
                is a new physics scale at which a new $B + B'$ conserving interaction is introduced that produces the effective interaction of Eq.~\ref{eq:neutronportal} when integrated out, 
                and it has an appropriate flavour structure. The need for flavour structure is apparent from the fact
                that the operator $\overline{u} \overline{d} \overline{d} u' d' d'$, if it were allowed, would cause the dark-neutron dark matter to decay to ordinary matter on a much shorter timescale than the age of the universe. 
                In addition to $udd$ in Eq.~\ref{eq:neutronportal}, the $uds$ and $uss$ combinations are also suitable. 
                Such interactions maintain kinetic equilibrium to $T \sim 1\ \text{GeV}$ for $M \leq 63\ \text{GeV}$.  
            
                Studying the effects of such an operator with quantitative rigour within the regime where one sector has become confining and the other has 
                not is difficult due to the complicated nature of the dark QHPT. Irrespective of that, however,
                it seems reasonable to argue that the dark QHPT itself can be held responsible for breaking the thermal contact 
                between the sectors. In such a scenario, once the dark quarks of the dark sector have formed bound dark-neutron states,
                these can then decay through the portal into SM quarks, which are still unconfined, transferring a sufficient amount of entropy density from the dark sector to 
                the visible sector. Since the dark neutrons have a mass greater than than the dark confinement scale
                they immediately begin being Boltzmann suppressed following the dark chiral-symmetry breaking phase transition, and by the time of the QHPT in the 
                visible sector, when quarks of the SM form bound baryonic states, the abundance of reactants is sufficiently small that thermal equilibrium
                between the sectors has ended. While this argument contains a natural explanation for a lower temperature dark sector, it is speculative. In addition, we have not
                attempted to construct an underlying renormalisable theory to produce the neutron-portal operator.

                \vspace{3mm}
                
                \noindent
                \emph{Lepton portal:}\ We can also consider a six-fermion leptonic interaction between the sectors that could
                accomplish a similar goal,
                 \begin{equation}\label{eq:lepportal}
                \frac{1}{M^5} \overline{l} \overline{l} \overline{e^c} \; {l}' {l}' {e^c}'     + h.c.,
                \end{equation}
                This type of operator can become ineffective at the temperature when the dark charged lepton $e'$ becomes Boltzmann suppressed.
                This requires that the mass of this species is tuned to be in the specific range between the confinement scales.
                In the large $\rho$ case we will see that we require an electron like species to have mass much smaller than this scale and 
                so we can again consider a particular flavour structure for the operator in Eq.~\ref{eq:lepportal} such as ${l}' {l}' {\mu^c}'$  for the dark sector.
                This allows us to choose dark muons to be the species that becomes suppressed between $\Lambda_{\text{QCD}}$ and $\Lambda_{\text{DM}}$.
                If, however, we are in the low $\rho$ regime, the original portal involving electrons can be involved
                which can allow all three light dark neutrinos and a dark photon to be consistent with the extra radiation bounds during BBN and the CMB formation.
                Such an operator could fall below the Hubble rate due to the lightest charged leptons becoming Boltzmann suppressed at the requisite temperature range.
                Efficient weak interactions of the dark sector can then remove the remaining asymmetric component of dark electrons along with the more massive dark protons and store the lepton asymmetry of the dark sector
                in the dark cosmic neutrino background.\footnote{Note that the kinetic mixing portal is not suitable for us. Setting the $Z-Z'$ mixing to a suitable value forces the photon kinetic mixing to keep the two sectors in 
                thermal contact to the present day, which is unacceptable for our theory.}
                
                \vspace{3mm}
                
                \emph{$CP$-odd Higgs-mediated interactions:}\ In the relativistic decoupling case it is natural to examine the interactions between the sectors that our model already contains 
                to observe if any of these can maintain contact until the
                appropriate region. Since the scaling of the Hubble rate is of the form
                \begin{equation}
                H(T)= \sqrt{\frac{4 \pi^3 g^*}{45}}\frac{T^2}{M_{\text{pl}}},
                \end{equation}
                we require that such reactions scale with temperature at a faster rate. 
                We can consider Higgs-mediated interactions that survive below the EWSB scales driven by the mixing between the heavy neutral $CP$-odd scalars of the two sectors. 
                The four-fermion interaction rate can be written as 
                \begin{equation}
                 \Gamma_{AA'} \simeq {\eta_2}^2 {\eta_1}^2 z_{10} \frac{v_1^2 w_2^2 T^5}{m_{A_1}^8}.
                \end{equation}
                This term can can provide a way to couple the two sectors without altering the global minima of the potential, thus without removing the asymmetric VEV configuration, or mixing the weak scales of the two sectors.
                This facility is unique to the $CP$-odd scalars.
                Such Higgs-mediated scatterings can thermally couple the two sectors down to $\sim 1\ \text{GeV}$ while an ordinary-mirror Higgs portal cannot, simply because the latter is inoperative once the Higgs bosons
                disappear from the thermal bath. 
                Now, for the Higgs-mediated interactions
                to maintain thermal equilibrium to a low enough temperature, the Yukawa couplings need to be sufficiently large. 
                However if these same Yukawa couplings were to give mass to the fermions involved in the cross-sector scatterings, then
                they would be Boltzmann suppressed prior to  $T \sim 1\ \text{GeV}$ and hence the interaction could remain above the Hubble rate to the required low temperature range.
                But, in our model, since the Higgs states providing the mediation utilise the Higgs doublets that are \emph{not} responsible for mass generation in their respective sectors, 
                such a portal coupling can work if the mass hierarchies permit the dark partners of some low mass fermions of the SM to be heavy, and vice-versa. 
                \begin{figure}[h!]
                \centering
                \includegraphics[angle=0,width=0.78\textwidth]{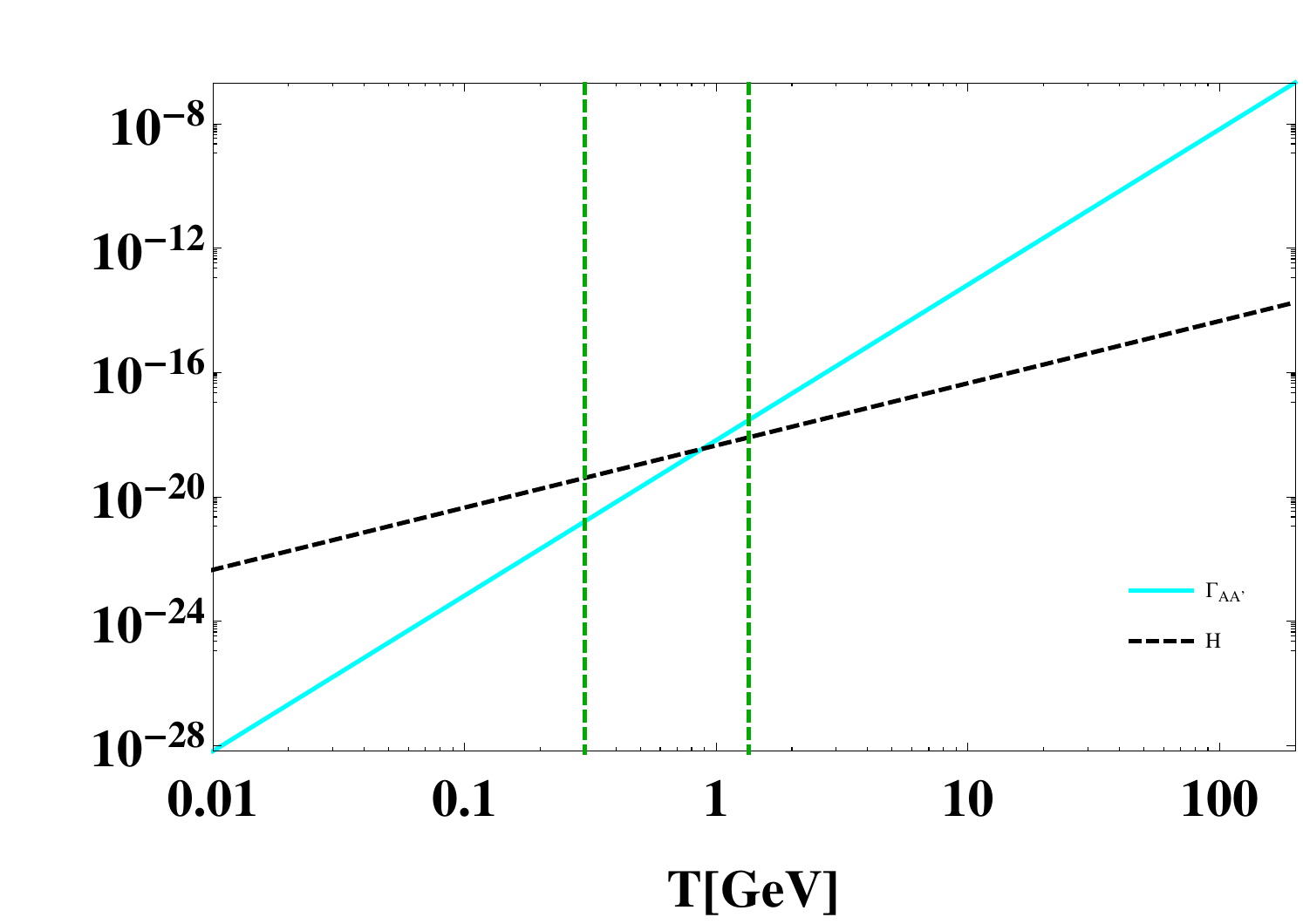}
                \caption{The $CP$-odd Higgs-exchange scattering rate as a function of temperature. In this example, it falls below the Hubble rate between the confinement scales, depicted in green, of the two sectors.
                The scattering rate uses $\eta_1=0.6,\eta_2=0.6,z_{10}=0.1,m_{A_1}=3525 \, \text{GeV}, v_1=246 \,\text{GeV}, w_2=7380\, \text{GeV}$.}
                \label{fig:Higgsexchange}
                \end{figure}
                In Fig.~\ref{fig:Higgsexchange}, we show the scaling of such a Higgs-mediated scattering channel with temperature in the small $\rho$ regime for a parameter choice that has rate fall below the Hubble rate
                between the visible and dark quark-hadron phase transitions.  
                
                This possibility has the advantage of using renormalisable interactions that are already an intrinsic part of the theory, but it must be admitted that having the interactions fall below the Hubble rate
                immediately after the dark QHPT constitutes a fine-tuning of parameters and introduces an unwanted coincidence.

                  \subsection{Dark big bang nucleosynthesis} \label{sec:BBN}
                  
                  In the dark sector, the confinement scale is approximately five times higher than its visible sector counterpart while the electroweak scale can span a much broader allowed parameter space given by $\rho$.
                  With the larger EW scale in the dark sector we also have larger masses for the $W'$ and $Z'$ bosons. 
                  The independence of the Yukawa couplings that generate mass for fermions in each sector 
                  also allows for the lightest baryon of the dark sector to be neutral under $U(1)'_Q$.  
                  For ordinary baryons, the mass difference of the proton and neutron, 
                  $\delta m = m_n-m_p$, can be considered as the sum of two components. 
                  The first is from QCD which can be approximated by the mass difference of the up 
                  and down quarks $\delta_{QCD}\simeq m_d-m_u \approx 2.51\ \text{MeV}$. 
                  The second is from EM interactions and is given by $ \delta_{EM} \approx -1.00\ \text{MeV}$. 
                  We then have $\delta_m = \delta_{QCD} +\delta_{EM} \approx 1.51\ \text{MeV}$. 
                  By reducing the difference in mass of the lightest of up and down type dark quarks, we can reduce the QCD contribution 
                  until $\left|\delta_{QCD}\right| < \left|\delta_{EM}\right|$ for dark baryons. The variation of the dark electroweak scale does 
                  impact the size of the QED correction to the proton through the change in $\alpha_{em}$, however this correction is insignificant. 
                  This can be seen by the matching condition $\frac{1}{\alpha_{em}} = \frac{1}{\alpha_{1}}+\frac{1}{\alpha_{2}}$,
                  and the fact that the $U(1)$ and $SU(2)$ gauge couplings run in opposing directions such that 
                  variations in the symmetry breaking scale only minimally alter the value of $\alpha_{EM}$.
                  
                  The lightest baryon of the dark sector is then the dark neutron $n'$ and the decay rate of the dark proton $p'$ into the dark neutron is suppressed by the larger mass of the dark sector $W'$ bosons. 
                  If the dark weak interaction rate now falls below the Hubble rate at an earlier point, it is possible to 
                  find a situation in which the dark BBN produces a maximal amount of dark helium, by contrast with the visible sector,
                  despite the dark neutron being the most stable baryon. This is because the $n'$-$p'$ ratio will be sensitive to the interactions that are still in equilibrium in the dark sector as well as the mass difference between them.
                  However, the formation of nuclei through primordial synthesis may not take place at all if the dark deuteron is unbound.
                  It is a well known feature of conventional BBN that
                  the deuteron, the bound state of one proton and neutron, is only very weakly bound \cite{Kolb:1990vq}.
                  Its function in BBN is essential, however, in that it forms the intermediate step between hydrogen and heavier elements such 
                  as helium, but a sufficient abundance of deuterium only appears at 
                  temperatures lower than 2.2 MeV due to its low binding energy. 
                  The binding energy of the deuteron can be related to the ratio of $\frac{m_\pi}{m_p}$ which has been estimated to need to be $<0.16$ 
                  for the deuteron to be stable. We can reconsider this ratio
                  in terms of the factor $x_D=\frac{(m_u + m_d)}{\Lambda_{QCD}}$. 
                  In Ref.~\cite{Golowich:2008wn}, a conservative estimate of the required ratio to make the deuteron unstable was found to be that $x_D$  should increase by a factor 
                  of 2.5. Such an increase can be readily achieved in our model provided that the  low mass dark quark masses are 
                  larger than $\sim 35\ \text{MeV}$, where $\Lambda'_{QCD}$ is $\sim 5$ times the SM value.
                  If the dark deuteron were to be unbound, its absence in the primordial era of statistical nuclear equilibrium would disallow the 
                  formation of helium and heavier elements. The makeup of dark matter is then dependent on a number of free parameters in our theory. We summarise these distinct cases below.
                  
                  \vspace{3mm}
                  
                  \noindent
                \emph{Large $\rho$ case.}\ \ In this case, with $\rho > 200$, we have the fact that dark weak interactions are suppressed by the time of
                  the dark QHPT and the light dark neutrinos have a mass scale $\sim 100 \,\text{MeV}$. 
                  The dark lepton and baryon asymmetries are fixed at their values following the sphaleron conversion and dark weak interactions prior to $1$ GeV similar to \cite{Barbieri:2016zxn}.
                  This yields the initial relations $\mathcal{N}'_B= (8/23) \mathcal{N}'_{B-L}$ and $\mathcal{N}'_L= -(15/23) \mathcal{N}'_{B-L}$. 
                  If we assume that the lepton asymmetry in this phase is evenly divided among relativistic lepton flavours then the distribution 
                  between charged and neutral leptons can be found by solving the series of equations that relate the chemical potentials after the dark EWPT 
                  until dark EW interactions freeze. Using the conservation of lepton and baryon number,
                  conservation of charge and the enforced relations from weak interactions such as $\mu_d=\mu_u + \mu_e- \mu_{\mu_e}$ we 
                  solve these equations at the point just prior to when this last weak interaction condition is removed and the charge distribution between the 
                  baryon and lepton sectors is set \cite{Schwarz:2009ii}. This is dependent on how many fermion species have been removed from the plasma when the dark weak freeze-out occurs.
                  For all dark fermions in the plasma at freeze-out we have that the ratio of the asymmetry stored in charged leptons compared to neutral leptons of $82:83$, while removing the two heaviest quarks yields $13:17$. 
                  Additionally removing a third quark and a single charged lepton yields $2:3$.
                  Dark QED charge conservation guarantees that the dark proton asymmetry equals the dark charged lepton asymmetry. This yields an abundance of charged leptons equal to that of charged protons such that 
                  we have a mix of dark neutral hydrogen atoms and dark neutrons after the dark QHPT. This asymmetry is then equal to the value of the lepton asymmetry stored in charged leptons just above the dark weak freeze-out temperature.
                  The lepton asymmetry stored in the light neutrinos of the dark sector at the point of dark weak freeze-out will ultimately be transferred to the visible sector's neutrino background following the decays of these states to visible species.
                  We can then write the number densities of dark species in terms of the $B'$, $L'$ and $B'-L'$ number densities with $n'_p=n'_e$ and $n'_p +n_n' = (8/23) n'_{B-L}$ such that $n'_n=(8/23) n'_{B-L}-n'_p$. 
                  The cases listed above then give neutral atoms and dark neutrons in ratios of  15:31, 13:29 and 3:7 respectively.
                  This assumes that the dark charged-lepton that is part of dark atoms has both a small enough mass to survive to low temperatures and to not be an important contributor to the overall dark matter mass density.
                  We can thus have a component of mirror atoms among the dark neutrons. 
                  
                  The later distribution between dark atoms and dark neutrons depends significantly on a number of further assumptions we make for dark sector parameters.
                  If the mass difference between dark protons and dark neutrons, $Q=m'_p-m'_n$, is larger than the sum of the masses of dark electrons and dark neutrinos, dark protons may decay with a lifetime dependent on
                  both the increased EW scale of the dark sector and a phase space factor which can be compared to that of the SM neutron. We can consider the case where the value of $Q$ becomes larger than $100 \,\text{MeV}$, 
                  while keeping the dark electron mass negligible and the dark neutrino mass fixed at $100\, \text{MeV}$ \cite{Griffiths:1987tj}. 
                  Dark protons can then all decay prior to the matter-dominated era.
                  Increasing $\rho$ necessitates larger values of Q to achieve this as the weak suppression increases the lifetime by a factor of $\sim \rho^4$.
                  
                  In the case that dark protons are stable due to $m_e'+m_{\nu}'$ exceeding $Q$ we must also ensure that by the onset of structure formation at matter-radiation equality, $T_V \approx 1\ \text{eV}$, the dark matter
                  is no longer undergoing long range interactions and can form the early inhomogeneities. This requires that the dark matter-radiation decoupling have taken place prior to the moment of matter-radiation equality.
                  While the lower temperature of the dark sector can bring the moment that free-streaming charged dark particle numbers are suppressed by recombination earlier in time, the dynamic temperature ratio
                  we considered in the large $\rho$ timeline is insufficient alone. However matter-radiation decoupling in the dark sector may take place at a higher value of $T_D$ compared to
                  the visible sector case. With the Saha equation's exponential dependence on the binding energy of hydrogen we can see that raising the ionisation energy of dark hydrogen through the
                  increased mass of dark electrons directly increases the temperature at which dark hydrogen is no longer ionised \cite{Berezhiani:2000gw, Ignatiev:2003js, Petraki:2011mv}. The ratio of
                  $T_D/T_V$ then satisfies the condition that dark matter-radiation decoupling takes place in the radiation dominated era of the universe provided that $T_D/T_V < 0.336 \, r$ where
                  $r$ is the ratio of photon decoupling temperatures $r=T_D^{\gamma'}/T_V^{\gamma}$ and is dependent on the ratio of 
                  binding energies $r\simeq B_H'/B_H = m_e'/m_e$ if one uses the same condition for sufficiently small fractional ionisation in each sector. 
                  
                  While protons may be stable in this latter case it remains possible for the dark hydrogen atoms to undergo electron capture and decay via $e' + p' \rightarrow n'+ e^+ +e^- +\nu $ due 
                  to the mixing in the neutrino sector considered in Eq.~\ref{eq:para1}. This can be compared to muon capture in protons in short-lived $\mu$-hydrogen atoms. Such decays 
                  are suppressed by the electron capture probability and the large dark EW scale and have lifetimes approximated by
                  \begin{equation}
                   \tau(e' + p' \rightarrow n'+ e^+ +e^- +\nu) \sim \left( |\psi(0)|^2 \frac{{G'_F}^2 |V'_{ud}|^2}{2 \pi} \frac{E^2_{VS}}{M^2}(M-m_n')^2 \right)^{-1},
                  \end{equation}
                  where $|\psi(0)|^2=m_r^3 \alpha'^3/\pi$ measures the wave function at the origin with $m_r$ the reduced mass \cite{Ando:2000zw, Czarnecki:2007th}. 
                  The Fermi constant, $G'_F$, scales with $W'$, $E_{VS}$ is the combined energy of visible sector decay products, $V'_{ud}$ is a dark-sector CKM element
                  and $M$ is the mass of the dark hydrogen atom. With a dark electron $\sim 10$ times the SM electron, and dark protons and dark neutrons at $\sim 5 \ \text{GeV}$, this suggests a lifetime
                  for these dark atoms between $\sim 10^{14}$ seconds for $\rho=200$, that is around the time of reionisation in the visible sector, to beyond the current age of the universe for $\rho > 2500$.
                  The decays of such states could be a promising source of indirect detection for dark matter if their lifetime is between these values. The above formula holds for an on-shell dark neutrino $\nu'$ which then decays to SM species $e^+,e^-,\nu,$ and thus
                  is an approximate lower bound on the lifetime for the general case where $\nu'$ is off-shell.
                  
                  \vspace{3mm}
                  
\begin{figure}[!ht]
                  \begin{minipage}{\textwidth}
                  \centering
                  \includegraphics[width=.78\textwidth]{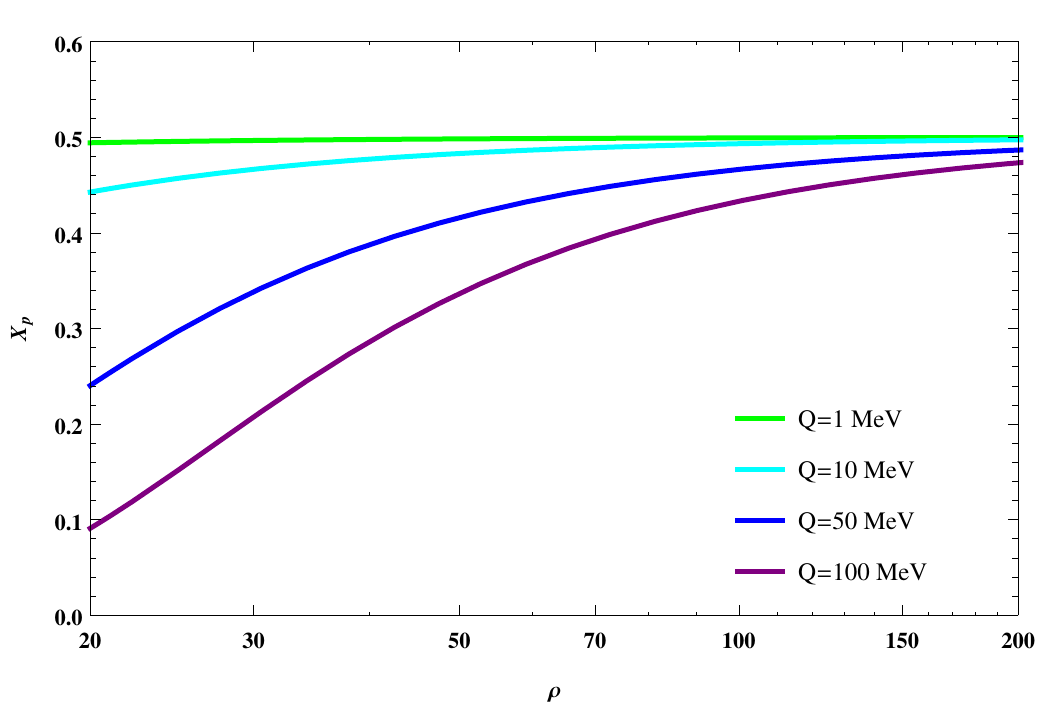}
                  \caption{The ratio of dark protons to dark neutrons, $X_p$, as a function of $\rho$ for a given mass difference, $Q = m'_p-m'_n$.
                  In the large $\rho$ limit $X_p \rightarrow 0.5$. }
                  \label{fig:pnratio}
                  \end{minipage}\\[1em]
                  \end{figure}

                  \noindent
                  \emph{Small $\rho$ case.}\ \ In this regime with efficient dark weak interactions, a light relativistic species of neutrino and a sufficiently light associated dark charged lepton we would have
                  charged baryons of the dark sector all decay into neutral single baryon states, due to $m_{p'}> m_{n'}$, which are then the dark matter component 
                  of our universe and so we have a complete population of dark neutrons.
                  In order to calculate the relative abundance of $n'$ to $p'$ in the dark sector during BBN we begin by taking the initial abundance of 
                  these species after the dark QHPT to be near equal, given the small mass difference between them. 
                  The value of $\rho$ will determine the ratio of dark protons to neutrons. With the mass difference, $Q=m'_p-m'_n$, we can write
                  \begin{equation}
                   X_p= n'_p/(n_p'+n_n')= e^{-Q/T'}/(1+ e^{-Q/T'}).
                  \end{equation}
                  Then the thermal freeze out of protons occurs when dark weak interactions freeze out at a temperature of $ T_{DWF} \sim 0.8 \rho^{4/3} \text{MeV}$. 
                  Figure~\ref{fig:pnratio} shows the how close the ratio is to 1:1 as $\rho$ increases.
                  
                  The lifetime of any remaining dark protons will depend on the mass difference Q and the value of $\rho$. Assuming a light charged lepton mass
                  and a value of $Q\sim 1\ \text{MeV}$  we have that in the limiting case  of $\rho=200$ the remaining dark protons will have a lifetime of $\sim 10^{12} \, s$ such that 
                  $X_p$ will approach zero at the time of structure formation.

                \subsection{Dark matter self-interaction constraints}\label{sec:Pheno}
                
                The possibility of self-interacting dark matter is constrained by a number of sources.
                In particular, ellipticity, substructure mergers and cluster collisions~\cite{Roos:2010wb, Markevitch:2003at}, impose upper bounds ranging from
                $0.7 -2.0 \, cm^2/g$. On the other hand, issues in simulations of ellipticity of galaxy structures, and the core-cusp problem can be 
                solved by interacting dark matter, though the required cross section may in some cases be in conflict with upper bounds \cite{Tulin:2017ara}.
                If dark matter interacts according to nucleon scattering where the cross section is set approximately by the size of 
                the nucleon itself, then we can consider this model in the context of other models of dark neutron dark matter.
                As noted in Ref.~\cite{Tulin:2017ara}, if the length scale and mass, $m_n$, of the nucleonic state is scaling with increasing $\Lambda_{DM}$ then the dark self interaction per unit mass can be considered as
                \begin{equation}
                 \sigma/m \sim 3\,  cm^2/g \, \times \left(\frac{\Lambda_{DM}}{m_n}\right)\left(\frac{\Lambda_{DM}}{a^{-1}}\right)^2\left(\frac{100 \text{MeV}}{\Lambda_{DM}}\right)^3,
                \end{equation}  
                where $a$ is the scattering length which should obey $a\sim\Lambda_{DM}^{-1}$. If the dark sector is made up of 
                both dark neutron and neutral dark atoms we can consider the dark neutron-atom interactions and dark atom-atom interactions as well.
                Since these are neutral we can consider possible magnetic self-interaction according to the analysis in Ref.~\cite{Lonsdale:2017mzg} where dipole-dipole 
                interactions were shown to be below the self-interaction bounds for a dark $\alpha'_{EM} = \alpha_{EM}$ and dark matter particles of mass 1 GeV. Compact object formation
                such as dark stars may also reduce the density of objects in galaxy clusters such that the diffuse gas assumptions which set limits on cross sections are not applicable.
                The details of dark first generation star formation will depend on the thermal history assumptions and critically on the composition of the dark sector, i.e. the ratio of dark hydrogen to dark helium to stable dark neutrons.
                In the case of a pure population of dark neutrons, we expect that with small self-interactions compact object formation will not take place. With a mixture of neutral dark atoms and dark neutrons, it is likely that 
                including an unbound deuteron, dark stars would not be able to survive without this link in the nucleosynthesis chain and so would still not be able to achieve the production of heavier dark elements just as in the BBN era.

                 \section{\bf Conclusion}\label{sec:Conclusion}
                 
                 We have shown how the 1:5 mass-density ratio of matter to dark matter can arise as a natural consequence of a high scale mirror symmetry that is spontaneously broken. 
                 The mechanism of asymmetric symmetry breaking, previously only explored in GUT contexts, has
                 been used to demonstrate a comprehensive model of early universe cosmology that has spontaneously broken mirror parity and creates the matter of the visible universe simultaneously 
                 with dark matter with a fully explained reason for both the number density
                 and mass scale of dark matter particles. It thus takes an important step beyond most models of asymmetric dark matter, and builds on earlier explorations of broken mirror models.
                 
                 The bounds on dark radiation from BBN and CMB necessitate a way to reduce the temperature of the dark sector in such scenarios.
                 While we have discussed a number of possible ways to arrange this in a tuned manner, we would highlight the unique and original mechanism that may accomplish this in a natural way 
                 in the form of an effective neutron-portal type operator. This interaction becomes ineffective between 
                 the dark and visible confinement scales, an epoch during which there is naturally a much larger degree of freedom count in the visible sector compared to the dark sector.                
                 Critically, this interaction becomes ineffective as a direct result of the dark quark-hadron phase transition, and thus provides a way to avoid the unwanted fine-tuning of the kinetic decoupling temperature.
                 The precise dynamics of this mechanism is, however, not amenable to straightforward calculation and is thus left to possible future work. This would include an examination
                 of underlying renormalisable interactions that can give rise to the effective operator.
                 
                 If we consider that the SM is an effective field theory then the possibility remains that not only is 
                 there much more to be found at higher energy scales, but also that the SM could be just one of a number of low energy sectors
                 that interact with each other predominantly only through gravitation in the present-day universe. The standard model does remain, however, our
                 greatest road map for developing theories that seek to 
                 explain the currently unexplained features of the cosmos.
                 This work follows that path by demonstrating 
                 a class of models of asymmetric dark matter that generate equal matter-antimatter asymmetries in separate sectors in which dimensional transmutation naturally creates
                 similar stable mass scales for matter and dark matter.
                 When spontaneous symmetry breaking generates fermion masses, it simultaneously breaks mirror symmetry and creates a dark sector that is distinct from the visible 
                 sector but retains the same amount of $B-L$ asymmetry. Rapid sphaleron processes in each sector transfer the equal amounts of lepton asymmetry into nearly equal baryon asymmetries. 
                 The common origin of gauge couplings in the UV guarantees a similar scale for the dark baryons and ordinary protons which can readily explain the 
                 abundance of dark matter in the universe relative to matter.
                 This dark matter can be made up of either a mixture of neutral dark hydrogen atoms and stable dark neutrons in the case of a large dark electroweak scale, or a pure population of dark neutrons in the case of $\rho < 200$.
                 This is a particular case of a more general class of asymmetric mirror dark matter models that utilise high scale discrete 
                 symmetry breaking to generate dark sectors which manifest differently from the visible sector at low energies and temperatures. 
                 These can include different gauge symmetries from parent GUT groups as well as differences in symmetry breaking scales, fermion masses, and as we
                 have explored in this work, the possibility of significant differences in the thermal histories between the two sectors.
                 Drawing on the concept of a $\Bbb{Z}_2$ mirror symmetry connecting just two copies of the SM gauge group we have seen 
                 how symmetric potentials can break mirror symmetry to create two markedly different sectors: one which sets the abundance of visible matter in the universe, 
                 and a dark copy with a high electroweak scale, slightly higher confinement and unique fermion flavour mass hierarchy, which determines the larger mass-abundance of 
                 dark matter.
                 
                 \section*{Acknowledgments}
                 This work was supported in part by the Australian Research Council. SJL thanks T. Dutka for helpful discussions.

\newpage           
\section*{Appendix A}
\setcounter{equation}{0}
\renewcommand{\theequation}{A\arabic{equation}}
                 
The neutral squared mass matrix in the mirror symmetric $\Phi$ basis among states, $(\phi_1,\phi_2, a_2, \phi_1',\phi_2',a_1)$, is
            
 \begin{equation}
 \begin{pmatrix}
m_{\phi_1  \phi_1 } & m_{\phi_1  \phi_2 } & m_{\phi_1  a_2} & m_{\phi_1  \phi'_1 } & m_{\phi_1  \phi'_2 } & m_{\phi_1  a_1 }  \\
m_{  \phi_2 \phi_1} & m_{\phi_2  \phi_2 } & m_{\phi_2  a_2} & m_{\phi_2  \phi'_1 } & m_{\phi_2  \phi'_2 } & m_{\phi_2  a_1 }  \\
m_{ a_2\phi_1 } & m_{a_2 \phi_2 } & m_{a_2 a_2} & m_{a_2 \phi'_1 } & m_{a_2 \phi'_2 } & m_{a_2 a_1 } \\
m_{  \phi'_1 \phi_1} & m_{\phi'_1  \phi_2 } & m_{\phi'_1  a_2} & m_{\phi'_1  \phi'_1 } & m_{\phi'_1  \phi'_2 } & m_{\phi'_1  a_1 } \\
m_{  \phi'_2 \phi_1} & m_{\phi'_2  \phi_2 } & m_{\phi'_2  a_2} & m_{\phi'_2  \phi'_1 } & m_{\phi'_2  \phi'_2 } & m_{\phi'_2  a_1 }  \\
m_{  a_1 \phi_1} & m_{a_1  \phi_2 } & m_{{a_1}   a_2} & m_{{a_1}   \phi'_1 } & m_{a_1   \phi'_2 } & m_{a_1   a_1 } \\
 \end{pmatrix}.
 \end{equation} 

This consists of the 21 elements, which in the asymmetric limit become 

\begin{align}
&m_{\phi_1  \phi_1}\simeq m^2_{11} + (3 v_1^2 z_1)\frac{1}{2} + (w_2^2 z_{12})\frac{1}{2} + (v_2^2 z_3)\frac{1}{2}+(v_2^2 z_4)\frac{1}{2} + (w_1^2 z_8)\frac{1}{2} + w_1 w_2 \text{Re}[z_{13}] \nonumber  \\ &+ \frac{1}{2} v_2^2 \text{Re}[z_5] + 3 v_1 v_2 \text{Re}[z_6],  \nonumber\\
 &m_{\phi_2  \phi_2}\simeq  m^2_{22} + (w_1^2 z_{12})\frac{1}{2} + (3 v_2^2 z_2)\frac{1}{2} + (v_1^2 z_3)\frac{1}{2}+ (v_1^2 z_4)\frac{1}{2} + (w_2^2 z_9)\frac{1}{2} + w_1 w_2 \text{Re}[z_{14}] \nonumber  \\ &+ \frac{1}{2} v_1^2 \text{Re}[z_5] + 3 v_1 v_2 \text{Re}[z_7] ,  \nonumber\\
 &m_{a_2 a_2}\simeq m^2_{22} + (w_1^2 z_{12})\frac{1}{2} + (v_2^2 z_2)\frac{1}{2} + (v_1^2 z_3)\frac{1}{2}+ (v_1^2 z_4)\frac{1}{2} + (w_2^2 z_9)\frac{1}{2} + w_1 w_2 \text{Re}[z_{14}] - \frac{1}{2} v_1^2 \text{Re}[z_5]\nonumber \\ &+ v_1 v_2 \text{Re}[z_7],  \nonumber \\
 &m_{\phi'_1  \phi'_1}\simeq  m^2_{11} + (3 w_1^2 z_1)\frac{1}{2} + (v_2^2 z_{12})\frac{1}{2} + (w_2^2 z_3)\frac{1}{2}+ (w_2^2 z_4)\frac{1}{2} + (v_1^2 z_8)\frac{1}{2} + v_1 v_2 \text{Re}[z_{13}]  \nonumber \\ &+ \frac{1}{2} w_2^2 \text{Re}[z_5] + 3 w_1 w_2 \text{Re}[z_6],  \nonumber\\
 &m_{\phi'_2  \phi'_2}\simeq  m^2_{22} + (v_1^2 z_{12})\frac{1}{2} + (3 w_2^2 z_2)\frac{1}{2} + (w_1^2 z_3)\frac{1}{2}+ (w_1^2 z_4)\frac{1}{2} + (v_2^2 z_9)\frac{1}{2} + v_1 v_2 \text{Re}[z_{14}]\nonumber  \\&+ \frac{1}{2} w_1^2 \text{Re}[z_5] + 3 w_1 w_2 \text{Re}[z_7],  \nonumber \\
 &m_{ a_1  a_1}\simeq    m_{22} + (v_1^2 z_{12})\frac{1}{2} + (w_2^2 z_2)\frac{1}{2} + (w_1^2 z_3)\frac{1}{2}+ (w_1^2 z_4)\frac{1}{2} + (v_2^2 z_9)\frac{1}{2} + v_1 v_2 \text{Re}[z_{14}]\nonumber  \\ & - \frac{1}{2} w_1^2 \text{Re}[z_5] + w_1 w_2 \text{Re}[z_7], \nonumber \\
 &m_{\phi_1  \phi_2}\simeq    v_1 v_2 z_3 + v_1 v_2 z_4 + \text{Re}[Y_3] + \frac{1}{2} w_1 w_2 \text{Re}[z_{10}] + \frac{1}{2} w_1 w_2 \text{Re}[z_{11}]+ \frac{1}{2} w_1^2 \text{Re}[z_{13}] + \frac{1}{2} w_2^2 \text{Re}[z_{14}]  \nonumber  \\  &+ v_1 v_2 \text{Re}[z_5]+ \frac{3}{2} v_1^2 \text{Re}[z_6] + \frac{3}{2} v_2^2 \text{Re}[z_7], \nonumber \\
 &m_{\phi_1  a_2}\simeq  \text{Im}[m^2_{12}] + \frac{1}{2} w_1 w_2 \text{Im}[z_{10}] + \frac{1}{2} w_1 w_2 \text{Im}[z_{11}] + \frac{1}{2} w_1^2 \text{Im}[z_{13}]+ \frac{1}{2} w_2^2 \text{Im}[z_{14}] - v_1 v_2 \text{Im}[z_5]  \nonumber  \\ &- \frac{3}{2} v_1^2 \text{Im}[z_6]- \frac{1}{2} v_2^2 \text{Im}[z_7],  \nonumber \\
 &m_{\phi_1  \phi'_1}\simeq  v_1 w_1 z_8 + \frac{1}{2} v_2 w_2 \text{Re}[z_{10}] + \frac{1}{2} v_2 w_2 \text{Re}[z_{11}] + v_2 w_1 \text{Re}[z_{13}]  + v_1 w_2 \text{Re}[z_{13}],  \nonumber\\
 &m_{\phi_1  \phi'_2}\simeq  v_1 w_2 z_{12} + \frac{1}{2} v_2 w_1 \text{Re}[z_{10}] + \frac{1}{2} v_2 w_1 \text{Re}[z_{11}] + v_1 w_1 \text{Re}[z_{13}] + v_2 w_2 \text{Re}[z_{14}], \nonumber\\ 
 &m_{\phi_1  a_1}\simeq \frac{1}{2} v_2 w_1 \text{Im}[z_{10}] - \frac{1}{2} v_2 w_1 \text{Im}[z_{11}] + v_1 w_1 \text{Im}[z_{13}],  \nonumber \\
 &m_{\phi_2  a_2}\simeq -(\frac{1}{2}) v_1^2 \text{Im}[z_5] - v_1 v_2 \text{Im}[z_7], \\ 
 &m_{\phi_2 \phi'_1}\simeq   v_2 w_1 z_{12} + \frac{1}{2} v_1 w_2 \text{Re}[z_{10}] + \frac{1}{2} v_1 w_2 \text{Re}[z_{11}] + v_1 w_1 \text{Re}[z_{13}] + v_2 w_2 \text{Re}[z_{14}],  \nonumber\\
 &m_{\phi_2 \phi'_2}\simeq  v_2 w_2 z_9 + \frac{1}{2} v_1 w_1 \text{Re}[z_{10}] + \frac{1}{2} v_1 w_1 \text{Re}[z_{11}] + v_2 w_1 \text{Re}[z_{14}] + v_1 w_2 \text{Re}[z_{14}], \nonumber\\ 
 &m_{\phi_2  a_1 }\simeq  \frac{1}{2} v_1 w_1 \text{Im}[z_{10}] - \frac{1}{2} v_1 w_1 \text{Im}[z_{11}] + v_2 w_1 \text{Im}[z_{14}], \nonumber \\ 
 &m_{a_2 \phi'_1 }\simeq    \frac{1}{2} v_1 w_2 \text{Im}[z_{10}] + \frac{1}{2} v_1 w_2 \text{Im}[z_{11}] + v_1 w_1 \text{Im}[z_{13}], \nonumber \\ 
 &m_{a_2 \phi'_2 }\simeq  \frac{1}{2} v_1 w_1 \text{Im}[z_{10}] + \frac{1}{2} v_1 w_1 \text{Im}[z_{11}] + v_1 w_2 \text{Im}[z_{14}],  \nonumber \\
 &m_{a_2 a_1 }   \simeq   -(\frac{1}{2}) v_1 w_1 \text{Re}[z_{10}] + \frac{1}{2} v_1 w_1 \text{Re}[z_{11}], \nonumber  \\
 &m_{\phi'_1 \phi'_2}\simeq  w_1 w_2 z_3 + w_1 w_2 z_4 + \text{Re}[Y3] + \frac{1}{2} v_1 v_2 \text{Re}[z_{10}] + \frac{1}{2} v_1 v_2 \text{Re}[z_{11}] + \frac{1}{2} v_1^2 \text{Re}[z_{13}] + \frac{1}{2} v_2^2 \text{Re}[z_{14}]  \nonumber \\ &+ w_1 w_2 \text{Re}[z_5]+ \frac{3}{2} w_1^2 \text{Re}[z_6] + \frac{3}{2} w_2^2 \text{Re}[z_7], \nonumber \\
 &m_{\phi'_1  a_1}\simeq  \text{Im}[m^2_{12}] + \frac{1}{2} v_1 v_2 \text{Im}[z_{10}] - \frac{1}{2} v_1 v_2 \text{Im}[z_{11}] + \frac{1}{2} v_1^2 \text{Im}[z_{13}]+ \frac{1}{2} v_2^2 \text{Im}[z_{14}] - w_1 w_2 \text{Im}[z_5] \nonumber  \\ &- \frac{3}{2} w_1^2 \text{Im}[z_6] - \frac{1}{2} w_2^2 \text{Im}[z_7],  \nonumber \\
 &m_{\phi'_2  a_1}\simeq  -(\frac{1}{2}) w_1^2 \text{Im}[z_5] - w_1 w_2 \text{Im}[z_7].  \nonumber
 \end{align}
In the dual Higgs basis we can express a new set of six fields that are in terms of only v and w. We can then move to the mass eigenstate basis by rotating these fields or, working from the initial basis, rotate the $8 \times 8$ matrix including $G_1$ and $G_2$. 
This results in two zero eigenvalues for the solution in Tables II-III as expected. Either case yields the same physical mass eigenstates which we find numerically.
These solutions to the mass eigenstates, in terms of the parameters of the mirror symmetric basis, are given in Tables II-III for particular choices of the parameter space that produces the asymmetric VEV pattern.
        \newpage
\section*{Appendix B}
\setcounter{equation}{0}
\renewcommand{\theequation}{B\arabic{equation}}
 \begin{figure}[!ht]
                \begin{minipage}{\textwidth}
                \centering
                \includegraphics[width=.41\textwidth]{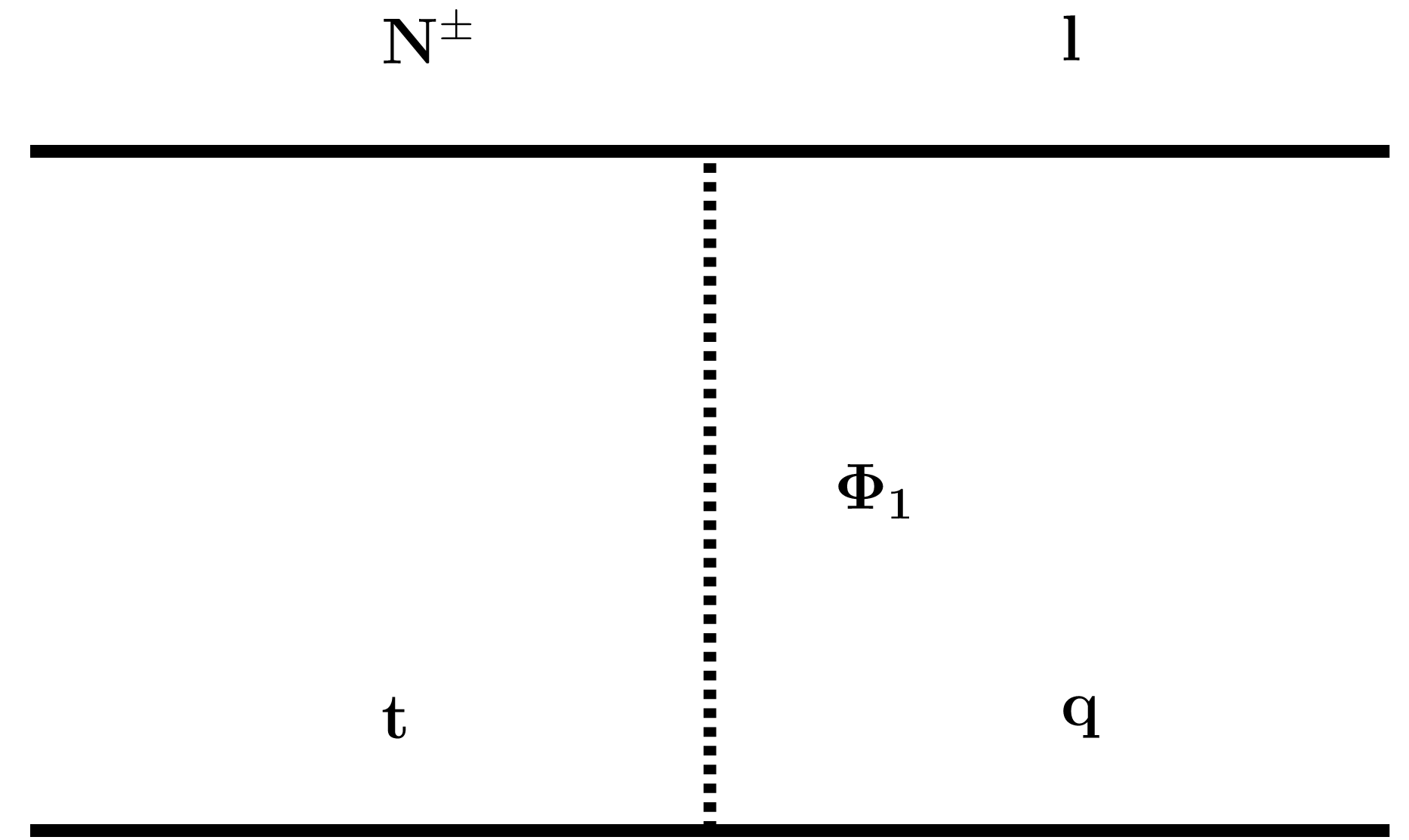}\hspace{1cm}
                            \includegraphics[width=.41\textwidth]{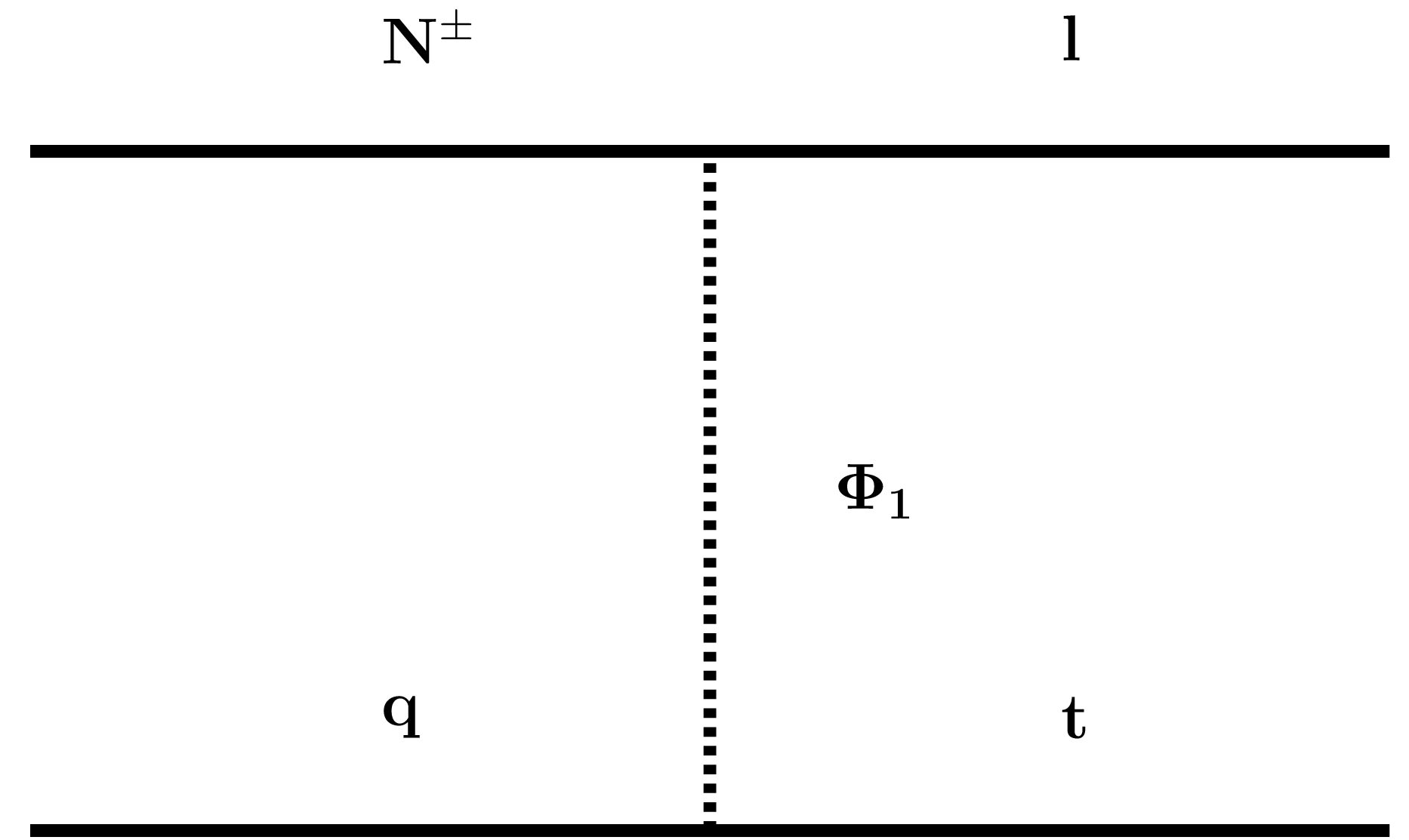}\\ \vspace{1cm}
                            \includegraphics[width=.41\textwidth]{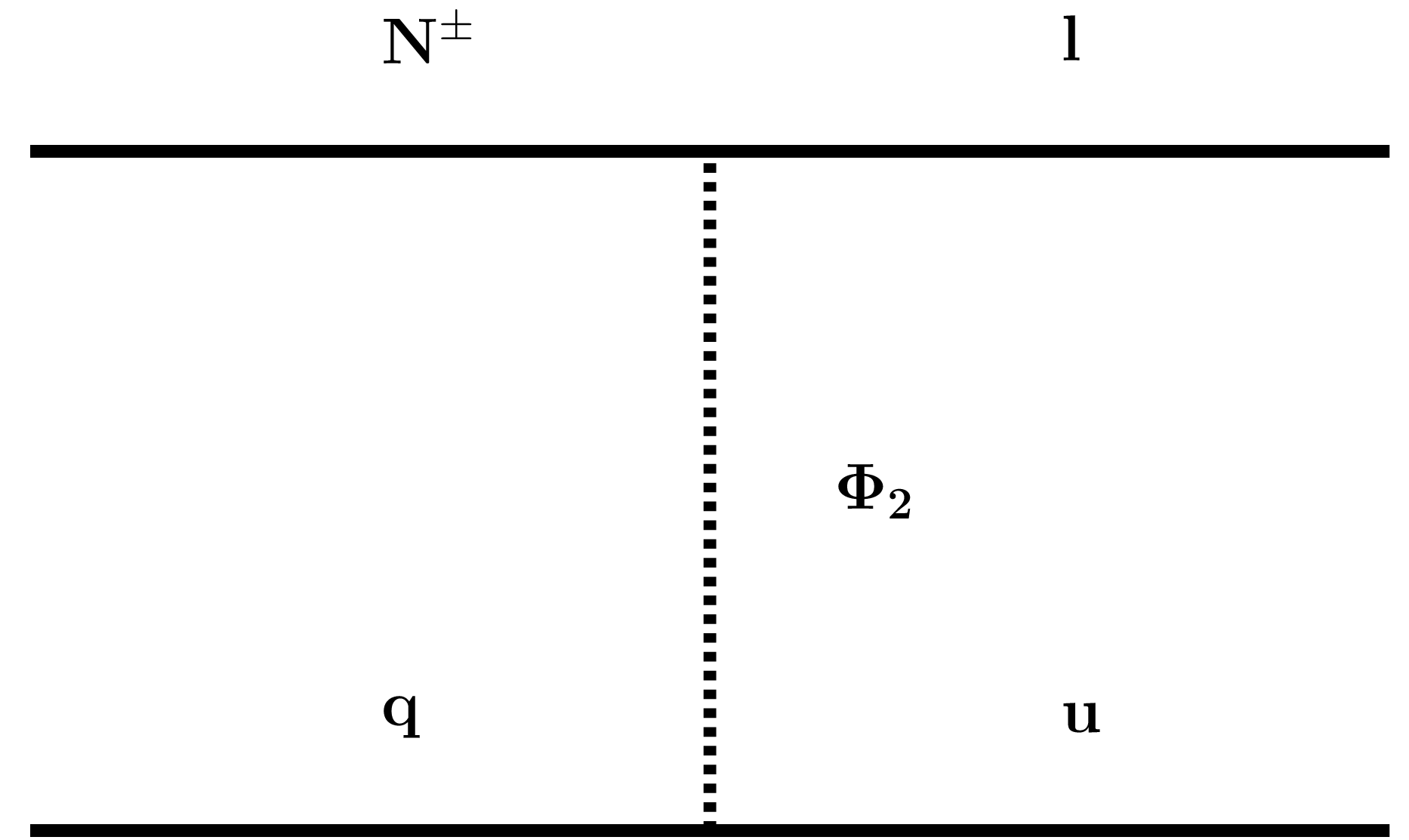}\hspace{1cm}
                            \includegraphics[width=.41\textwidth]{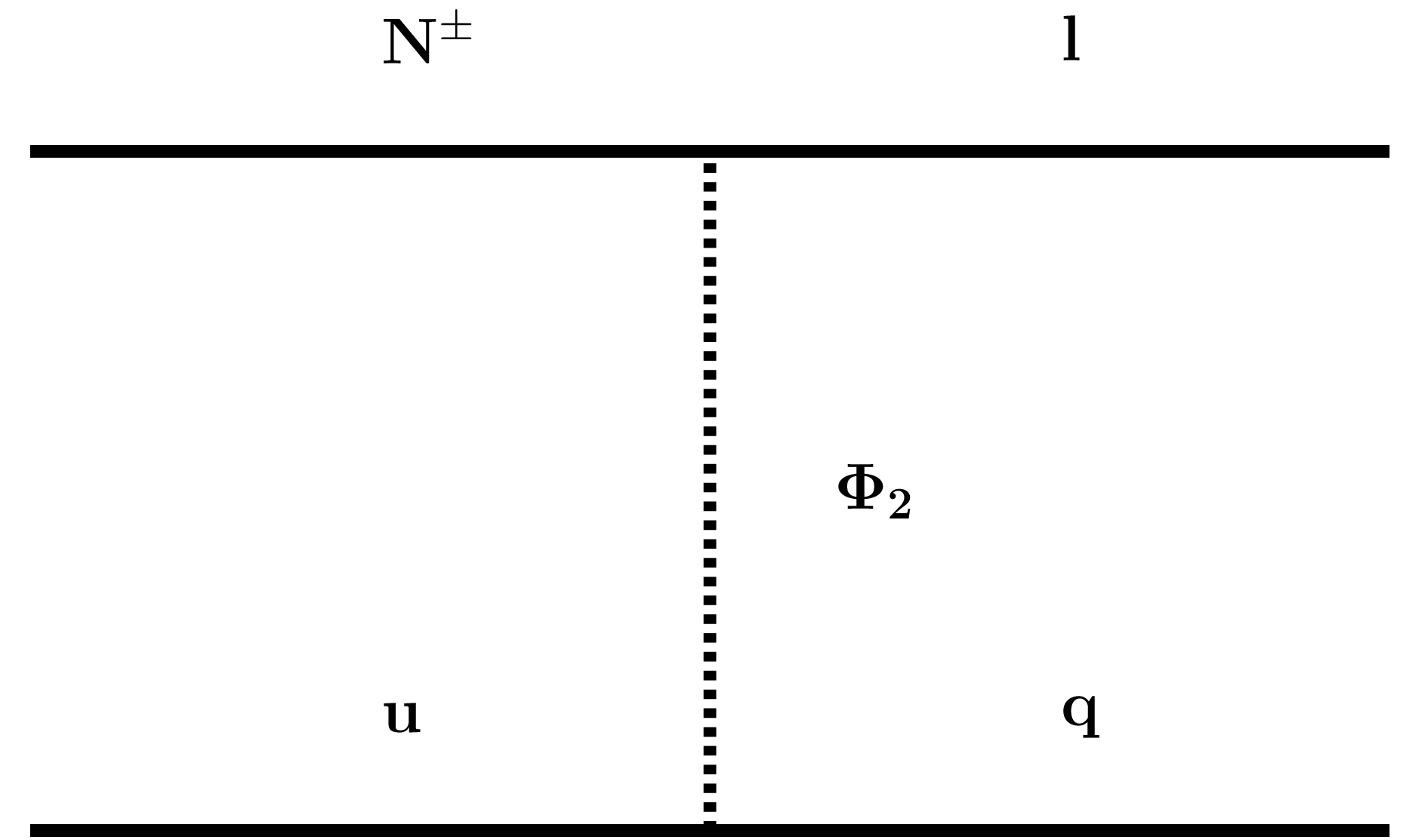}\\ \vspace{1cm}
                           \includegraphics[width=.41\textwidth]{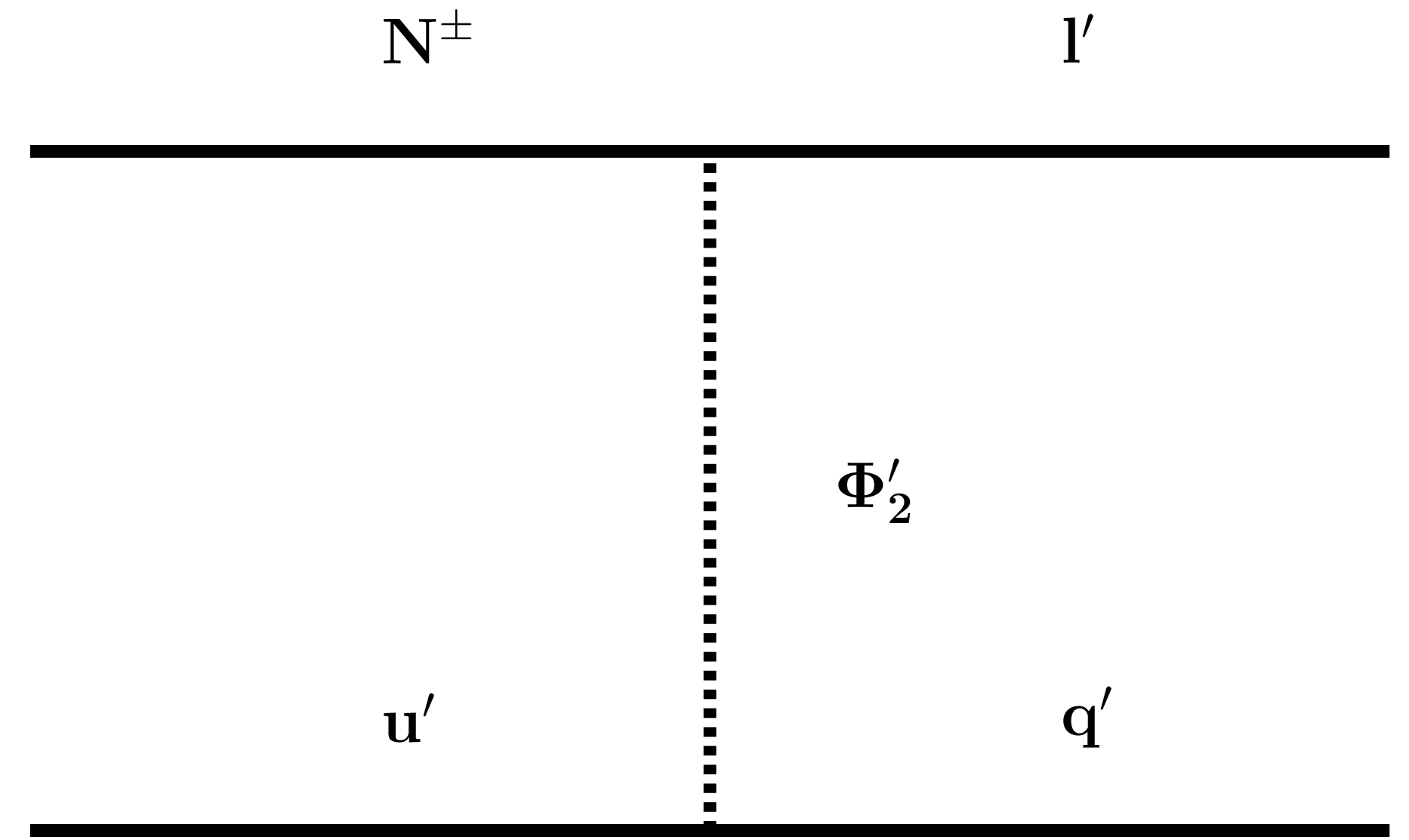}\hspace{1cm}
                            \includegraphics[width=.41\textwidth]{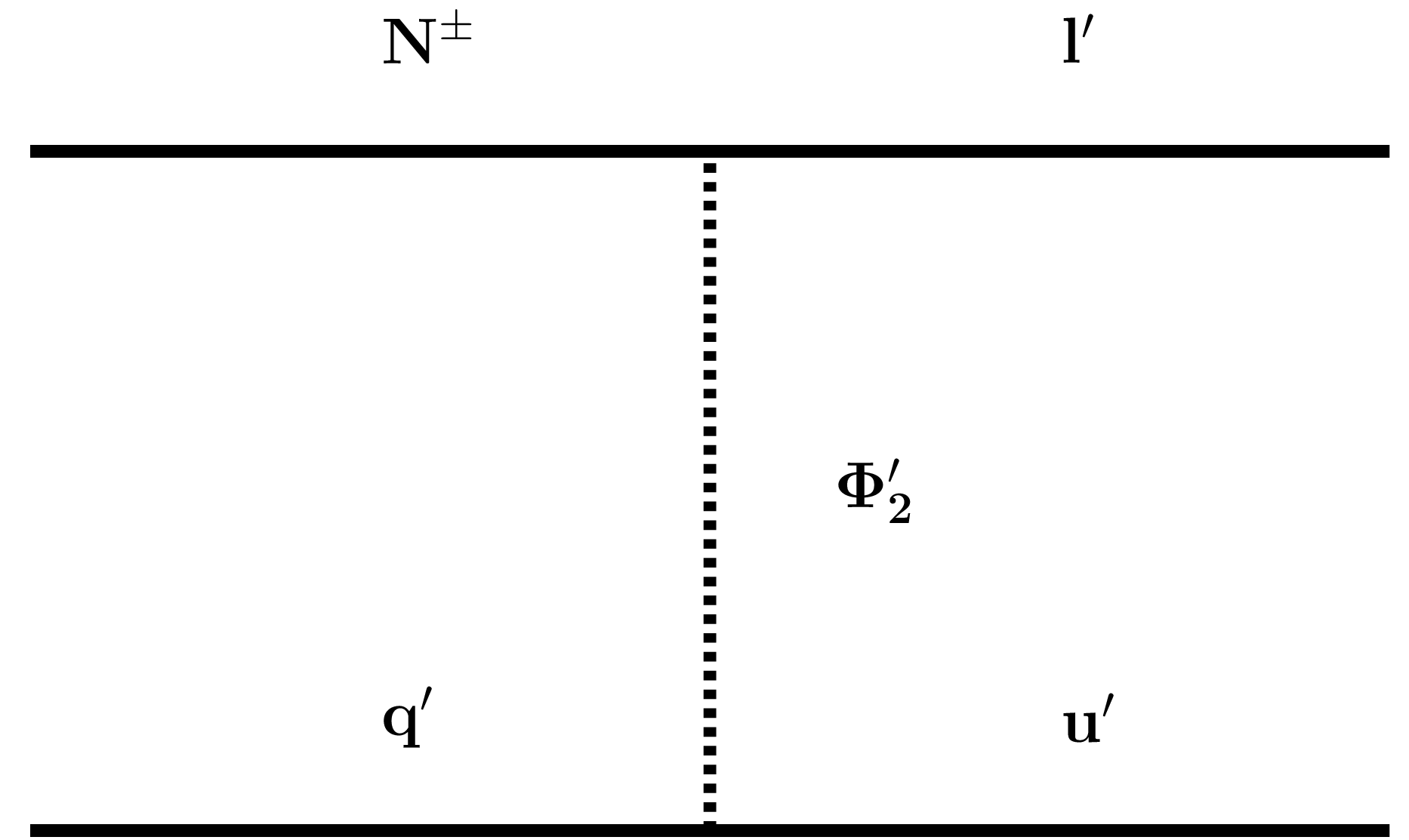}\\ \vspace{1cm}
                             \includegraphics[width=.41\textwidth]{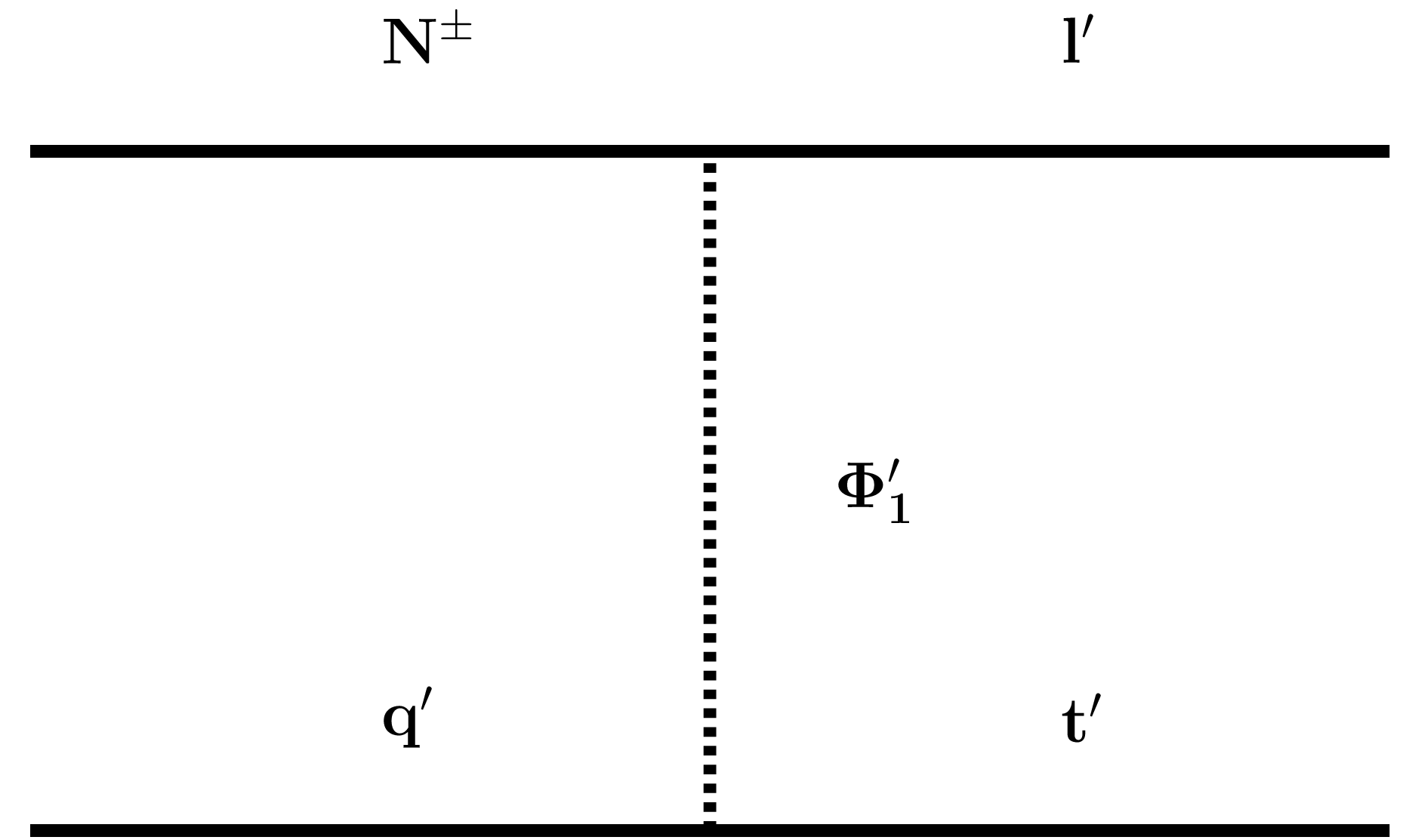}\hspace{1cm}
                            \includegraphics[width=.41\textwidth]{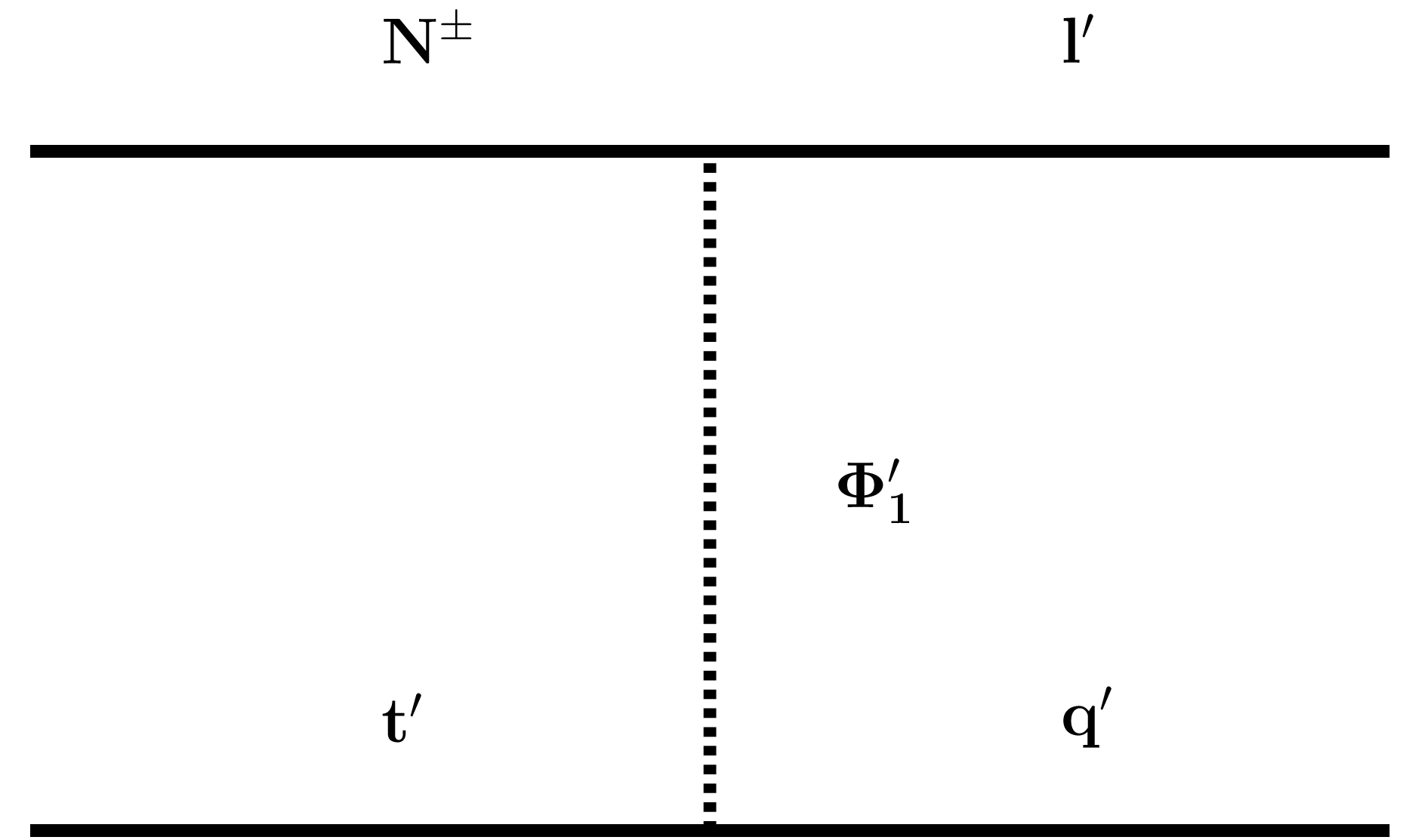}\\ \vspace{1cm}
                \caption{Scattering channels considered in the Boltzmann equations of Section 5.}
                \label{fig:ss1}
                \end{minipage}\\[1em]
                \end{figure}
                
                 \begin{figure}[!ht]
                \begin{minipage}{\textwidth}
                \centering
                      \includegraphics[width=.41\textwidth]{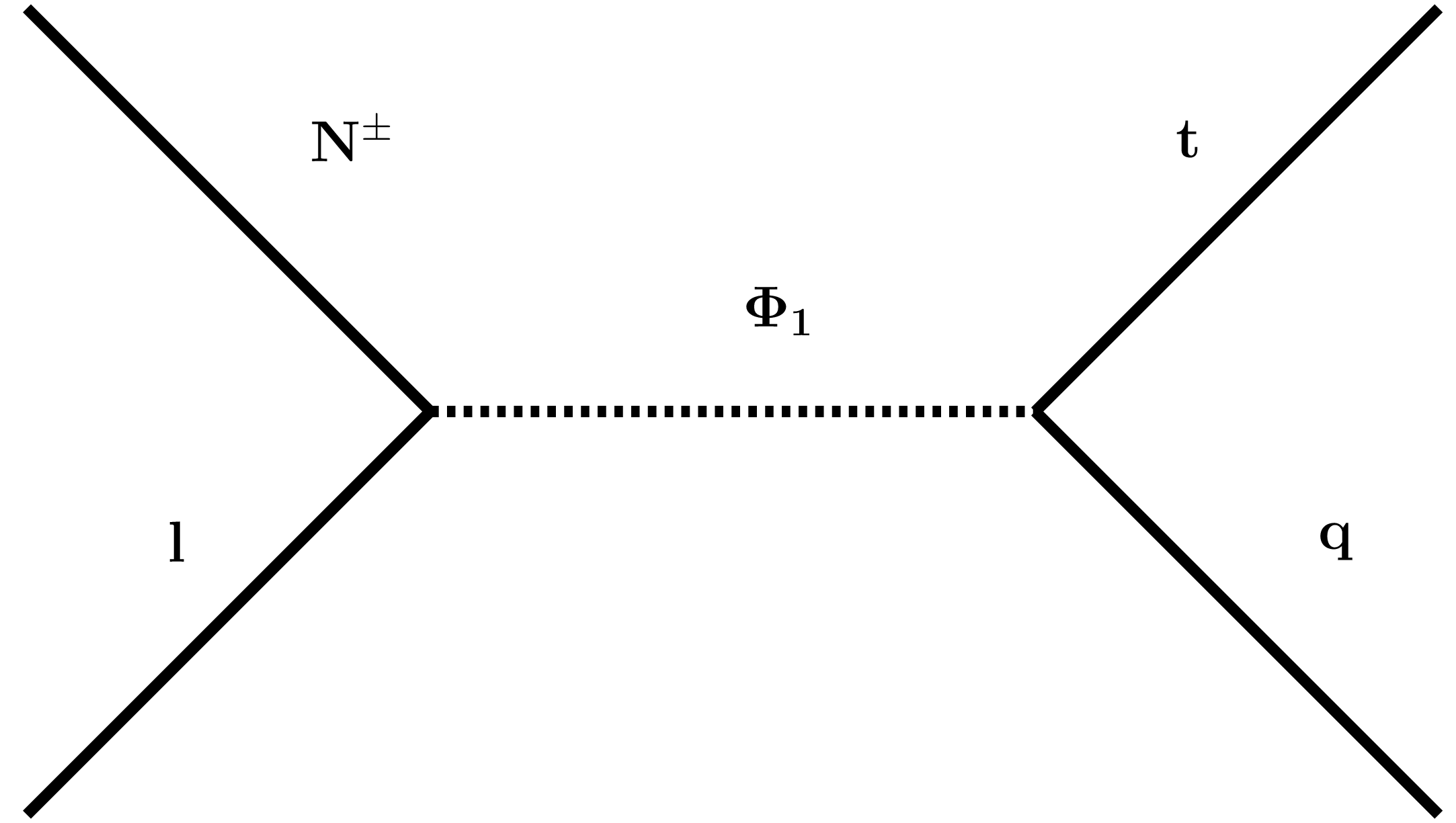}\hspace{1cm}
                        \includegraphics[width=.41\textwidth]{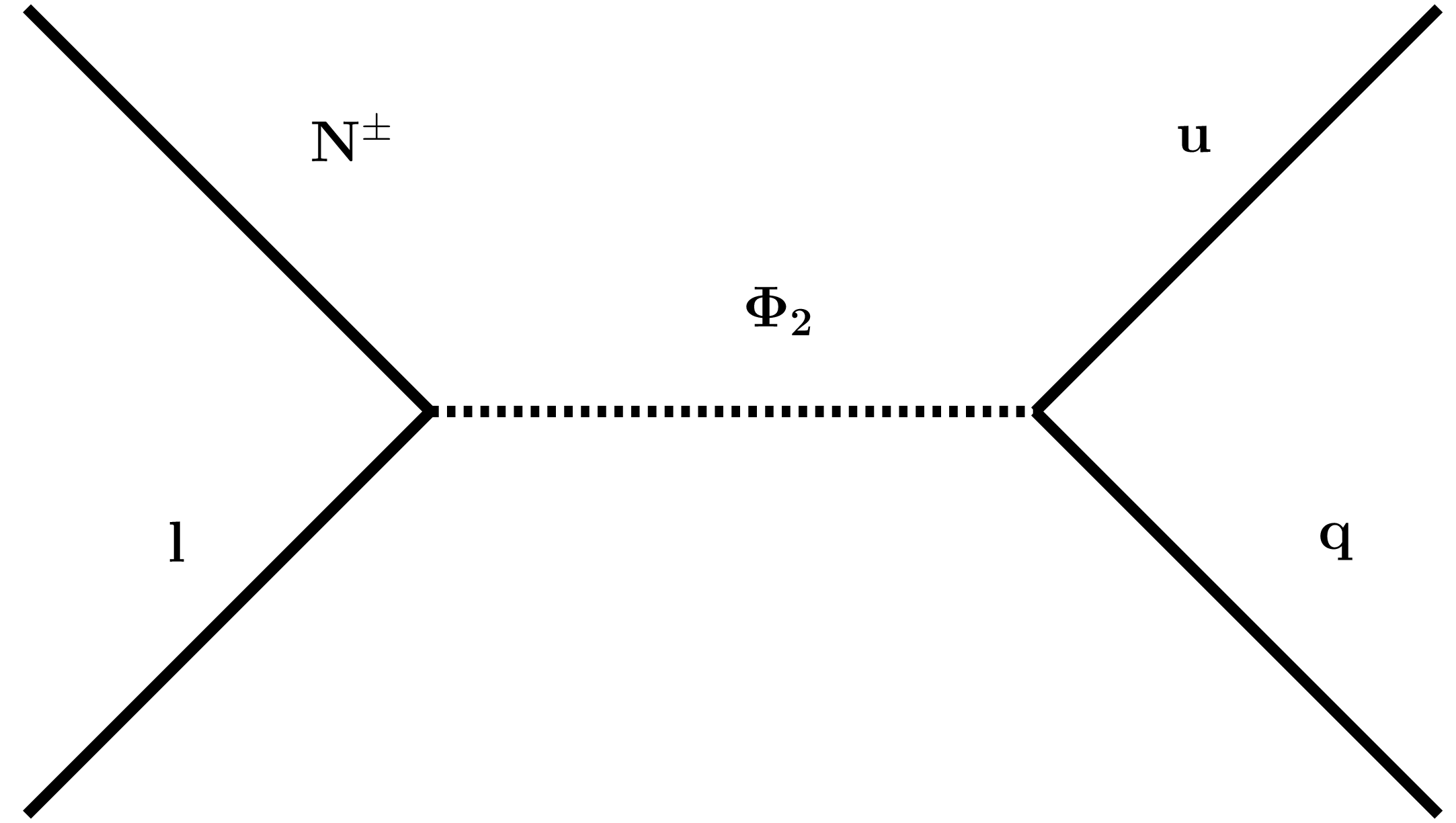}\\ \vspace{1cm}
                          \includegraphics[width=.41\textwidth]{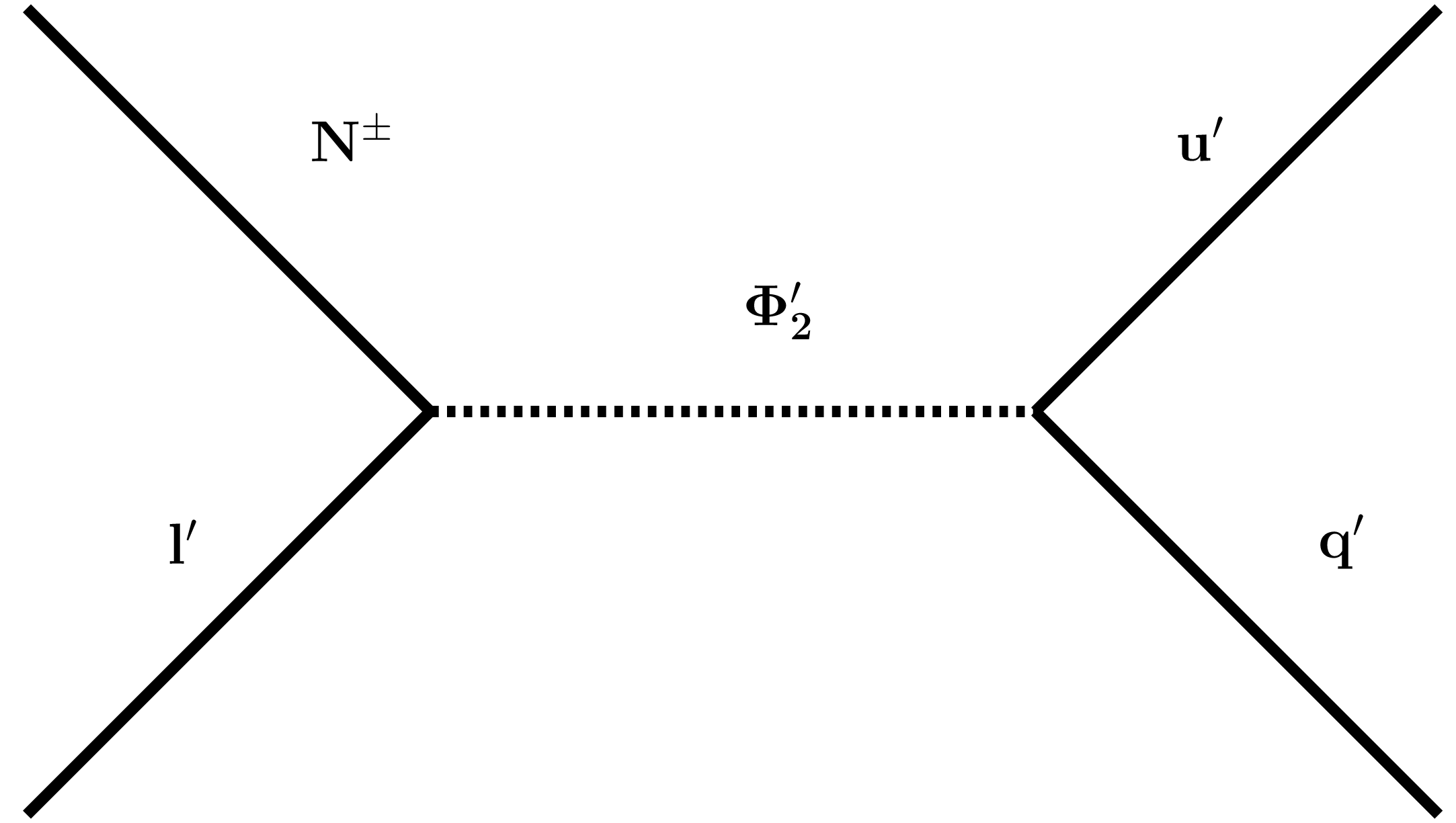}\hspace{1cm}
                            \includegraphics[width=.41\textwidth]{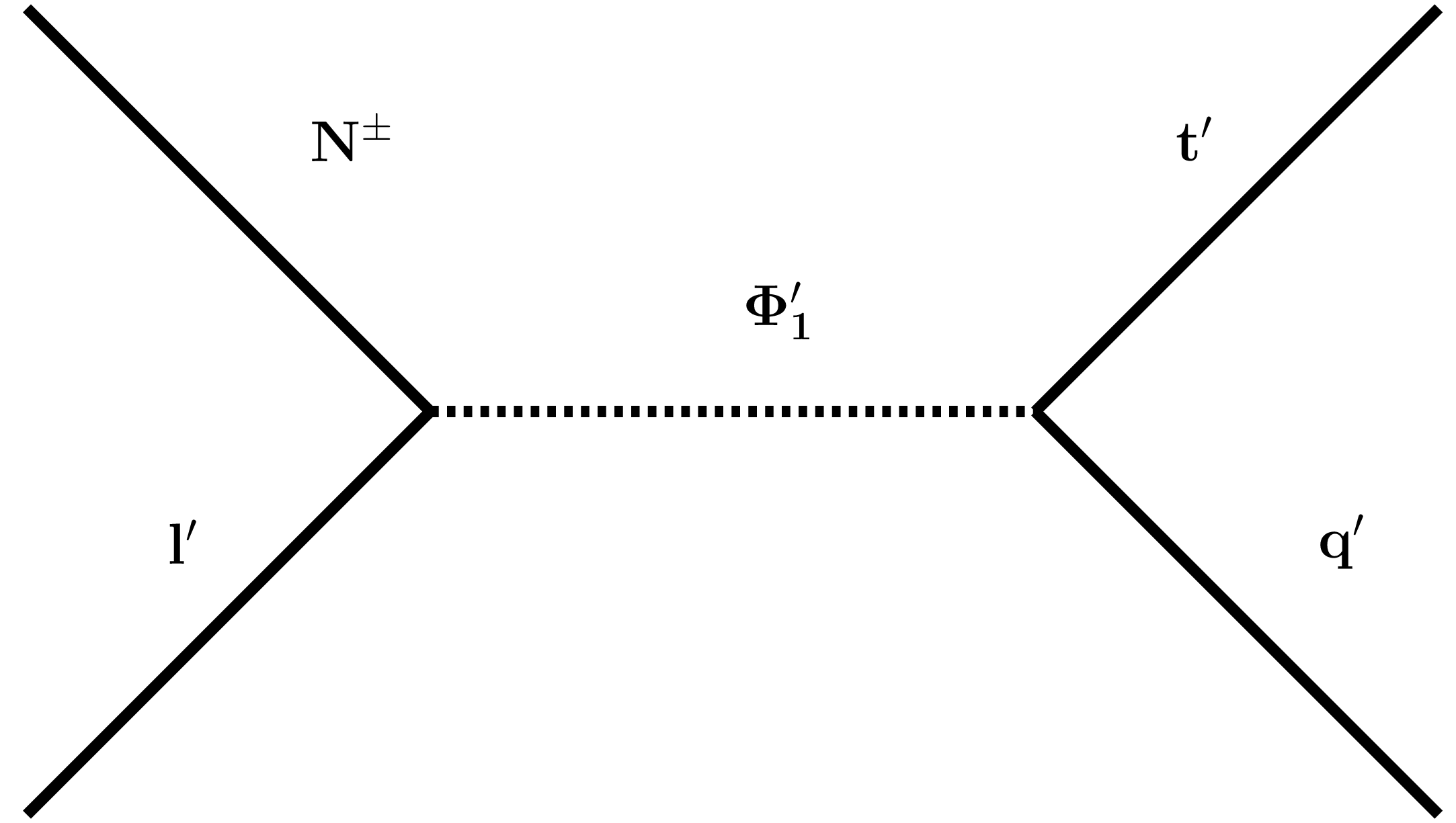}
                            \caption{Scattering channels considered in the Boltzmann equations of Section 5.}
                \label{fig:ss2}
                \end{minipage}\\[1em]
                \end{figure}


\bibliography{mainbib.bib}


\end{document}